\documentclass{bmcart}

\usepackage{amsthm,amsmath,amssymb,amsfonts}
\usepackage{mathtools}
\usepackage{bm}
\usepackage[utf8]{inputenc} 
\usepackage{graphicx}
\usepackage{tikz}
\usepackage{multirow}
\usepackage{algorithm}
\usepackage{algpseudocode}
\usepackage{hyperref}

\newlength\myindent
\newcommand{\dt}{\Delta t}

\makeatother
\newcommand{\T}{{\text{T}}}
\newcommand{\tv}[1]{\textbf{#1}}

\newcommand{\bg}[1]{\boldsymbol#1}

\newcommand{\RN}[1]{\textup{\uppercase\expandafter{\romannumeral#1}}}

\startlocaldefs
\endlocaldefs

\begin{document}

\begin{frontmatter}

\begin{fmbox}
\dochead{Research}


\title{A multiscale model of terrain dynamics for real-time earthmoving simulation}


\author[
   addressref={aff1,aff2},                   
   corref={aff1},                       
   email={martin.servin@umu.se}   
]{\inits{MS}\fnm{Martin} \snm{Servin}}
\author[
   addressref={aff2},
   email={tomas.berglund@algoryx.se}
]{\inits{TB}\fnm{Tomas} \snm{Berglund}}
\author[
]{\inits{SU}\fnm{Samuel} \snm{Nystedt}}

\address[id=aff1]{
  \orgname{Department of Physics, Ume\aa\ University}, 
  \postcode{SE-90187}                                
  \city{Ume\aa},                              
  \cny{Sweden}                                    
}
\address[id=aff2]{%
  \orgname{Algoryx Simulation AB},
  \street{Kuratorv\"agen 2B},
  \postcode{SE-90736}
  \city{Ume\aa},
  \cny{Sweden}
}


\begin{artnotes}
\end{artnotes}

\end{fmbox}


\begin{abstractbox}

\begin{abstract} 
A multiscale model for real-time simulation of terrain dynamics is explored.
To represent the dynamics on different scales the model combines the description of soil as a continuous solid, as distinct particles and as rigid multibodies.
The models are dynamically coupled to each other and to the earthmoving equipment.
Agitated soil is represented by a hybrid of contacting particles and continuum solid, with the moving equipment and resting soil as geometric boundaries.
Each zone of active soil is aggregated into distinct bodies, with the proper mass, momentum and frictional-cohesive properties,
which constrain the equipment's multibody dynamics.
The particle model parameters are pre-calibrated to the bulk mechanical parameters for a wide range of different soils.
The result is a computationally efficient model for earthmoving operations that resolve the motion of the soil, using a fast iterative solver, 
and provide realistic forces and dynamic for the equipment, using a direct solver for high numerical precision.
Numerical simulations of excavation and bulldozing operations are performed to test the model and measure the computational performance.
Reference data is produced using coupled discrete element and multibody dynamics simulations at relatively high resolution.
The digging resistance and soil displacements with the real-time multiscale model agree with the reference model up to 10-25\%, and run more than three orders of magnitude faster.
\end{abstract}




\end{abstractbox}
%

\end{frontmatter}



\section*{Introduction}


Physics-based simulation of earthmoving equipment and soil is an important tool for developing smarter systems that meet the increasing demands for energy efficiency, productivity and safety in the agriculture, construction, and mining industries.
Simulation with real-time performance is essential when developing new control systems, human-machine interfaces or training operators using interactive simulators with hardware or human in the loop.
Fast simulation is vital also for applying artificial intelligence to motion planning and control of earthmoving equipment.
These techniques are data-hungry, requiring many repeated simulations of a wide range of normal operating conditions as well as many, potentially hazardous, edge cases.
Although the simulations can be run mostly in parallel, real-time performance, or faster, is necessary for covering a sufficiently large parameter space within reasonable time.
Simulating heavy equipment alone at real-time with good accuracy is challenging but feasible using rigid multibody dynamics and a well-optimized solver \cite{bender2014,haug:1989:cak}. 
It may seem out of reach to include also the environment, involving vast amount of soil with complex dynamics on scales that cross six orders of magnitude \cite{Fang2016}. 

The first physics-based models for real-time simulation of heavy equipment and deformable terrain appeared in mid 1990, starting with Li and Moshell assuming the Mohr-Coulomb theory and volume preserving deformations \cite{Li1993}. 
Park \cite{Park2002} developed an elaborate model of the digging resistance in soil for the purpose of construction excavator simulators.
The model starts from the fundamental earthmoving equation (FEE) \cite{reece:1964:fee} for a blade cutting a horizontal soil bed, assuming an active zone in the shape of a wedge \cite{mckyes:1985:sct}.
The interface between the wedge and the passive soil represent the failure surface, stretching from the cutting edge of the blade to the free surface of the terrain.
Park extends the model to a 3D bucket digging in sloped terrain, including the formation of a secondary separation plate by the deadload of material in the bucket and penetration resistance from the tool's teeth.
The FEE-based models take into account the strength of the soil, expressed in terms of its internal friction and cohesion, but are limited to stationary conditions and do not describe the motion of the soil.
Holz et al \cite{Holz2009,Holz2013,Holz2015} combined the fundamental earthmoving equation (FEE) with contacting particle dynamics to model the cutting resistance and motion of the soil.
In this approach, static soil is adaptively converted into particles as the tool cut the terrain.
Digging resistance, based on the FEE, is applied as a six-degree of freedom kinematic constraint with force limits on the digging tool.
A wedge-shaped active zone is assumed.
The portion of terrain in the active zone that has not yet undergone failure and particle conversion provide mass to the FEE.
The particle contacts on top of the wedge contribute as surcharge mass, and the contacts with the tool provide additional digging resistance.
Later, a similar approach is taken in \cite{Jaiswal2019}.

Unfortunately, relying on the FEE for the digging resistance has serious drawbacks.
It is valid only at steady state and it suffers from unphysical singularities, as pointed out in \cite{Holz2015}.
Furthermore, the active zone possesses no inertia or momentum. 
When accelerated motion is involved, the FEE can clearly not provide correct dynamics and proper reaction forces.
On the other hand, resolving the active zone with contacting particles challenges the computational performance.
If the particles are not too many, real-time simulation is possible using nonsmooth contact dynamics \cite{renouf:2005:3df} or position-based dynamics \cite{Holz2014}, with a stationary iterative solver like the Gauss-Seidel algorithm.
This is associated with numerical errors that manifest as artificial elasticity or insufficient friction \cite{servin:2014:esn,wang:2016:wsp}.
If the errors become too large the soil will behave more like a compressive fluid rather than a stiff soil that yield only if the shear stress reach critical values
The errors grow with the number of particles and applied stress and decrease with the time-step and increased number of solver iterations.
In the other limit, of a few coarse particles, the solver error decrease but the spatial discretization errors grow large. 

To cope efficiently with the disparate length and time scales, we explore a multiscale model for the soil dynamics. 
The model targets real-time simulation of earthmoving equipment and terrain with realistic reaction forces and soil deformations.
A \emph{macroscale model}, describing the rigid multibody dynamics, is simulated using a direct solver for high numeric precision.
The active soil, resolved as particles in a \emph{mesoscale model}, is simulated using an iterative solver at high-speed with large error tolerance.
The key idea is that the soil dynamics is represented in \emph{both} models, with a coupling that filters out the discretization and solver errors from the mesoscale model but capture the bulk dynamics. 
When an earthmoving object come in contact with the terrain, the potential failure surfaces and zones of active soil are predicted. 
The soil inside the active zones is represented using a hybrid model of contacting particles and continuum solid, which support smooth transitioning between the resting solid, liquid and gas phases.
The coupling back to the macroscale model is mediated through \emph{aggregate bodies}.
They have the momentary shape, mass, and velocity of the resolved soil in each active zone.
The motion of the aggregates are constrained relative to the equipment and the terrain failure surface, in accordance with the Mohr-Coulomb model.
For blunt objects contacting the terrain, the subsoil stress distribution is estimated and the soil compacts if the stress reach critical values.
A \emph{microscale model}, simulated using relatively small particles and high numerical precision, is used for pre-calibration of the particle parameters to match the bulk mechanical properties of a wide range of soils.
The primary model parameters of the multiscale model are the bulk mechanical properties of the soil at a given bank state: the internal friction, cohesion, dilatancy, elasticity, and mass density; as well as the equipment's geometry, surface friction and cohesion.
To test the model, simulations of excavation and bulldozing operations in various soils are performed.
The digging resistance and soil displacements from the real-time multiscale model are compared with those from the microscale reference model.

\section*{Modelling and simulation of soil and earthmoving equipment}

\subsection*{Length- and timescales}
Soil and granular media are strongly dissipative multiphase materials with multiple length- and timescales \cite{Fang2016,andreotti:2013:gmb}.  
They consist of contacting grains with size ranging from clay at $10^{-6}$ m to cobble and boulders at $10^{0}$ m.
Natural soil has a certain moisture content, that may significantly increase the strength of the soil by cohesive forces.
At sufficient moisture levels, however, a pore pressure develops that lower the inter-particle normal forces and, consequently, the internal friction. 
Large deformations are commonly localized in shear bands, sometimes as narrow as a few particle diameters. 
The soil outside the shear bands is displaced rigidly.  
Soil is strongly dissipative. 
Therefore, the solid phase is the natural state.
If it is agitated, for instance by an earthmoving equipment, it may transition to the liquid or gaseous phases.
For sand and gravel in typical earthmoving operating conditions, the grain collision timescale is in the microsecond regime while the liquid timescale is around a millisecond \footnote{ The liquid phase time scale is the characteristic time scale of particle rearrangements, $d/\sqrt{p/\rho}$, with particle diameter $d$, mean stress $p$, and specific mass density $\rho$. 
 The time scale in the gaseous state is the characteristic collision time $4 [v^{-1} (5m/4k)^2 ]^{1/5}$, assuming a strongly damped Hertzian contact model \cite{Antypov2011}, for impact velocity $v$, particle mass $m$, and contact elasticity $k = \sqrt{d} E /2(1 - \nu^2)$, with Young's modulus $E$ and Poisson's ratio $\nu$.}.
%
%
Earthmoving equipment, on the other hand, has characteristic size of 10 m, operating range of around 100 m and may displace several cubic meters of soil per second.  
An earthmoving tool can be controlled with a spatial precision of about 10 mm and its geometric features is also on this scale.
A typical loading cycle, with a bucket excavator or a wheel loader, has time duration in the range
between 15-30 s \cite{Filla2011,Du2016}. 
The natural timescales of the rigid body motion is about 1-10 Hz and the control systems typically operate at a frequency below 100 Hz \cite{Fernando2019}.
These multiple and separated length- and timescales are important to consider when modelling soil and granular media interacting with earthmoving equipment.

\subsection*{Distinct particles}
The discrete element method (DEM) \cite{Cundall:1979:DNM} describe granular media and soil as consisting of distinct contacting particles of finite size.
It is a versatile model that automatically describe soil in the solid, liquid, and gaseous phases and the transitions between them.
DEM is clearly applicable for simulating the soil dynamics in earthmoving operations \cite{coetzee:2017:rcd,shmulevich:2007:ibs,obermayr:2014}.
It is, however, very computationally intensive.
A common solution is to not represent the true grains with their actual distribution of size, shape, and mechanical properties.  
Instead, the soil is represented by a collection of large, often spherical, \emph{pseudo particles} with contact parameters \-- elasticity, friction, cohesion and rolling resistance \-- that are calibrated such that the bulk mechanical properties match the ones observed in the soil that the model is meant to represent.
The mapping between the particle parameters and macroscopic soil parameters can be carried out using the triaxial test or cone penetration test, as described in \cite{wiberg:2020:dem}.
But even with pseudo-particles of size 10 to 100 mm, the number of particles for representing a terrain the size of earthmoving equipment exceeds what is currently possible to simulate in real-time using DEM.

\subsection*{Continuum}
On length scales larger than the particle size, soil may be modelled using continuum mechanics.  
The state of the soil is represented with scalar, vector and tensor fields for mass density $\rho(\tv{x},t)$, displacement $\bm{u}(\tv{x},t)$, velocity $\bm{v}(\tv{x},t)$, stress $\bm{\sigma}(\tv{x},t)$ and strain $\bm{\epsilon}(\tv{x},t)$.
The fields obey the equations of mass continuity
\begin{equation}\label{eq:mass_conservation}
    \partial_t \rho + \bm{\nabla}\cdot(\rho \bm{v}) = 0.
\end{equation}
and momentum balance
\begin{equation}\label{eq:mass_balance}
    \left[ \partial_t + \bm{v} \cdot \bm{\nabla} \right] \bm{v}  = \rho^{-1}\bm{\nabla} \cdot \bm{\sigma} + \bm{f}_\text{ext},
\end{equation}
where we use the short notation for partial time derivative $\partial_t = \frac{\partial}{\partial t}$, and $\bm{f}_\text{ext}$ denote any external force, like gravity.
In the dense regime, and when the stress is below a certain yield strength condition, the soil behaves as an elastic solid with some constitutive law relating stress to strain, e.g., Hooke's law for small deformations $\bm{\sigma}(\tv{x},t) = \bm{C}:\bm{\epsilon}(\tv{x},t)$.
For isotropic and homogenous materials the stiffness tensor $\bm{C}$ has only two independent parameters, often represented by the Young's modulus $E$ and the Poisson's ratio $\nu$.
When the stresses reach the yield condition the solid fails and deforms plastically, rupture or flows rapidly.
The simplest yield condition for soil is the Mohr-Coulomb criteria.  
It predicts that the material will fail along any plane with normal $\bm{n}$ where the shear stress, $\tau_{\bm{n}} = \sqrt{ (\bg{\sigma}\cdot\bm{n})^2 - \sigma_{\bm{n}}^2}$, reaches the critial value of
\begin{equation}\label{eq:mohr_coulomb}
	\tau_{\bm{n}} =  \mu \sigma_{\bm{n}} + c ,
\end{equation}
where the normal stress is $\sigma_{\bm{n}} = \sigma_{\alpha\beta} n_\beta$.  
The model has two parameters for the strength of the material: the internal friction, $\mu $, and the cohesion, $c$.  
The internal friction is often represented by the angle of internal friction $\phi = \arctan(\mu)$. 
Analogously with Coulomb friction, the critical shear strength grows linearly with the normal stress (pressure).  
The shear stress must also overcome any cohesive strength of the material, which is independent of the stress and strain in the Mohr-Coulomb model.
If the material yields quasistatically it may be modelled as an elastoplastic solid, with a plastic flow rule that accounts for strain-hardening (or softening) and the volume expansion during shear.
The latter is known as dilatancy and is defined as $\text{tr}(\dot{\bm{\epsilon}})/3\lVert\dot{\bar{\bm{\epsilon}}}\rVert$, where $\bar{\bm{\epsilon}}$ is the deviatoric strain.
The hardening can be incorporated in the Mohr-Coulomb law by \cite{andreotti:2013:gmb,Wood1990} by an effective internal friction 
\begin{equation}\label{eq:effective_internal_friction}
	\mu = \tan(\phi + \psi)
\end{equation}
where $\psi = \arcsin\left[\text{tr}(\bm{\epsilon})/3\lVert\bar{\bm{\epsilon}}\rVert\right]$ is the dilatancy angle. 
Soil may compact plastically under uniform compression if the applied stress is large enough to cause particle rearrangements.
This is captured by a soil's compression index
\begin{equation}\label{eq:compression_index}
    C_\text{c} \equiv \frac{ \Delta e}{\Delta\ln(\sigma)}
\end{equation}
which measure the change in void ratio, $e$, by a change in confining stress.

The finite element method is the most widely used and versatile technique for simulating deformable solids. 
The popular integration schemes for elastoplastic solids \cite{Neto2008} first compute trial strain and stress fields, and then use Newton's method for searching the plastic strain increment that fulfill the constitutive model and plastic flow rule at each time-step. 
In the regime of loose soil and large shear rate, the material is better described as a viscoplastic solid or a non-Newtonian fluid with some constitutive law between the stress and strain-rate, like the $\mu(I)$-rheology for granular media \cite{Forterre2008}.
Cemented soil, with clay or silt particles, may fracture by brittle failure.
Mesh-based numerical methods for solid dynamics have various difficulties with large deformations and topological changes associated with soil that undergo plastic or brittle failure, and transitions between the dense solid, dilute liquid and gaseous phases.
Meshfree methods, such as the material point method, are showing promising results on 3D problems, coupled with multibody dynamics, but real-time simulation of earthmoving operations remain out of reach \cite{Daviet2016,Hu2018}.

\begin{figure}[h]
    \centering
    \includegraphics[width=0.9\textwidth]{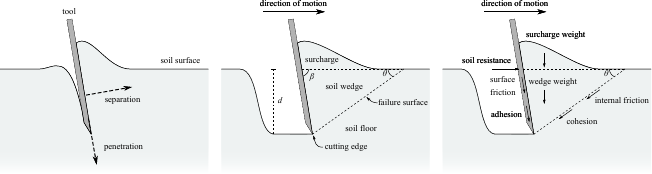}
    \caption{Illustration of a blade interacting with a soil bed. There are two modes of operation, penetration, and separation (left).
    The failure surface form a wedge-shaped active zone (middle). When a blade cut the soil there are many forces contributing to the soil resistance (right).}
    \label{fig:soil_tool_interaction}
\end{figure}

Analytical and semi-analytical solutions, derived from the continuum theory, are valuable for both insight and for creating fast simulation models. 
This include the Boussinesq type-of-equations for the stress and deformation fields underneath a load applied on the surface of a semi-infinite domain.
The reaction forces on a thin blade cutting a soil bed has been thoroughly analysed \cite{mckyes:1985:sct}.
A blade, or \emph{separating plate}, has two basic modes of operation, \emph{penetration} and \emph{separation}.  
Penetration is the motion straight into the soil with relative velocity in tangential direction of the plate only.  
Soil cavities at the tip of the blade, often equipped with teeth, are thus forced to expand \cite{Park2002,Yu1991}.  The penetration resistance may be considerable although the deformations are relatively small.
Separation corresponds to movement normal to the plate and is the main cause of soil failure and large displacements.  The edge where the blade meet the material is referred to as the \emph{cutting edge}.  See Fig.~\ref{fig:soil_tool_interaction}
The shape of the failure surface can be computed analytically in the two dimensional case, applicable for a wide blade, using the method of stress characteristics and assuming the Mohr-Coulomb criteria.  The failure surface is often approximated with a plane.  This defines an \emph{active zone} with the shape of a wedge.  Rankine's theory for a flat soil with a blade pushing on it in the horizontal direction predicts that the soil fails at an angle $\theta = \pi / 4 - \phi / 2$ against the horizontal.  In three dimensions, the failure surface extends sideways also, which cause long berms along the sides of a pushing blade.
The separation force acting on a blade moving horizontally at a constant speed is well described by the Fundamental Earthmoving Equation (FEE) \cite{reece:1964:fee}.  The FEE is motivated by wedge model of the soil failure.  The force resistance per tool width $L$, in the FEE is composed of four terms
\begin{equation}\label{eq:FEE}
    \frac{F}{L} = \rho g d^2 N_\gamma + Q N_Q +  c d N_\text{c} + c_\text{a} d N_\text{a},
\end{equation}
with specific soil mass density $\rho$, tool penetration depth $d$, soil cohesion $c$, surcharge force $Q$ (per tool width) and soil-tool adhesion $c_\text{a}$.  
The first term is due to the weight of the wedge, the second term is the additional (vertical) surcharge, the third term the cohesive force in the failure surface, and the fourth term is the resistance due the adhesion between the blade and the soil, see Fig.~\ref{fig:soil_tool_interaction}.  
The four $N$-factors (found in \cite{reece:1964:fee,mckyes:1985:sct}) depend on the geometry of the tool and the failure zone, the internal friction and cohesion, as well as of the surface friction and adhesion.  
The quadratic dependency on the cutting depth $d$ reflect that the weight depends linearly on the cross-section area of the failure zone.  
Note that the cohesive and adhesive force terms are proportional to the area of the failure surface and blade contact surface, respectively. 
One key limitation of the FEE is that it assumes steady state and low speed of the blade and soil flow.
Also, the $N$-factors suffer from singularities at certain geoemtric configurations.

\subsection*{Contacting rigid multibodies}
The earthmoving equipment can be simulated efficiently using rigid multibody dynamics \cite{bender2014,haug:1989:cak}.  
Articulated and actuated mechanisms are thus modelled as rigid bodies with kinematic constraints that represent the mechanical joints, motors, and driveline. 
Dynamic contacts between the bodies are best modelled as kinematic constraints and complementarity conditions to express unilaterality, impacts and Coulomb friction.
In the current paper, we apply rigid multibody dynamics with contact dynamics for both the equipment and the soil, and in all the levels of the multiscale model.  
Therefore, we describe the computational framework at greater level of detail.

The state of a rigid multibody system with $N_\text{b}$ bodies, $N_\text{j}$ joints and actuators and $N_\text{c}$ contacts, is represented on descriptor form in terms of the position, $\bm{x}(t) \in \mathbb{R}^{6N_\text{b}}$, velocity, $\bm{v}(t)\in \mathbb{R}^{6N_\text{b}}$, and Lagrange multipliers, $\bm{\lambda}_{\mathrm{j}}(t)\in \mathbb{R}^{6N_\text{j}}$ and $\bm{\lambda}_{\mathrm{c}}(t)\in \mathbb{R}^{6N_\text{c}}$, that are responsible for the constraint forces in joints and contacts.  
The system position variable is a concatenation of the spatial and rotational coordinates of the $N_\text{b}$ bodies, $\bm{x} = [\tv{x}, \tv{e}]$, and the velocity variable holds the linear and angular velocities, $\bm{v} = [\tv{v}, \bg{\omega}]$.  
The time evolution of the  system state variables $[\bm{x}, \bm{v}, \bm{\lambda}$] is given by the following set of equations
\begin{align}
    \label{mdb:eqOfm1}
    & \bm{M}\dot{\bm{v}} = \bm{f}_{\mathrm{ext}} + 
        \bm{G}_{\mathrm{j}}^\mathrm{T} \bm{\lambda}_{\mathrm{j}} +
        \bm{G}_{\mathrm{c}}^\mathrm{T} \bm{\lambda}_{\mathrm{c}}\\
    \label{mdb:eqOfm2}
    & \varepsilon_{\mathrm{j}} \bm{\lambda}_{\mathrm{j}} + 
        \eta_{\mathrm{j}} \bm{g}_{\mathrm{j}} + 
        \tau_{\mathrm{j}} \bm{G}_{\mathrm{j}} \bm{v} = \bm{u}_{\mathrm{j}}, \\
    \label{mdb:eqOfm3}
    & \mathtt{contact\_law}(\bm{v},\bm{\lambda}_{\mathrm{c}},\bm{g}_{\mathrm{c}}, \bm{G}_{\mathrm{c}}),
\end{align}
where $\bm{M} \in \mathbb{R}^{6N_\text{b} \times 6N_\text{b}}$ is the system mass matrix, $\bm{f}_{\mathrm{ext}}$ is the external force, and $\bm{G}_{\mathrm{j}}^\mathrm{T} \bm{\lambda}_{\mathrm{j}}$ and $\bm{G}_{\mathrm{c}}^\mathrm{T} \bm{\lambda}_{\mathrm{c}}$ are constraint forces for joints (and motors) and for contacts, respectively.
The forces have dimension $\mathbb{R}^{6N_\text{b}}$ and is composed of linear force and torque.  
Eq. \eqref{mdb:eqOfm2} is a generic constraint equation, with constraint function $\bm{g}_{\mathrm{j}}(\bm{x})$, Jacobian $\bm{G} = \partial\bm{g}/\partial\bm{x}$, compliance $\varepsilon_{\mathrm{j}}$ and viscous damping rate $ \tau_{\mathrm{j}} $.
An ideal joint is represented with $\varepsilon_{\mathrm{j}} = \tau_{\mathrm{j}} = \bm{u}_{\mathrm{j}} = 0$, in which case Eq.~(\ref{mdb:eqOfm2}) express a holonomic constraint, $\bm{g}_{\mathrm{j}}(\bm{x}) = 0$.  
A linear or angular motor may be represented by a velocity constraint (setting $\varepsilon_{\mathrm{j}} = \eta_{\mathrm{j}} = 0$ and $ \tau_{\mathrm{j}} = 1 $), $ \bm{G}_{\mathrm{j}} \bm{v} = \bm{u}_{\mathrm{j}}(t)$, with set speed $\bm{u}_{\mathrm{j}}(t)$.  
The holonomic and nonholonomic constraints can be seen as the limit of a stiff potential, $\mathcal{U}_\varepsilon = \tfrac{1}{2\varepsilon}\bm{g}^T\bm{g}$, or a Rayleigh dissipation function, $\mathcal{R}_\tau = \tfrac{1}{2\tau}(\bm{G}\bm{v})^T\bm{G}\bm{v}$, respectively \cite{lacoursiere:2007:rsv}.  This offer the possibility of mapping known models of viscoelasticity to the compliant constraints.  Descriptor form means that no coordinate reduction is made.  The system is represented explicitly with its full degrees of freedom, although with the presence of constraints.

We consider the system to have \emph{nonsmooth dynamics} \cite{acary:2008:nmn}.  That means that the velocity and Lagrange multipliers are allowed to be time-discontinuous, reflecting instantaneous changes from impacts, frictional stick-slip transitions or joints and actuators reaching their limits.  This is unavoidable when using an implicit integration scheme\footnote{The alternative is to resolve the contact events using smooth trajectories, stiff potentials and small time-step explicit time integration.} because of the coupling between the state variables trough unilateral and frictional contacts.

As contact law between particles we use a model that include cohesive-viscoelastic normal contacts (n), tangential Coulomb friction (t) and rolling resistance (r).  These are formulated in terms of inequality and complementarity conditions for the velocities, Lagrange multipliers and constraint functions.  The resulting model can be seen as a time-implicit version of conventional DEM and is therefore referred to as \emph{nonsmooth DEM} \cite{radjai:2011:dem,servin:2014:esn}.  We use the following conditions on $\bm{v}$ and $\bm{\lambda}_\text{c} = [\lambda_\text{n},\bm{\lambda}_\text{t},\bm{\lambda}_\text{r}]$ as $\mathtt{contact\_law}$:
\begin{align}
    \label{eq:theory:mB:signCoul1}
    &0 \leq \varepsilon_{\mathrm{n}} \lambda_{\mathrm{n}} + g_{\mathrm{n}}  +
    \tau_{\mathrm{n}} \bm{G}_{\mathrm{n}} \bm{v} \perp ( \lambda_{\mathrm{n}} + f_{\mathrm{c}} )\geq 0 , \quad
    f_{\mathrm{c}} \equiv c_{\mathrm{p}} A_{\mathrm{p}} / \lVert\bm{G}_{\mathrm{n}}^{\mathrm{T}}\rVert \\
    \label{eq:theory:mB:signCoul2}
    &\gamma_{\mathrm{t}} \bm{\lambda}_{\mathrm{t}} + \bm{G}_{\mathrm{t}} \bm{v} = 0, \quad
    \lVert\bm{\lambda}_{\mathrm{t}}\rVert \leq \mu_{\mathrm{t}} \lVert\bm{G}_{\mathrm{n}}^{\mathrm{T}} \lambda_{\mathrm{n}}\rVert \\
    \label{eq:theory:mB:signCoul3}
    &\gamma_{\mathrm{r}} \bm{\lambda}_{\mathrm{r}} + \bm{G}_{\mathrm{r}} \bm{v} = 0, \quad
    \lVert\bm{\lambda}_{\mathrm{r}}\rVert \leq r \mu_{\mathrm{r}} \lVert\bm{G}_{\mathrm{n}}^{\mathrm{T}} \lambda_{\mathrm{n}}\rVert,
\end{align}
where $\bm{g}_{\mathrm{n}}$ is a function of the contact overlap and the Jacobians, $\bm{G}_{\mathrm{n}}$, $\bm{G}_{\mathrm{t}}$ and $\bm{G}_{\mathrm{r}}$, are the normal, tangent and rotational directions of the contact forces \cite{servin:2014:esn}.  
The parameters $\varepsilon_{\mathrm{n}}$, $\tau_{\mathrm{n}}$, $\gamma_{\mathrm{t}}$ in Eq.~(\ref{eq:theory:mB:signCoul1}) control the contact stiffness and damping, and $f_{\mathrm{c}}$ the cohesion.
Setting these parameters to zero means that no penetration should occur, $\bm{g}_{\mathrm{n}}(\bm x) \geq 0$, and if so the normal force should be repulsive, $\bm{\lambda}_{\mathrm{n}} \geq 0$.
The inclusion of $ f_{\mathrm{c}}$ enables cohesive normal force with maximum value $c_{\mbox{\tiny{P}}} A_{\mbox{\tiny{P}}}$, where $c_{\mbox{\tiny{P}}}$ is the particle cohesion and $A_{\mbox{\tiny{P}}}$ is the particle cross section area.  
The cohesion is active when the contact overlap is smaller than a certain \emph{cohesive overlap}, set to a fraction of the particle size, e.g., $\delta_\mathrm{c} = 0.025 d$.
Eq.~(\ref{eq:theory:mB:signCoul2}) state that contacts should have zero slide velocity, $\bm{G}_{\mathrm{t}} \bm{v} = 0$, giving rise to a friction force that is bounded by the Coulomb friction law with friction coefficient $\mu_{\mathrm{t}}$.  
Similarly, Eq. \eqref{eq:theory:mB:signCoul3} states that, as long as the constraint torque is no greater than the rolling resistance law, the contacting bodies are constrained to zero relative rotational motion, $\bm{G}_\text{r} \bm{v} = 0$. 
Here, $\mu_{\mathrm{r}}$ is the rolling resistance coefficient and $r$ is the particle radius.  
It is a well-known fact that the effect of particle angularity, on internal friction and angle of repose, can be captured using spherical particles with rolling resistance.
As explained in \cite{Estrada:2011:IRR}, a $n$-sided polygon can be assigned a rolling resistance coefficient $\mu_\text{r} = (1/4) \tan (\pi/2 n)$, which gives
$\mu_\text{r} = 0.05$ for an eight-sided polygon,  $\mu_\text{r} = 0.1$ for a square and $\mu_\text{r} = 0$ for a sphere ($n = \infty$).  
We map the normal contact law, Eq.~(\ref{eq:theory:mB:signCoul1}), to the  Hertz-Mindlin model for contacting viscoelastic spheres, 
$
    \bm f_{\mathrm{n}} = 
        k_{\mathrm{n}} \delta^{3/2} \bm n + 
        k_{\mathrm{n}} c_\text{d} \delta^{1/2}\dot{\delta} \bm n
$,
where $\delta(\bm{x})$ is the contact overlap, $k_{\mathrm{n}} = \tfrac{1}{3} E^*
\sqrt{d^*}$ is the contact stiffness, $c_\text{d}$ is a damping coefficient,
$E^* = [(1 - \nu^2_{a})/E_a + (1 -
\nu^2_{b})/E_b]^{-1}$ is the effective Young's modulus, $d^* = (d^{-1}_a + d^{-1}_b)^{-1}$ is the effective diameter for two contacting spheres, $a$ and
$b$, with Young's modulus $E_a$, diameter $d_a$ and Poisson's ratio $\nu_a$ etc.
The mapping to Eq.~(\ref{eq:theory:mB:signCoul1}) is accomplished by $g_\text{n} = \delta^{5/4}$, $\varepsilon_\text{n} = 5/4 k_\text{n}$ and $\tau_\text{n} =5 c_\text{d}/4$.

We separate collisions into resting contacts and impacts using an impact threshold velocity $v_{\mathrm{imp}}$.  If the relative contact velocity is smaller than this value the contacts are modelled as described above.  In case of impacts we apply the Newton impact law	$\bm{G}_{\mathrm{n}} \bm{v}^+ = - e \bm{G}_{\mathrm{n}} \bm{v}^-$
with restitution coefficient $e$, while preserving all other constraints in the system on the velocity level, $\bm{G} \bm{v}^+ = 0$. This is carried out in an impact stage solve, prior to the main solve for the constrained equations of motions (\ref{mdb:eqOfm1})-(\ref{mdb:eqOfm3}).  With this division, the restitution coefficient become the key parameter for modelling the dissipative part of the normal force.  For the resting contacts we can simply enforce numerical stability using $\tau_\text{n} = 2 \dt$ with little consequence of the damping being artificially strong \cite{Wang2015a}.

For numerical integration we employ the SPOOK stepper \cite{lacoursiere:2007:rsv}.
It is a first order accurate discrete variational integrator, developed particularly for fixed time-step real-time simulation of multibody systems with non-ideal constraints and non-smooth dynamics.  It has been proven to be linearly stable and found stable on contact problems in empricial studies \cite{lacoursiere2011}.    
The numerical time integration scheme for advancing the system's position and velocity from $[\bm{x}_n, \bm{v}_n]$ at time $t_{n}$ to $[\bm{x}_{n+1}, \bm{v}_{n+1}]$ at time $t_{n+1} = t_n + \dt$ consist of a position update $\bm{x}_{n+ 1} = \bm{x}_{n} + \dt \bm{v}_{n + 1}$
after having computed the new velocity $\bm{x}_{n+ 1}$ and corresponding Lagrange multiplier $\bm{\lambda}$.  
This is done by solving the following mixed complementarity problem (MCP) \cite{murty:1988:lcl}
\begin{align}
    \left[
		\begin{array}
	      		[c]{ccc}%
	      		\bm{M} & - \bm{G}_n & - \bar{\bm{G}}_n\\
	      		\bm{G}_n & \bg{\Sigma} & 0 \\
	      		\bar{\bm{G}}_n & 0 & \bar{\bg{\Sigma}}  \\
	    \end{array}
        \right]
        & 
		\begin{bmatrix} 
		{\bm{v}_{n+1}} \\ \bm{\lambda}_{} \\ \bar{\bm{\lambda}}_{}
		\end{bmatrix}
         -
        \begin{bmatrix} 
            \bm{p}_{n} \\ \bm{q}_n \\ \bm{u}_n
            \end{bmatrix}
        = \bm{w}_{l} - \bm{w}_{u}, \\
        0 \leq \bm{z} - \bm{l} & \perp \bm{w}_{l}\geq 0\\
        0 \leq \bm{u} - \bm{z} & \perp \bm{w}_{u}\geq 0
\end{align}
where the constraints have been grouped into position constraints (no bar) and velocity constraints (with bar), 
$\bm{p}_{n} = \bm{M}\bm{v}_{n} + \dt \bm{f}_n$, $\bm{q}_n =  - \tfrac{4}{\dt} \Upsilon\bm{g} + \Upsilon\bm{G}\bm{v}_{n}$, and $\bm{u}_n$ is the target speeds of the velocity constraints (zero for frictional contacts).
The regularization and stabilization terms are $\bg{\Sigma} = \tfrac{4}{\dt^2}\text{diag}[ \varepsilon_i/(1 + 4\tau/\dt) ]$, $\bar{\bg{\Sigma}} = \tfrac{1}{\dt}\text{diag}\left( \gamma \right)$ and $\bg{\Upsilon} = \text{diag}[ 1/(1 + 4\tau/\dt) ]$.
The slack variables $\bm{w}_{l}$ and $\bm{w}_{u}$ are only used internally by the MCP solver.
For details about the solver, see Sec.~Implementation. 

\subsection*{Coarse-graining}
The process of averaging particle kinematics and contacts forces into continuous and differentiable fields is referred to as homogenization. \emph{Coarse-graining} is one particular way of doing homogenization.
This is useful when combining particle and continuum models of granular media. 
The fields are computed by sampling the particle variables using a smoothing kernel, $\zeta(\tv{x})$, that is normalized $\int \zeta(\bm{x}) \mathrm{d}\bm{x}^3 = 1$ and approaches zero on a smoothing length $R$.
The fields of mass and momentum density are computed $ \rho(\bm{x},t) = \sum_a m^a \zeta(\bm{x} - \tv{x}^a(t))$ and $\bm{p}(\bm{x},t) = \sum_a m^a \tv{v}^a \zeta(\bm{x} - \tv{x}^a(t))$, respectively, and the velocity field is simply 
\begin{equation}\label{eq:cg_velocity}
    \bm{u}(\bm{x},t) = \bm{p}(\bm{x},t)/\rho(\bm{x},t).
\end{equation}

The rate of strain tensor can thus be computed as $\dot{\varepsilon}_{\alpha\beta}(\bm{x},t) 
\equiv \frac{1}{2} \left( \frac{\partial u_\alpha}{\partial x_\beta} + \frac{\partial u_\alpha}{\partial x_\beta}\right)$.
The stress tensor is the sum of the kinetic stress $\sigma_{\alpha\beta}^\text{k}(\bm{x},t) = -\sum_a m^a u^a_\alpha v^a_\beta \phi(\bm{x} - \bm{x}^a(t))$ and the contact stress $\sigma_{\alpha\beta}^\text{c}(\bm{x},t) = -\sum_k f^{ab}_{\alpha,k} x^{ab}_\beta \int_0^1 \phi(\bm{x} - \tv{x}^a(t) + s\tv{x}^{ab}(t) )\mathrm{d}s$, where the summation is over the set of contacts, $\tv{f}^{ab}_k$ is the contact force between particle $a$ and $b$ with branch vector $\tv{x}^{ab} = \bm{x}^{b} - \bm{x}^{a}$.
Different smoothing kernels can be used for different purposes.
The Gaussian kernel, $\zeta(\tv{x}) = (\sqrt{2\pi}R)^{-3} \exp (-\lVert\bm{x}\rVert^2/2R^2)$, make the fields differentiable.
The Heaviside function is faster to evaluate and can be used for a discrete representation of the fields that preserve mass precisely.
From analytical studies and numerical experiments with contacting elastic spheres \cite{andreotti:2013:gmb}, it is found that the elastic bulk modulus is non-linear and change with bulk volume as $\partial \sigma /\partial \varepsilon = Z E \sqrt{\Delta V / V}$, where $Z$ is the average number of contacts per grain. 
This suggest an effective Young's modulus of the form
\begin{equation}\label{eq:effective_elasticity}
    E = E_\text{0} \left( 1 \pm \left\lVert \rho/\rho_\text{0} - 1\right\rVert^{1/2} \right),
\end{equation}
for a small change in density $\rho/\rho_0 = 1 + \Delta V/V$ relative to a reference state with mass density $\rho_0$ and Young's modulus $E_0$. 
The sign $\pm$ is positive for compaction and negative for expansion.
The dilatancy angle also increase with the level of compaction, and with it the internal friction by Eq.~(\ref{eq:effective_internal_friction}). 
Based on numerical simulations and coarse-graining of dense granular media, Roux and Radjai \cite{Roux1998} proposed 
\begin{equation} \label{eq:dilatancy}
    \psi = c_\varphi (\varphi - \varphi_\text{c}),
\end{equation} 
where $\varphi$ and $\varphi_\text{c}$ are the current and critical porosity, at which the soil switch between positive dilation (volume expansion) and negative dilation (volume shrinkage) upon shearing, and $c_\varphi$ is a constant that depends on the particle shape.

\section*{A multiscale model of terrain dynamics}
\begin{figure}[h]
    \includegraphics[width=0.95\columnwidth]{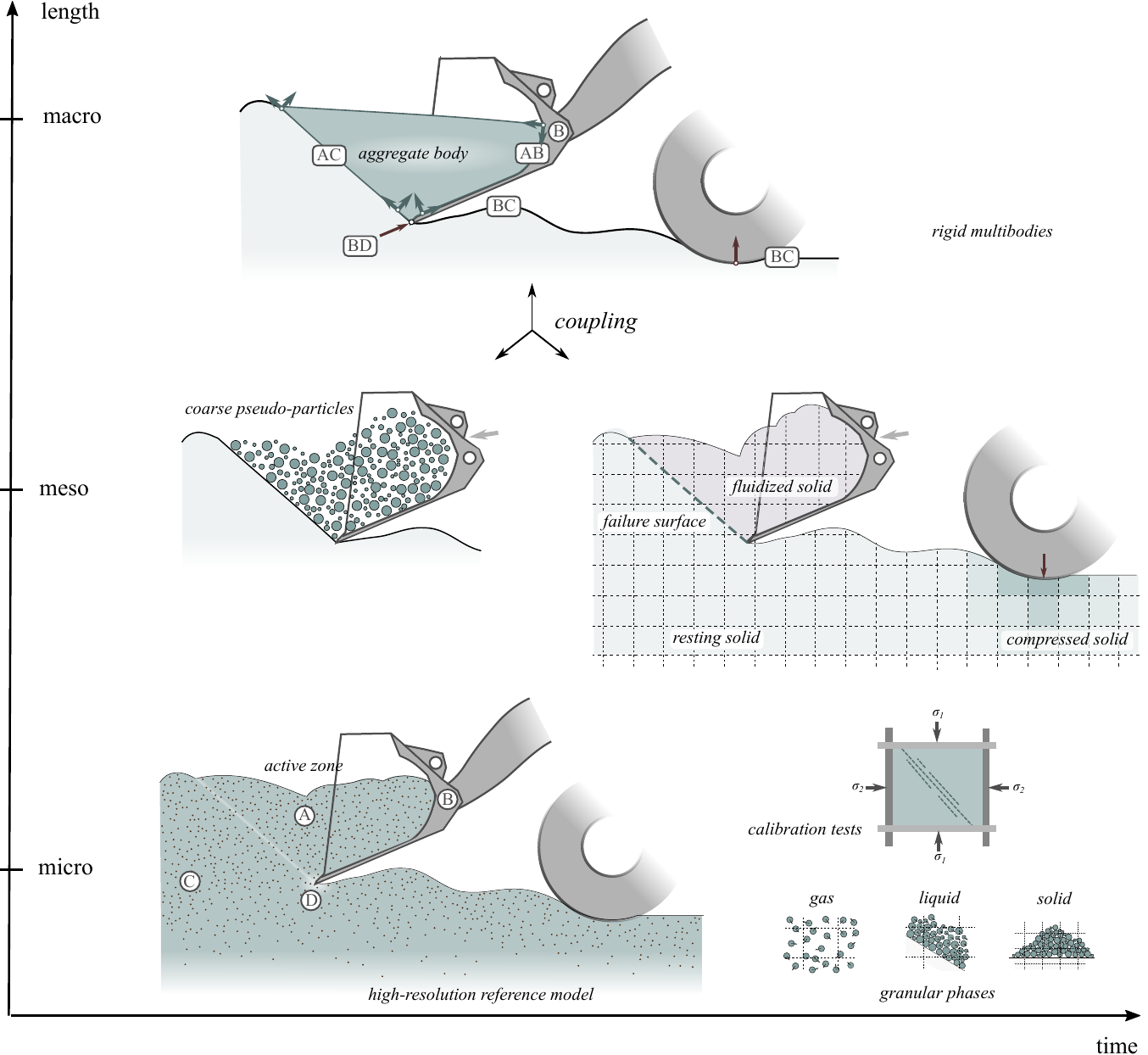}
    \caption{Overview if the multiscale model.
        The microscale model is simulated off-line for validation and calibration of the mesoscale and macroscale models, that are simulated in real-time coupled to each other. 
        The mesoscale model combines a coarse particle and continuum representation of the soil inside the active zone.
        The active soil is aggregated into a single body with frictional-cohesive couplings with the resting terrain and the rigid multibody equipment.}
        \label{fig:overview}
\end{figure}

The multiscale model has three levels of abstraction, illustrated in Fig.~\ref{fig:overview}, that we refer to as \emph{micro-}, \emph{meso-} and \emph{macroscale}.
In the microscopic model the terrain is fully resolved by relatively fine-grained particles with contact properties pre-calibrated to represent various type of soil, e.g., dirt, gravel, or sand.
It serves as a ground-truth reference model, simulated off-line in advance with relatively small time-step and high numerical accuracy. 
The output is used for validation of the coarser scale models and, if necessary, for calibration of parameters not known from theory or experiments.
The meso- and macroscale models are simulated in real-time and synchronously coupled to each other.
The terrain deformations and soil flow is integrated using the mesoscale model, with a hybrid representation of the soil that combine coarse particles and fields discretised with a regular grid of voxels. 
Input to the mesoscale model is the motion and contact forces of the equipment at the interfaces to the terrain. 
This is provided by the macroscale model, which focus on the rigid multibody dynamics of the equipment, and any other objects interacting with the terrain. 
The equipment experience the resting terrain as a quasistatic surface and each region of agitated soil as distinct dynamic bodies, whose shape, mass velocity and mechanical strength is aggregated from the mesoscale model.

It is a significant feature that each of the three models use the same parameters to characterize the soil properties. 
These are mapped to the model parameters for the particle, voxel, and aggregate body dynamics.
The parameter mapping relies in part on the theory of continuum soil mechanics and in part on parameter calibration using the microscale model.
The primary soil parameters include mass density $\rho_\text{b}$, internal friction angle $\phi_\text{b}$, cohesion $c_\text{b}$, dilatancy angle $\psi_\text{b}$, bulk elasticity modulus $E_\text{b}$, that characterise the mechanics of the soil as found in a natural \emph{bank state}.
These are collected in a bulk parameter vector $\bm{p}_\text{b}  =  [\rho_\text{b}, \phi_\text{b},  c_\text{b}, \psi_\text{b}, E_\text{b}] $.
The bank state refers to the state at which the true soil is found in nature.
It may be the result of geological, meteorological and machined processes, that leave the soil at a particular packing density and moisture content.
Two soils with identical (real) particles may thus have different bank state properties and are considered as two distinct soil with different bulk parameters $\bm{p}_\text{b}$.

\subsection*{Microscopic model}
At highest resolution we represent the terrain as a set of relatively fine-grain frictional-cohesive particles, $\mathcal{N}_{\mbox{\tiny{P}}}$, and the equipment as a set of rigid multibodies, $\mathcal{N}_{\mbox{\tiny{RB}}}$, with some set of joints and actuators. 
Let $\mathcal{N}^{\mbox{\tiny{B}}}_{\mbox{\tiny{P}}}$ denote the set of particles in direct contact with a body $B \in \mathcal{N}_{\mbox{\tiny{RB}}}$ of the equipment.
The equations of motion for the earthmoving body and the terrain particles, labelled $a$, are
\begin{align}
  \label{mdb:eqOfma}
  \bm{m}_a\dot{\bm{v}}_a & = \bm{f}^{\mathrm{ext}}_a + 
  \bm{G}_{a\mbox{\tiny{B}}}^\T \bm{\lambda}_{a\mbox{\tiny{B}}} + \sum_{a'\in \mathcal{N}^{\mbox{\tiny{a}}}_{\mbox{\tiny{P}}}} \bm{G}_{aa'}^\T \bm{\lambda}_{\mathrm{aa'}},\\
  \label{mdb:eqOfmB}
  \bm{m}_{\mbox{\tiny{B}}}\dot{\bm{v}}_{\mbox{\tiny{B}}} & = \bm{f}^{\mathrm{ext}}_{\mbox{\tiny{B}}} + 
      \bm{G}_{\mbox{\tiny{B}}}^\T \bm{\lambda}_{\mbox{\tiny{B}}} + \sum_{a\in \mathcal{N}^{\mbox{\tiny{B}}}_{\mbox{\tiny{P}}}} \bm{G}_{a\mbox{\tiny{B}}}^\T \bm{\lambda}_{a\mbox{\tiny{B}}},
    \end{align}
where $\bm{G}_{\mbox{\tiny{B}}}^\T \bm{\lambda}_{\mbox{\tiny{B}}}$ is the constraint force coupling the earthmoving body to the rest of the equipment, $\bm{G}_{\mbox{\tiny{B}}a}^\T \bm{\lambda}_{a\mbox{\tiny{B}}}$ is the contact force on the body from particle $a$, and $\bm{G}_{aa'}^\T \bm{\lambda}_{\mathrm{aa'}}$ is the inter-particle contact force between particle $a$ and $a'$.
The body-particle interfacial force is the source for the digging resistance perceived by the equipment. The interfacial stress alters also the internal stress in the soil, causing shear failure or brittle rupture if it reach the critical stress.

The particles can roughly be divided into two domains. Either they are part of the active zone, $\mathcal{N}^{A}_{\mbox{\tiny{P}}}$, which is sheared or rigidly displaced by the earthmoving body, or they remain part of the resting bed of particles, $\mathcal{N}^{C}_{\mbox{\tiny{P}}}$.  
The domains are separated by a failure surface $AC$.
If the earthmoving body is a tool with penetrating teeth or a sharp edge, the particles in direct contact with that constitute a third domain, $\mathcal{N}^{D}_{\mbox{\tiny{P}}}$.  The particle domains are illustrated in Fig.~\ref{fig:overview}.

We use a \emph{pseudo-particle} representation of the soil.  That means that the particles do not represent the true grains with their actual distribution of size, shape, and mechanical properties.  Instead, the soil is represented by a collection of relatively large spherical particles with specific mass density and contact parameters \-- elasticity, friction, cohesion and rolling resistance \-- $\bm{p}_{\text{p}} = [ \rho_\text{p}, \mu_\text{t}, \mu_\text{r},  c_\text{p},  E_\text{p} ]$.
These parameters are calibrated to numerical values that give the particle soil the desired bulk mechanical properties $\bm{p}_\text{b}$. 
The size of the pseudo-particles is chosen small enough to resolve the important features of the equipment and precision at which it can be controlled, i.e., around $10 \-- 50$ mm in earthmoving applications.

\subsection*{Mesoscale model}
The mesoscale model is a medium-resolution model of the soil dynamics using a hybrid particle-continuum approach.  
The soil in the active zone is primarily represented by coarse particles.
The continuous soil model has two phases, resting \emph{solid mass} and \emph{fluidized mass}.
The former represents resting soil outside the active zone and is considered an elastoplastic solid.
The latter complement the use of particles for representing soil displacement in the active zone. The fluidized mass is convected with the coarse-grained velocity field of the particles, and subject to gravity.
The macroscale model supplies the equipment's motion and contact forces at the terrain interface as input to mesoscale model.
This is the basis for predicting the active zones and provide boundary conditions that drives the mesoscale soil dynamics.

A regular grid is used for the discrete representation of the continuum model.  
The grid cells, or \emph{voxels}, are labelled $\bm{i} = (i,j,k)$ by the triplet of integer positions in the grid, aligned with the global coordinate axes.  
Each voxel has a centre point $\bm{x}_{\bm{i}} = [x_{\bm{i}}, y_{\bm{i}}, z_{\bm{i}}]$ and constant volume $V_0 = l_0^3$.  
Each voxel has a variable mass $m_{\bm{i}}$, velocity $\bm{v}_{\bm{i}}$, compaction $w_{\bm{i}}$ and occupancy $\varphi_{\bm{i}}$.
These are collected in a state vector $\bm{s}_{\bm{i}} = 
[ m_{\bm{i}}, \bm{v}_{\bm{i}}, w_{\bm{i}}, \varphi_{\bm{i}}]$.
The evolution of $\bm{s}_{\bm{i}}$ is treated by a cellular automata, with transition rules that implements convection of fluidized mass, plastic compaction at critical sub-soil stress, conversion between the particle and continuum representation, and surface flow if the local slope exceeds the soil's angle of repose $\delta_\text{b}$ \cite{Bouchaud1994,Li1993}.
For cohesion-free materials it coincides well with the internal friction angle, given by Eq.~(\ref{eq:effective_internal_friction}).  With cohesion it may be larger \cite{mitarai:2006:wgm}.
Beyond this slope the terrain is not in a stable equilibrium, and will fail by avalanching into a valid state. 
The transitions are constructed to preserve the total mass to machine precision. 

The voxel mass density, $\rho_{\bm{i}} = \rho^\text{s}_{\bm{i}} + \rho^\text{f}_{\bm{i}}$, is composed by the density of resting solid, $\rho^\text{s}_{\bm{i}}$, and fluidized mass, $\rho^\text{f}_{\bm{i}}$.  
All voxels in the terrain are fully occupied, $\varphi_{\bm{i}} = 1$, with solid mass and empty of fluidized mass, except for surface voxels, that may have occupancy less than unity, $ \varphi_{\bm{i}} = V_{\bm{i}}^\text{s}/V_0 \in[0,1]$.
These voxels, and the ones above the surface, may also contain fluidized mass.
The solid mass density has a natural bank state value $\rho_\text{b}$ but can vary locally within the range $\rho^{\bm{i}}_\text{s} \in [\rho_\text{min},\rho_\text{max}]$ if subject to compaction or swelling, $w_{\bm{i}} \equiv \rho^{\bm{i}}_\text{s}/\rho_\text{b}$.
The amount of solid mass in a voxel $\bm{i}$ is consequently $m^{\bm{i}}_s = \rho^{\bm{i}}_\text{s} V^{\bm{i}}_\text{s} = w_{\bm{i}} \varphi_{\bm{i}} \rho_\text{b} V_0$.  
The upper and lower limit on the mass density imply that the compaction is bounded by the upper and lower values $w_\text{max} = \rho_\text{max}/\rho_\text{b}$ and $w_\text{min} = \rho_\text{min}/\rho_\text{b}$.
We identify $S \equiv w^{-1}_\text{min}$ as the \emph{swell factor} of a soil that is transformed from bank state to its maximally loose packed state.

\begin{figure}[h]
  \centering
  \includegraphics[width=0.85\textwidth]{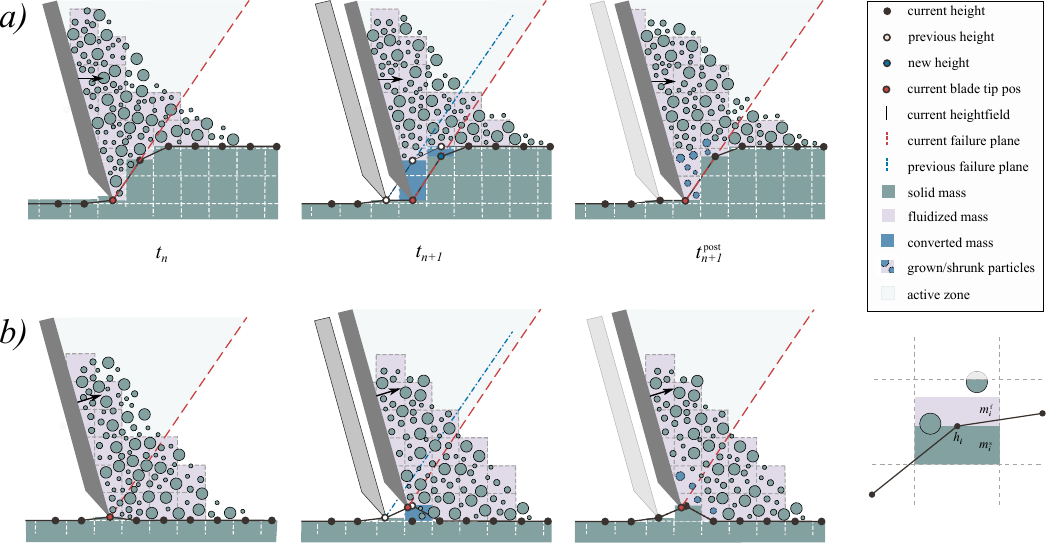}
  \caption{As a blade and its active zone moves into the terrain (a), new solid voxels are resolved into particles or fluidized mass that form the aggregate body.  The voxel height value, corresponding to the solid occupancy, is found by projection to the failure plane of the active zone.
  When the blade is raised (b), the gap is filled by solid mass converted from fluidized or particle mass in the vicinity of the edge.}
  \label{fig:moving_active_zone}
\end{figure}

The surface voxels define a surface heightfield, $z = h(x,y)$, that is used for contacts with the particles, the equipment, or any other objects in the macroscale model. 
It has a discrete representation $h_{ij} = h(x_{i},y_j)$.  The height value in a column with index $(i,j)$ is the centre position, $z_{\bm{i}'}$ of the top-most non-empty voxel, $\bm{i}' = (i,j,k')$, plus the local mass fill ratio relative to that voxel centre, i.e., 
\begin{equation}\label{eq:surface}
    h_{ij} = z_{\bm{i}'} + (\varphi_{\bm{i}'} - 1/2) l_0,
\end{equation}
This make the surface heightfield a continuous function of the solid occupancy, see Fig.~\ref{fig:moving_active_zone}.  
Between the grid points the surface height field is interpolated linearly.  

The response by the terrain is different for contacts with sharp and blunt geometries.
The former lead to shear failure with a localized failure surface while the latter cause soil compaction.
If the contacting body has a sharp \emph{cutting edge}, a co-moving active zone is predicted.
The motion of the terrain inside the active zone is resolved by particles and fluidized mass. 
See Fig.~\ref{fig:moving_active_zone} for illustration.
The basic shape of the active zone is that of a wedge, defined by the cutting edge, the soil failure surface that extends from the edge to the free surface of the terrain, and the \emph{separation plane} of the cutting body.
The slope of the failure surface, $\theta$, depends on the soil's internal friction, $\phi$, and the orientation of the separation plane relative to the to the terrain surface, $\beta$.
From simulations with the microscale model we identify the following model 
\begin{equation}
    \theta (\phi,\beta) = \frac{\pi}{2} - \left(  \frac{\phi + \beta}{2} \right),
\label{eq:failure_plane}
\end{equation}
This extends the classical Rankine failure angle $\frac{\pi}{4} - \frac{\phi}{2}$ to sloped terrain.
Furthermore, to handle nonuniform distributions of material, the active zone is discretised in a set of parallel wedges in the lateral direction.

As the cutting body moves through the soil, the active zone overlap with new voxels.
Overlapping soil is converted from solid mass, to particles or fluidized mass.
The amount of converted mass is computed using Eq.~(\ref{eq:surface}), such that the surface voxel's fill ratio take the \emph{exact value} where the surface heightfield coincides with the failure surface, see Fig.~\ref{fig:moving_active_zone}. 
The liberated mass is converted into new particles or growth of existing particles
in the vicinity of the voxel, provided the local void ratio allows it. 
To avoid ending up with too small particles (poor performance) or too large particles (discretization error) the particle size is restricted to a given range $[d_\text{min}, d_\text{max}]$. 
The size range is chosen such that the number of particles remain within the limits of real-time simulation.
Any residual mass, that cannot be converted into particle mass, is converted into fluidized mass in the voxel.
At the next time-step, the fluidized mass is again candidate for formation and growth of particles.
If it reaches a surface voxel outside the active zone, it is converted back to resting solid mass, whereby the surface heightfield is raised accordingly.
The inertia of the coarse particles is assumed to dominate over the dilute fluidized mass.
Therefore, we simply model the flow of fluidized mass by conserved advection 
\begin{equation}\label{eq:advection}
    \partial_t \rho_\text{f} + \bm{\nabla}\cdot(\rho_\text{f}\bm{u}) = r_\text{f}.
\end{equation}
where $\bm{u}(\bm{x})$ is the coarse-grained particle velocity field, using Eq.~(\ref{eq:cg_velocity}), and $r(\bm{x}, t)$ is a source (or sink) term for mass converted from (or to) particle mass or solid mass.
Any fluidized mass that is found in a voxel with no particles is projected in the direction of gravity towards the surface where it is converted into solid mass.

When a cutting body is raised from the terrain surface a gap may arise between the edge and the surface.
In reality, the gap is normally closed by fine-grain soil flowing from the active zone.
Mesoscale particles that are coarser than the gap cannot represent this.
Instead, this is modelled by particles (and any fluid mass) in the vicinity of the cutting edge losing mass to the surface voxels at the gap.
This is illustrated in Fig.~\ref{fig:moving_active_zone}.
The amount of mass necessary to fill the gap is computed using Eq.~(\ref{eq:surface}).
Particles losing mass shrink correspondingly until they reach the lower size limit, where they are converted to fluidized mass.
The process of converting soil from solid mass to particle and fluidised mass, and back, \emph{enable cutting and grading with high precision, despite the particle and voxel discretization being much coarser}.
It should also be noted that the operations for mass conversion and transport are by construction guaranteed to preserve the total mass to machine precision. 

The time evolution of the coarse particles, $\mathcal{N}_{\mbox{\tiny{CP}}}$, is governed by the equation of motion
\begin{equation} 
  \label{eq:eqmotb}
    \bm{m}_b\dot{\bm{v}}_b = \bm{f}^{\mathrm{ext}}_b 
    + \bm{G}_{b\mbox{\tiny{B}}}^\T \bm{\lambda}_{\mathrm{\mbox{\tiny{B}}b}} 
    + \bm{G}_{b\mbox{\tiny{C}}}^\T \bm{\lambda}_{\mathrm{\mbox{\tiny{C}}b}}
    + \sum_{b'\in \mathcal{N}^{\mbox{\tiny{b}}}_{\mbox{\tiny{CP}}}} \bm{G}_{bb'}^\T \bm{\lambda}_{\mathrm{bb'}}
\end{equation}
and mass balance at voxel $\bm{i}$ containing the particles $\mathcal{N}_{\mbox{\tiny{CP}}}^{\bm{i}}$
\begin{equation}
    \label{eq:eqmotmb}
    \sum_{b \in \mathcal{N}_{\mbox{\tiny{CP}}}^{\bm{i}}}
  \dot{m}_b =  - \dot{m}_s^{\bm{i}} - r_\text{f}^{\bm{i}}, \ \lVert\dot{m}_b\rVert \leq \tfrac{3 m_b}{d_b} d'_\text{max}, \ d\in[d_\text{min}, d_\text{max}].
\end{equation}
where $\bm{G}_{b\mbox{\tiny{B}}}^\T \bm{\lambda}_{\mbox{\tiny{B}}b}$ is the contact force from the cutting body and $\bm{G}_{b\mbox{\tiny{C}}}^\T \bm{\lambda}_{\mbox{\tiny{C}}b}$ from the terrain surface.
The change in solid mass, $\dot{m}_s^{\bm{i}} = w_{\bm{i}} \dot{\varphi_{\bm{i}}} \rho_\text{b} V_0$, is linked to the change in the surface heightfield imposed by the moving cutting body.
The change in particle mass depends on the change in solid mass, but is constrained by a maximum particle growth rate, $d'_\text{max}$, and size range.
If the change in solid and particle mass does not match, the residual mass is converted to or from fluidized mass at rate $r_\text{f}^{\bm{i}}$, provided $m_\text{f}^{\bm{i}} \geq 0$.
Since the contact model is scale-invariant by construction, the pre-calibrated parameter $\bm{p}_\text{p}$ will give the mesoscale model the desired bulk mechanical properties $\bm{p}_\text{b}$. 
Particles that come to rest outside the active zone are converted to solid mass, whereby the surface heightfield increase correspondingly.
Soil that has undergone rapid flow tend to be more loosely packed than the original bank state.
To model this, each soil is assigned a swell factor $S = \rho_\text{b}/\rho_\text{min}$.
Voxels with mass converted from particles to solid mass are thus given the compaction value $w_{\bm{i}} = 1/(1 + s)$ and the local solid mass density $\rho^\text{s}_{\bm{i}} =  \rho_\text{b}/(1 + S)$. 
The effect on the soil's strength by the change in compaction is addressed below.

We assume that contacting blunt bodies cause soil compaction if the subsoil stress reaches critical values.
Since the displacements are small it is not necessary to resolve these deformations using particles.
Instead, we operate directly on the local compaction of solid mass $w_{\bm{i}}$.
The contact forces $\bm{f}_k = \bm{G}^T_{\mbox{\tiny{BC}}} \bm{\lambda}_{\mbox{\tiny{BC}}}$ between the blunt body $B$ and the terrain surface $C$ are obtained from the macroscale model.
The forces act at some contact points $\bm{x}_k$ that enclose a contact patch of area $A_{b\mbox{\tiny{BC}}}$.
The subsoil stress $\sigma_{\bm{i}}$ in a voxel $\bm{i}$ at depth $z_{\bm{i}}$, and the resulting plastic deformation, can be estimated from analytical solutions and numerical extensions \cite{Keller2007,zhu2011}.
As a first order approximation following \cite{Madsen2014}, we use the model, $\sigma_{\bm{i}} =  \sigma_{b\mbox{\tiny{BC}}} \left[ 1 - \left(z_{\bm{i}}/\sqrt{A_{b\mbox{\tiny{BC}}} + z_{\bm{i}}^2}\right)^3 \right]$, for a circular distributed normal load, $\bm{\sigma}_{b\mbox{\tiny{BC}}} = \sum_k \bm{f}_k / A_{b\mbox{\tiny{BC}}}$, on a semi-infinite elastic solid.
The compaction can be estimated from the compression index in Eq.~(\ref{eq:compression_index}) and by noting the relation to the void ratio  $w \approx 1 + e_\text{b} - e$.
Noting that the void ratio and compaction are related by $w \approx 1 + e_\text{b} - e$ we obtain the following model for the compaction in a voxel $\bm{i}$ due to subsoil stress at a depth $z_{\bm{i}}$ beneath the contact 
\begin{equation}\label{eq:mass_compaction_voxel}
  w_{\bm{i}} =  1 + [1-\varphi_\text{c}]C_\text{c} \ln ( \sigma_{\bm{i}} / \sigma_\text{b} ),
\end{equation}
with clamping to the set bounds $w_{\bm{i}} \in [w_\text{min}, w_\text{max}]$.
Typical values for the compression index range between $0.001$ (dense sand) and $1$ (soft clay) and we use 
$\sigma_\text{b} = 1$ kPa as the consolidation stress for the bank state.
A change in compaction alters the mass density in the voxel to $\rho_{\bm{i}} =\rho_\text{b} w_{\bm{i}}$ and mass is shifted downwards column wise to preserve the total mass. 
This reduce the fill ratio of the surface voxels at the contact and the surface heightfield is reduced correspondingly.
The increased level of compaction also makes the material stiffer and stronger. From Eqs.~(\ref{eq:effective_elasticity}) and (\ref{eq:dilatancy}) we obtain the following models for the bulk elasticity, dilatancy angle and effective angle of internal friction
\begin{align}
    E_\text{bulk}^{\bm{i}} & = E^0_\text{bulk} \left( 1 \pm k_\text{E} \lVert w_{\bm{i}} - 1 \rVert ^{n_\text{E}} \right), \\
    \psi_{\bm{i}} & = c_\psi \left( w_{\bm{i}} - w_{\text{c}}  \right), \\
    \phi_{\bm{i}} & = \phi_\text{b} + \psi_{\bm{i}},
\end{align} 
where the stiffening parameters $k_\text{E}$, $n_\text{E}$ and have default values $1$ and $0.5$, respectively, but can be tuned to represent more complex soil.
The parameter $c_\psi = \partial \psi / \partial w$ control the rate of hardening, with $0.1 \-- 1.0$ radians being a typical range.
The critical compaction, $w_\text{c}$, determines whether the soil at bank state is expanding or compressing under shear.
It is related to the bank state dilatancy angle $\psi_\text{b}$ by $w_\text{c} = 1 - \psi_\text{b}/c_\psi$
The compacted state is permanent until the soil is disturbed, e.g., by earthmoving equipment interacting with it.  

\subsection*{Macroscale model}
The macroscale model focus on the rigid multibody dynamics of the equipment, $B$.
The resting terrain is perceived as a surface heightfield, $C$.
In each active zone, evolved in time by the mesoscale model, the soil is aggregated into a single body with the inertia, centre of mass and velocity of the particles and fluidized mass. 
These \emph{aggregate bodies}, labelled $A$, have the degrees of freedom of rigid bodies but finite mechanical strength.
This is accomplished by compliant frictional-cohesive contact constraints at the interfaces of the aggregate to the terrain surface, which currently coincide with the failure plane, and to the equipment.
The equations of motion are
\begin{align}
      \label{mdb:eqAgg}
      \bm{M}_{\mbox{\tiny{A}}}\dot{\bm{v}}_{\mbox{\tiny{A}}} & = \bm{f}^{\mathrm{ext}}_{\mbox{\tiny{A}}} 
      + \bm{G}_{\mbox{\tiny{AB}}}^\T \bm{\lambda}_{\mbox{\tiny{AB}}}
      + \bm{G}_{\mbox{\tiny{AC}}}^\T \bm{\lambda}_{\mbox{\tiny{AC}}},\\
      \label{mdb:eqEq}
      \bm{M}_{\mbox{\tiny{B}}}\dot{\bm{v}}_{\mbox{\tiny{B}}} & = \bm{f}^{\mathrm{ext}}_{\mbox{\tiny{B}}} + 
          \bm{G}_{\mbox{\tiny{B}}}^\T \bm{\lambda}_{\mbox{\tiny{B}}} 
          + \bm{G}_{\mbox{\tiny{BA}}}^\T \bm{\lambda}_{\mbox{\tiny{AB}}}
          + \bm{G}{\mbox{\tiny{BC}}}^\T \bm{\lambda}_{\mbox{\tiny{BC}}}
          + \bm{G}_{\mbox{\tiny{BD}}}^\T \bm{\lambda}_{\mbox{\tiny{BD}}}.
    \end{align}
where the constraint forces $\bm{G}_{\mbox{\tiny{AB}}}^\T \bm{\lambda}_{\mbox{\tiny{AB}}}$ and $\bm{G}_{\mbox{\tiny{AC}}}^\T \bm{\lambda}_{\mbox{\tiny{AC}}}$ act on the aggregate from the equipment and the terrain failure surface, respectively.
The equipment is held together and actuated by $\bm{G}_{\mbox{\tiny{B}}}^\T \bm{\lambda}_{\mbox{\tiny{B}}}$.
The separation resistance and the inertia of the soil in the active zone is mediated by $\bm{G}_{\mbox{\tiny{BA}}}^\T \bm{\lambda}_{\mbox{\tiny{AB}}}$.
Through $\bm{G}_{\mbox{\tiny{BC}}}^\T \bm{\lambda}_{\mbox{\tiny{BC}}}$, the equipment experience direct contact with the terrain surface, e.g., tyres, tracks or the exterior of the bucket.
Soil cutting objects may also be subject to a penetration resistance force $\bm{G}_{\mbox{\tiny{BD}}}^\T \bm{\lambda}_{\mbox{\tiny{BD}}}$.
See Fig.~\ref{fig:overview} for illustration of the interaction interfaces.

Each mesoscale active zone, $A$, is aggregated to a macroscale body with the following mass, inertia tensor, centre of mass, linear and angular velocity
\begin{align}
    \label{eq:aggregate_mass}
    M_{\mbox{\tiny{A}}} & = 
        \sum_{a\in\mathcal{N}^{\mbox{\tiny{A}}}_{\mbox{\tiny{P}}}} m_a 
        + \sum_{\bm{i}\in\mathcal{N}^{\mbox{\tiny{A}}}_{\mbox{\tiny{f}}}} m_{\bm{i}}, \\
    I^{\mbox{\tiny{A}}}_{\alpha\beta} & = 
        \sum_{a\in\mathcal{N}^{\mbox{\tiny{A}}}_{\mbox{\tiny{P}}}} m_a \left( \lVert\bm{x}_a - \bm{x}_{\tiny{A}}\rVert \delta_{\alpha\beta} + x^a_\alpha x^a_\beta \right) 
        + \sum_{\bm{i}\in\mathcal{N}^{\mbox{\tiny{A}}}_{\mbox{\tiny{f}}}} m_{\bm{i}} \left( \lVert\bm{x}_{\bm{i}} - \bm{x}_{\tiny{A}}\rVert \delta_{\alpha\beta} + x^{\bm{i}}_\alpha x^{\bm{i}}_\beta \right), \\
    \bm{x}_{\mbox{\tiny{A}}} & = 
        M_{\mbox{\tiny{A}}}^{-1}\sum_{a\in\mathcal{N}^{\mbox{\tiny{A}}}_{\mbox{\tiny{P}}}} m_a \bm{x}_a 
        + M_{\mbox{\tiny{A}}}^{-1}\sum_{\bm{i}\in\mathcal{N}^{\mbox{\tiny{A}}}_{\mbox{\tiny{f}}}} m_{\bm{i}}\bm{x}_{\bm{i}}, \\
    \bm{v}_{\mbox{\tiny{A}}} & = 
        M_{\mbox{\tiny{A}}}^{-1}\sum_{a\in\mathcal{N}^{\mbox{\tiny{A}}}_{\mbox{\tiny{P}}}} m_a \bm{v}_a 
        + M_{\mbox{\tiny{A}}}^{-1}\sum_{\bm{i}\in\mathcal{N}^{\mbox{\tiny{A}}}_{\mbox{\tiny{f}}}} m_{\bm{i}}\bm{v}_{\bm{i}}, \\
        \label{eq:aggregate_rot_vel}
    \bm{\omega}_{\mbox{\tiny{A}}} & = 
        \bm{I}^{-1}_{\mbox{\tiny{A}}} \sum_{a\in\mathcal{N}^{\mbox{\tiny{A}}}_{\mbox{\tiny{P}}}} m_a (\bm{x}_a - \bm{x}_{\mbox{\tiny{A}}}) \times \bm{v}_a 
        + \bm{I}^{-1}_{\mbox{\tiny{A}}} \sum_{\bm{i}\in\mathcal{N}^{\mbox{\tiny{A}}}_{\mbox{\tiny{f}}}} m_{\bm{i}} (\bm{x}_{\bm{i}} - \bm{x}_{\mbox{\tiny{A}}}) \times \bm{v}_{\bm{i}},
\end{align}
where $\mathcal{N}^{\mbox{\tiny{A}}}_{\mbox{\tiny{P}}}$ and $\mathcal{N}^{\mbox{\tiny{A}}}_{\mbox{\tiny{f}}}$ denote the set of particles and voxels with fluidized mass in active zone $A$.

The contacts aggregated from the $\tiny{AB}$ and $\tiny{AC}$ interfaces are turned into the following velocity constraint
\begin{align}
  \label{eq:aggregateNormal}
  &0 \leq \gamma_{\mathrm{n}} \lambda_{\mathrm{n}} + \bm{G}_{\mathrm{n}} \bm{v} 
  \perp ( \lambda_{\mathrm{n}} + f_{\mathrm{c}} )\geq 0 , \quad
  f_{\mathrm{c}} \equiv c_{\mathrm{b}} A  \\ 
  \label{eq:aggregateTangent}
  &\gamma_{\mathrm{t}} \bm{\lambda}_{\mathrm{t}} + \bm{G}_{\mathrm{t}} \bm{v} = 0, \quad
  \lVert\bm{\lambda}_{\mathrm{t}}\rVert \leq \mu\lVert \lambda_{\mathrm{n}}\rVert.
\end{align}
Eq.~(\ref{eq:aggregateNormal}) prevent relative motion in the normal direction, i.e., compressive or tensile deformations.
Note that, unlike model Eq.~(\ref{eq:theory:mB:signCoul1}), this model does not include a contact overlap.
This make the aggregate viscoplastic in nature, with no sense of any reference configuration. 
The viscous damping is controlled by the parameter $\gamma_{\mathrm{n}} = \Delta t/ (c_\text{as} E_\text{b})$, where $c_\text{as}$ is the aggregate's damping coefficient (referred to as \emph{aggregate stiffness multiplier} in the implementation).
If the force is tensile and reach the cohesive force limit, given by $c_{\mathrm{b}} A$, separation may occur at the interface.
Eq.~(\ref{eq:aggregateTangent}) prevent sliding motion in the failure surface and in the interface to the equipment, unless the forces reach the Coulomb condition.  
The friction coefficient in the failure surface is set to the soil's internal friction $\mu = \mu_{\mathrm{b}}$.  
At the equipment interface the friction coefficient is set by the tool's surface friction $\mu = \mu_{\mathrm{tool}}$.
The contact points between the aggregate body $\tiny{A}$, the equipment $\tiny{B}$ and the failure surface $\tiny{C}$ is the reduced set of contact points from the mesoscale model that best approximate the two contact areas, $A_{\mbox{\tiny{AB}}}$ and $A_{\mbox{\tiny{AC}}}$, with four contact points each.  
This is illustrated in 2D in Fig.~\ref{fig:overview}.

If the cutting edge has a thickness or is equipped with teeth, there may be a significant \emph{penetration force}, $\bm{G}_{\mbox{\tiny{BD}}}^\T \bm{\lambda}_{\mbox{\tiny{BD}}}$ in addition to the separation force.
This is modeled with the following velocity constraint
\begin{equation} \label{eq:penetration_resistance}
    \bm{G}_{\mbox{\tiny{BD}}} \bm{v}_{\mbox{\tiny{B}}} = 0, \quad \bm{G}_{\mbox{\tiny{BD}}}^\T \bm{\lambda}_{\mbox{\tiny{BD}}} \leq f_\text{pt} \equiv n_\text{t} \left[ p_\text{t} + (c_\text{t} + p_\text{t}  \mu_\text{t}) /\tan \theta_\text{t} \right] A_\text{t} ,
\end{equation}
where $ \bm{G}_{\mbox{\tiny{BD}}} = \bm{t}_B$ is a unit vector for the direction of penetration, e.g., pointing in the direction of the teeth.
For penetration to occur, the tool must overcome the force limit $f_\text{pt}$ \cite{Bennett2016}, which depend on the tool's friction coefficient $\mu_\text{t}$ and adhesion $c_\text{t}$, and on the number of teeth $n_\text{t}$, with cross-section area $A_\text{t}$, tooth angle $\theta_\text{t}$. 
The tooth pressure, $p_\text{t}$, is modelled using the finite cavity expansion theory \cite{Yu1991}.
In the plastic limit the tooth pressure become $p_\text{t} = p_0 + \left[\Upsilon + (\alpha - 1) p_0\right]/\left[ 2 + \alpha\right]$ under cylindrical expansion, where $\Upsilon = 2 c_\text{t} \cos \phi / (1 - \sin\phi)$, $\alpha = (1 + \sin\phi)/(1 - \sin\phi)$. 
The mean pressure is computed by $p_0 = \frac{1}{2} \rho g z \left[ (1 + K_0) + (1 - K_0) \cos (2 \beta) \right]$, taking the lateral earth pressure and tool inclination into account \cite{Bennett2016}, where $K_0$ is the coefficient of lateral earth pressure, $\rho$ is the specific soil mass density, $z$ is the penetration depth and $\beta$ is the insertion angle.  
A simple model for the coefficient of lateral earth pressure is given by Jaky \cite{Jaky1948} as $K_0 = 1 - \sin \phi$.

\subsection*{Complex digging tools}
\label{sec:diggingtols}
A bucket used for excavation or wheel loading is more complex than a blade. 
It can be seen as a bottom plate, with a curved back wall and sidewalls that allow material to accumulate in the bucket. 
This deadload form a \emph{secondary separation plate}, at an angle much steeper than the bottom plate. 
This affect the stress distribution, the shape of the active zone, by Eq.~(\ref{eq:failure_plane}), and ultimately also the digging resistance.
Furthermore, the forces from the soil that surrounds the bucket act both tangentially (friction) and normally. 
This cause additional digging resistance as well as strong resistance for motion in the lateral directions.
Buckets are also used for other purposes than digging. It is a common operation to do surface leveling and compaction using the bucket exterior.
To support these features we treat a bucket as a composite model of an elementary \emph{digging tool} and a set of \emph{soil deformers} as described below.

\begin{figure}[h]
    \centering
    \begin{minipage}{.25\textwidth}
	\includegraphics[width=0.95\textwidth]{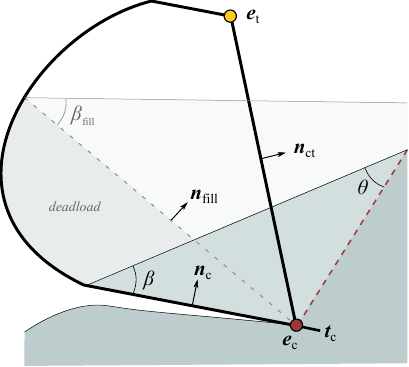}
    \end{minipage}
    \begin{minipage}{.3\textwidth}
        \includegraphics[width=0.95\textwidth]{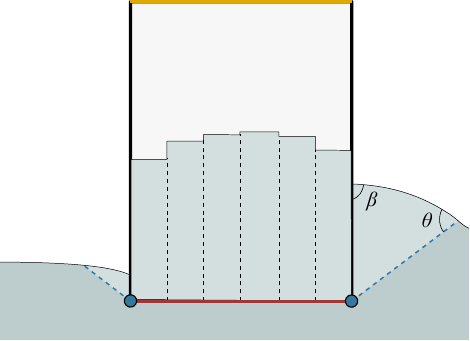}
    \end{minipage}
    \\
    \begin{minipage}{.25\textwidth}
        \hspace{3mm}
        \includegraphics[height=0.95\textwidth]{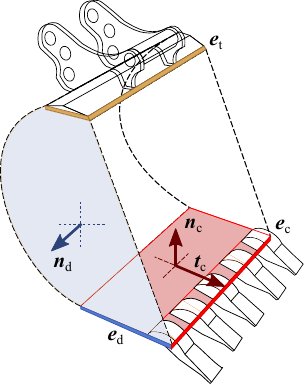}
    \end{minipage}
    \begin{minipage}{.3\textwidth}
        \includegraphics[width=0.95\textwidth]{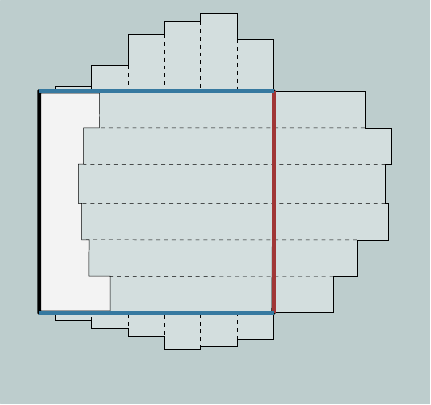}
    \end{minipage}
        \caption{The left images illustrate a digging tool with a cutting edge $\bm{e}_\text{c}$ (red line), separating plate $\bm{n}_\text{c}$ (red face), deformer edge $\bm{e}_\text{d}$ (blue line), deformer face $\bm{n}_\text{d}$ (blue face) and top edge $\bm{e}_\text{t}$ (yellow line).  The inner shape is indicated with the dashed lines.  The normal $\bm{n}_\text{fill}$ of the secondary separation plate range from $\bm{n}_\text{c}$ to $\bm{n}_\text{ct}$ depending on the fill ratio of the inner shape. 
        The three active zones, discretized by vertical wedges, is illustrated from the rear (upper right) and from above (lower right).}
    \label{fig:digging_tool}
\end{figure}

A \emph{digging tool} has a \emph{cutting edge} $\bm{e}_\text{c}$, a parallel \emph{top edge} $\bm{e}_\text{t}$, and a \emph{penetration direction} $\bm{t}_\text{c}$ orthogonal to these edges
and to the normal of the bottom plate $\bm{n}_\text{c}$.
This is the \emph{primary separation plate} of the digging tool.
See Fig.~\ref{fig:digging_tool} for an illustration.  
A convex digging tool has an \emph{inner shape} that is the void enclosed by the cutting edge, top edge and the concave tool surface connecting them.
The face that connect the cutting edge and the top edge has normal $\bm{n}_\text{tc}$.
If the bucket is full, this face act as a secondary separation plate.
If the digging tool is a simple blade there is no inner shape and $\bm{n}_\text{c} = \bm{n}_\text{tc}$.
For concave tools we track the amount of material accumulated in the inner shape (deadload) and do linear interpolation of the secondary separation plate $\bm{n}_\text{fill}$ between $\bm{n}_\text{c}$ and $\bm{n}_\text{ct}$.
This changes the angle $\beta$ in Eq.~(\ref{eq:failure_plane}) and thus the slope of the failure surface.

A \emph{soil deformer} is simply a separation plate with no penetration resistance and no inner shape.
It is defined by a \emph{deformation edge} $\bm{e}_\text{d}$ and a parallel \emph{top edge} $\bm{e}_\text{t}$.  
They form a face with normal $\bm{n}_\text{d}$.
The active zone from a soil deformer is not automatically resolved using particles, unless the velocity of the deformation edge exceed a given threshold.
This avoid using particles to simulate the terrain when it remains at steady state.

A bucket can thus be represented with soil deformers for the exterior faces and with a digging tool for the cutting blade and the bucket interior.
Fig.~\ref{fig:digging_tool} shows a bucket with two side walls, each assigned a soil deformer.
Together with the digging tool, this gives rise to a total of three active zones, which are discretised in the lateral direction.
The backside of the bucket can also assigned a deformer, in which case the number of active zones sum to four.
If not, the soil cannot be displaced by the backside of the bucket other than by pure compression.

The aggregate normal forces is modeled by Eq.~(\ref{eq:aggregateNormal}).
It is a velocity constraint, which are prone to numerical drift.
The drift has no significance in soil cutting but prohibit proper resistance of soil deformers  resting or being pressed onto a soil bed.  
A drift will cause them to sink unnaturally into the soil.
This is remedied by adding a stabilizing perturbation $\xi g_\text{n}$ to the constraint, i.e., $\gamma_{\mathrm{n}} \lambda_{\mathrm{n}} + \bm{G}_{\mathrm{n}} \bm{v} \to \gamma_{\mathrm{n}} \lambda_{\mathrm{n}} + \xi g_\text{n} + \bm{G}_{\mathrm{n}} \bm{v}$, which turn it into a ordinary contact constraint.
Transitioning between the velocity constraint (providing smooth soil cutting) and contact constraint (avoiding artificial sinkage) is achieved by making the stabilization coefficient $\xi$ a function of the orientation of the deformer relative to the terrain surface.

\subsection*{Algorithm}
The multiscale terrain model can be summarized by the Algorithm~\ref{alg:algorithm}.

\begin{algorithm}
    \caption{Multiscale terrain}domains
     \begin{algorithmic}[1] 
        \State initialize terrain
            \State \hskip1.0em  set surface heightfield $h(x,y)$ and bank state soil parameters $\bm{p}_\text{b} = [\phi_\text{b}, c_\text{b}, \psi_\text{b}, E_\text{b}]$
            \State \hskip1.0em  set initial voxel states $\bm{s}_{\bm{i}} = 
            [ m_{\bm{i}}, \bm{v}_{\bm{i}}, w_{\bm{i}}, \varphi_{\bm{i}}]$, 
            \State \hskip1.0em  apply cellular automata to ensure the surface heightfield is soil-consistent 
        \State initialize earthmoving equipment $[\bm{x}_,\bm{v},\bm{g}]_{\mbox{\tiny{B}}}$
            \State \hskip1.0em  set contact parameters for bodies and terrain surface $[\mu, c, e, E]_{\mbox{\tiny{BC}}}$
            \State \hskip1.0em  set digging tool edges and direction vectors $[\bm{e}_\text{c}, \bm{e}_\text{t}, \bm{t}_\text{c} ]$ and  $[n_\text{t},A_\text{t},\theta_\text{t}]$
            \State \hskip1.0em  set deformers $\to [\bm{e}_\text{d}, \bm{e}_\text{t}]$
        \While {running simulation}
            \State conversion of resting and active soil
                \State \hskip1.0em  for each body $B$ intersecting the terrain $C$
                    \State \hskip2.0em  compute active zones, discretise in wedges with failure angle $\theta(\phi,\beta)$
                    \State \hskip2.0em  convert resting soil in active zones to particle and fluidized mass 
                \State \hskip1.0em  convert resting particles outside zones to loose solid mass with $w_{\bm{i}} = w_\text{min}$
                \State \hskip1.0em  apply cellular automata to re-distribute resting soil violating the angle of repose
                \State \hskip1.0em  update the surface heightfield $h(x,y)$ from the voxel state $\bm{s}_{\bm{i}}$ 
            \State do collision detection
            \State add contact constraints in the macroscale and mesoscale simulations
            \For {each active zone}
                \State \hskip1.0em  aggregate body $A$ from voxels and particles using Eq.~(\ref{eq:aggregate_mass})-(\ref{eq:aggregate_rot_vel})
                \State \hskip1.0em  add aggregate velocity constraint (\ref{eq:aggregateNormal})-(\ref{eq:aggregateTangent}) with interfaces $AB$ and $AC$
            \EndFor
            \State penetration resistance
                \State \hskip1.0em  estimate pressure on edge/tooth and compute maximum penetration resistance
                \State \hskip1.0em  add the penetration constraint Eq.~(\ref{eq:penetration_resistance})
            \State time-step Eq.~(\ref{mdb:eqOfm1})-(\ref{mdb:eqOfm3})
                \State \hskip1.0em  solve the macroscale MCP using direct LDLT solver $\to [\bm{v}, \bm{\lambda}]_{\mbox{\tiny{B}}}$ 
                \State \hskip1.0em  solve the mesoscale MCP using iterative PGS solver $\to [\bm{v}, \bm{\lambda}]_{\mbox{\tiny{b}}}$
                \State \hskip1.0em  update positions $\bm{x}_{i + 1} = \bm{x}_{i} + \dt \bm{v}_{i + 1}$
            \State soil compaction 
                \State \hskip1.0em  estimate the sub-soil stress from the surface contact forces $BC$
                \State \hskip1.0em  update the soil compaction $w_{\bm{i}}$ using Eq.~(\ref{eq:mass_compaction_voxel})
                \State \hskip1.0em  displace resting solid mass vertically in compacted voxel columns
                \State \hskip1.0em  update the soil strength parameters $\bm{p}(w)$
        \EndWhile 
 \end{algorithmic}
\label{alg:algorithm}
\end{algorithm}

\section*{Implementation}
\label{sec:implementation}
The multiscale model was implemented in the C++ physics engine AGX Dynamics \cite{AGX20}.
It supports real-time simulation of rigid multibody and particle systems with contacts, articulation joints and non-smooth dynamics.
Numerical time integration is made with the SPOOK stepper.
A block-sparse LDLT solver with pivoting \cite{Lacoursiere2010} is used as direct solver for the macroscale model, and for the equipment subsystem in the meso- and microscale models.
The solutions are exact to machine precision.
The dynamics of the contacting particles is solved to lower precision using a projected Gauss-Seidel (PGS), which is accelerated using domain decomposition for parallel processing and warm-starting \cite{servin:2014:esn,wang:2016:wsp}.
The microscale reference simulations are run with $1$ ms time-step and $500$ PGS iterations.
The multiscale model is run with $ 16.7$ ms time-step, for real-time simulation at 60 Hz, and with $10$ PGS iterations for the mesoscale particle model.
This corresponds to an error tolerance of about 1 \% and 10 \%, respectively, according to the model in \cite{servin:2014:esn} and for the test systems in this paper.
%
%
%
The voxel data representations and operations are implemented using the Open VDB library \cite{Museth2013}.  
It is optimized for large, sparse, time-varying volumetric data discretised on a 3D grid and support hierarchical representations. 

\section*{Soil library}
It is important that the bulk and particle representations are dynamically consistent.
That means that the particle parameters, for each soil, need to be calibrated to the values that produce the desired bulk properties, e.g., internal friction, cohesion, angle of repose etc.
Otherwise they do not describe the same material and there is risk that the conversions between particle and continuum inject energy to the system, which can lead to numerical instability.
Therefore, 15 soils were calibrated in advance to the bulk mechanical properties. 
These are listed in Table \ref{table:soil_library}. 
The bulk parameters, at bank state, were chosen from tabulated values for different materials, e.g., gravel and sand. 
To narrow down the particle parameter search space, we were guided by friction measurements of sand grains.
The particles are given the Young's elasticity of $E = 10^9 Pa$, Poisson's ratio $0.15$, specific mass density $2200$ kg/m$^3$ and coefficient of restitution $0$.
Since there may exist many types of sand and gravel (with different distributions of size, shape, microscopic friction, packing density and moisture level), each with distinct bulk properties, an integer is added to the name to distinguish between them.
We also created a set of purely frictional soils (fs) and cohesive-frictional soils (cfs) with no particular real soil in mind.
For testing purposes a frictionless and a cohesive-frictionless soil (cs) were also created, but they are not expected to be of practical use.

\begin{figure}[h]
    \centering
	\includegraphics[width=0.45\textwidth, trim=0 0 32 40, clip]{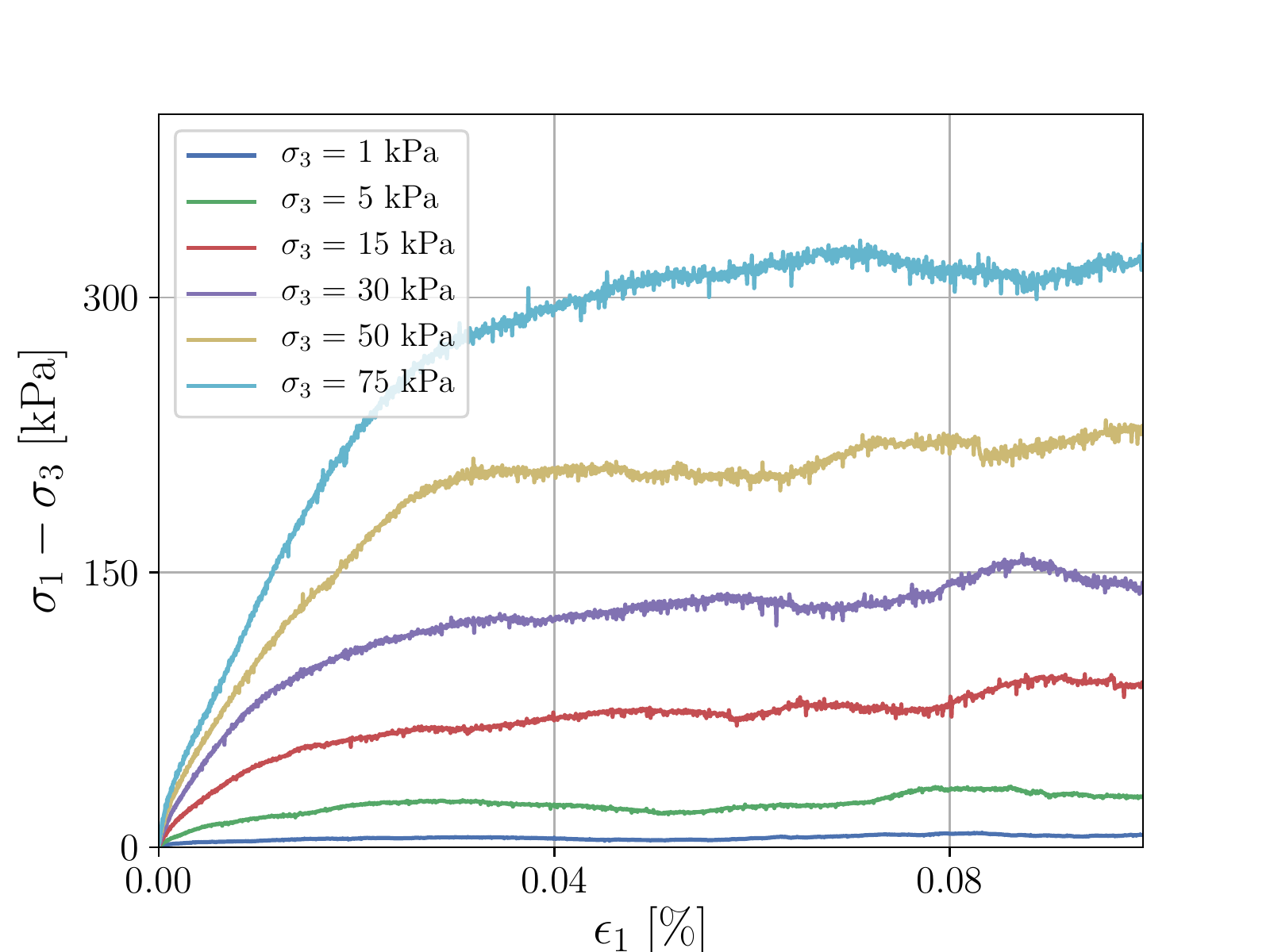}
    \hspace{3mm}
    \includegraphics[width=0.45\textwidth, trim=0 0 32 40, clip]{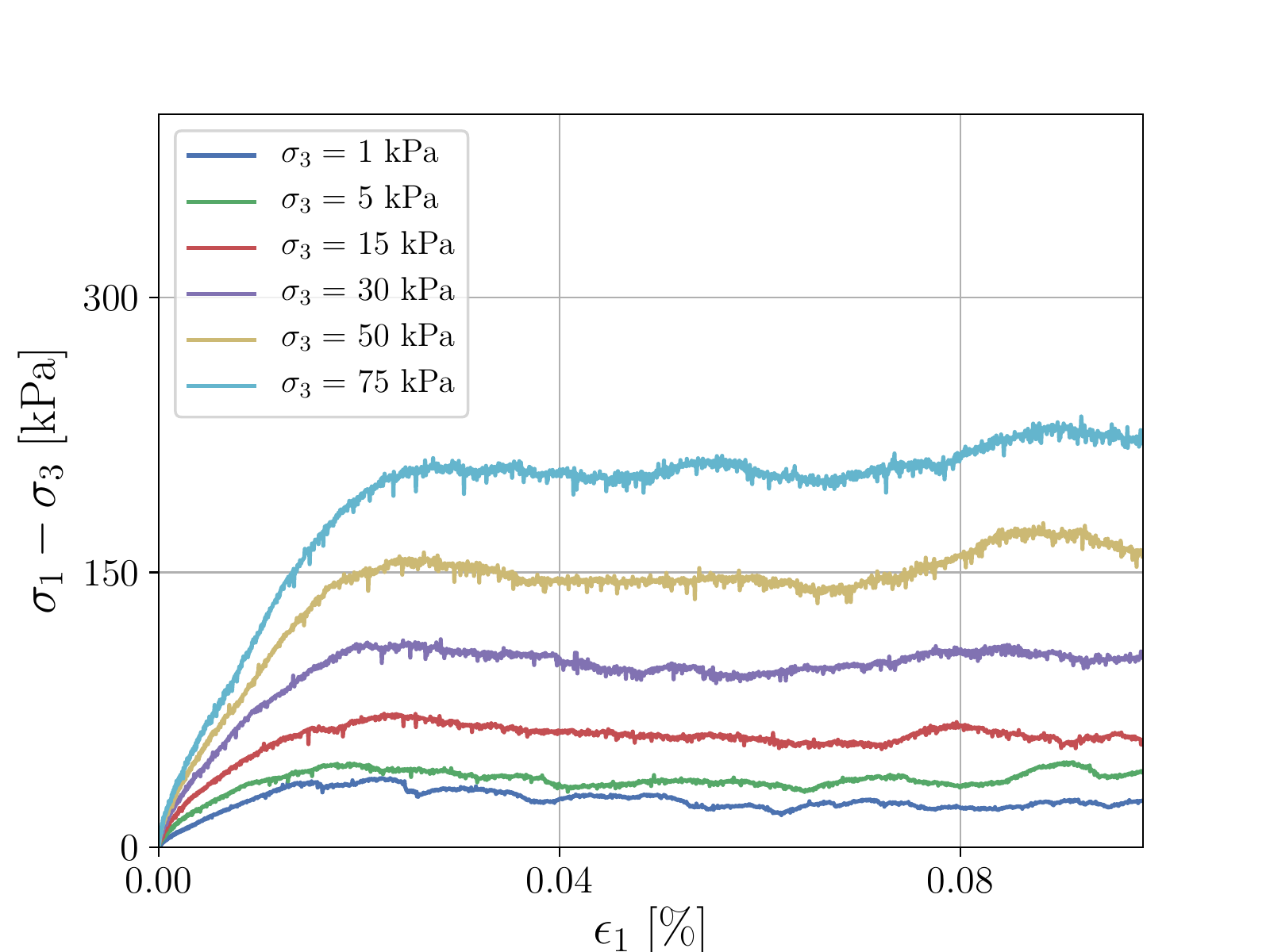}
    \caption{The principal stress as function of the vertical strain from triaxial test on \texttt{gravel-1} (left) and \texttt{wet-sand-1} (right). 
    Initially the principal stress grows nearly linearly, from which the secant modulus is determined, until shear failure occurs and the principal stress saturate.}
    \label{fig:triaxial_stress}
\end{figure}

\begin{figure}[h]
    \centering
    \includegraphics[width=0.45\textwidth, trim=0 40 0 40, clip]{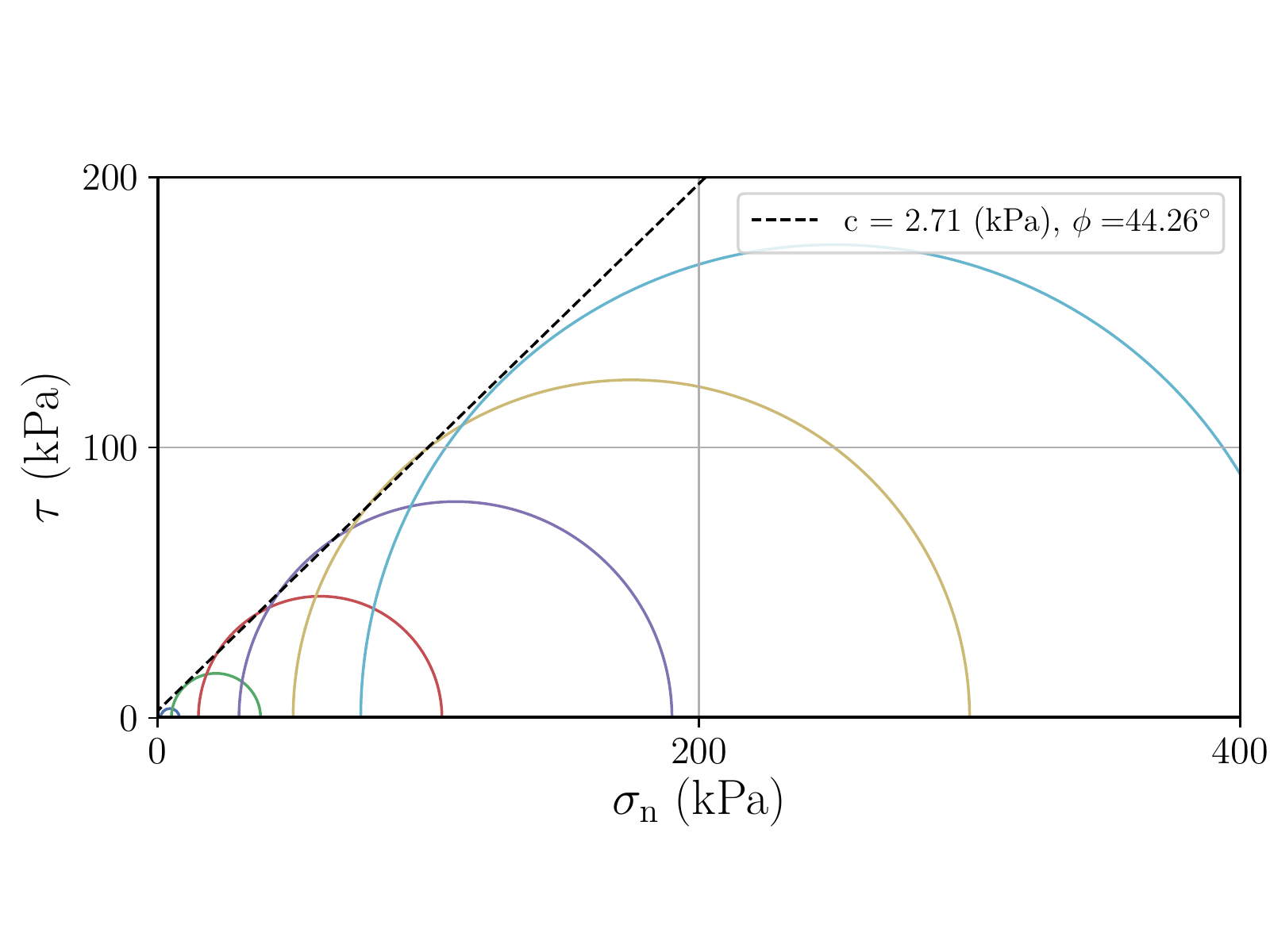}
    \hspace{3mm}
    \includegraphics[width=0.45\textwidth, trim=0 40 0 40, clip]{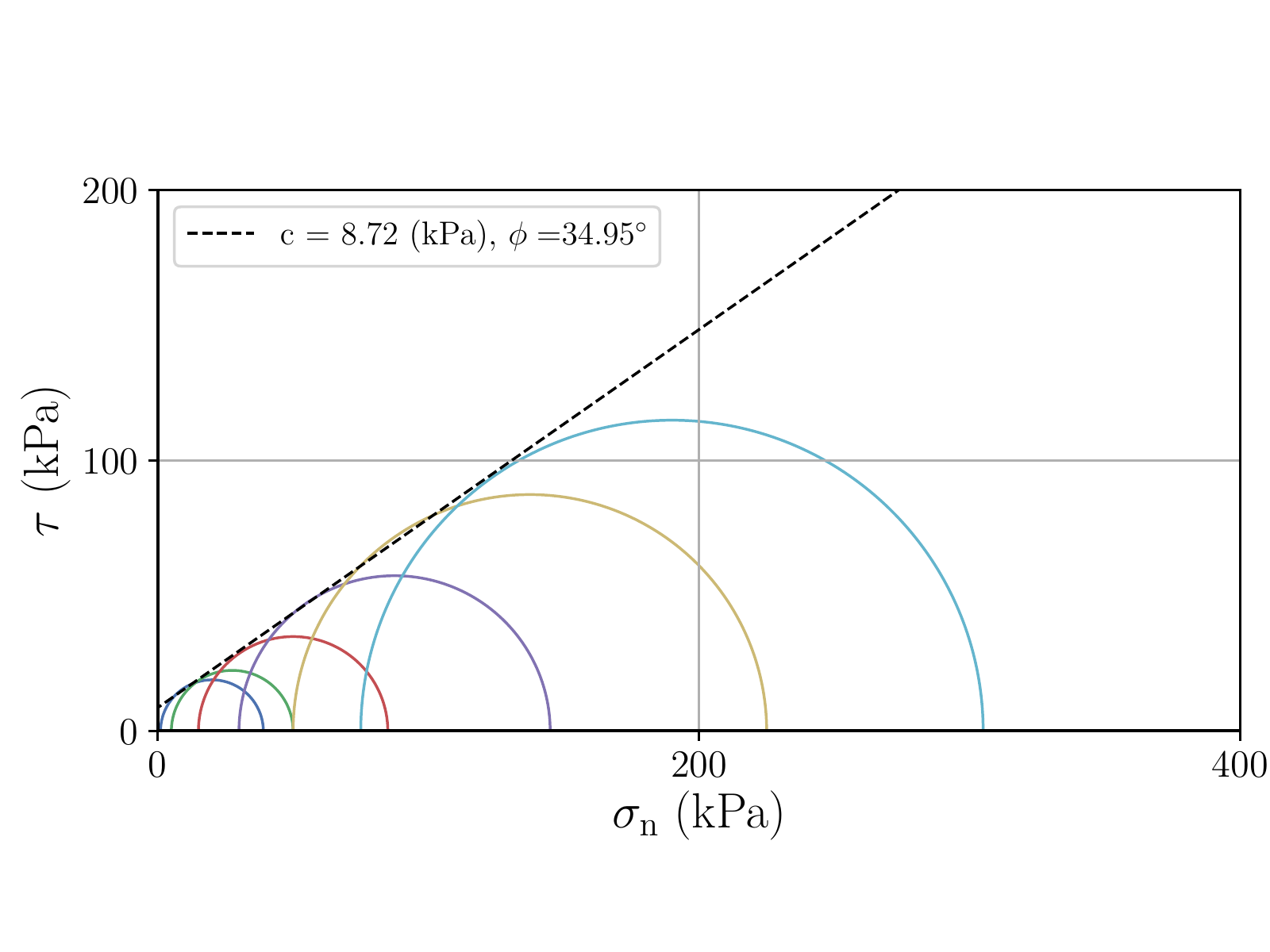}
    \caption{The Mohr circles for determining the internal friction and cohesion from triaxial test on \texttt{gravel-1} (left) and \texttt{wet-sand-1} (right). }  
    \label{fig:triaxial_MC}
\end{figure}

\begin{figure}[h]
    \centering
	\includegraphics[width=0.45\textwidth, trim=0 0 32 40, clip]{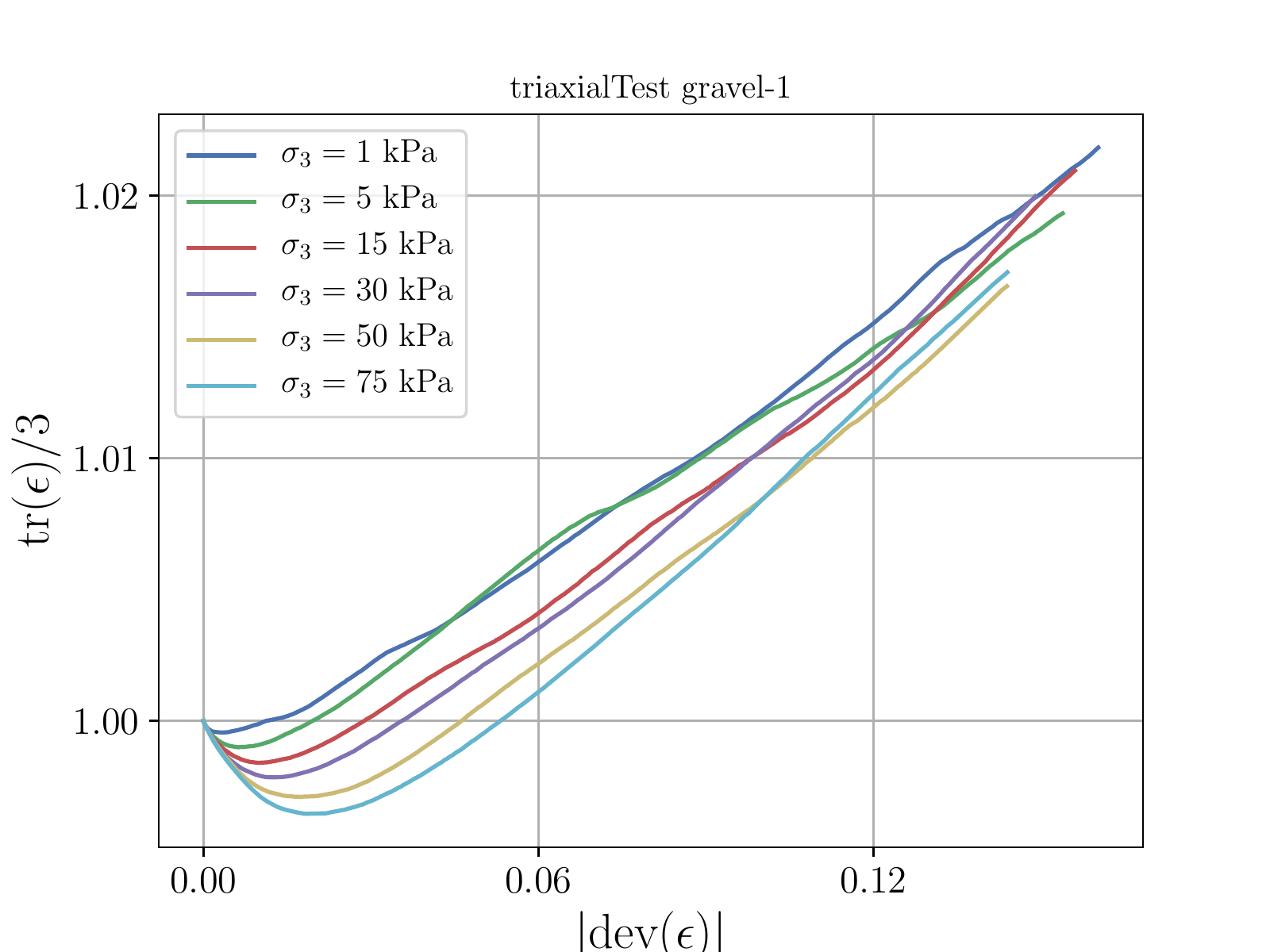}
    \hspace{3mm}
    \includegraphics[width=0.45\textwidth, trim=0 0 32 40, clip]{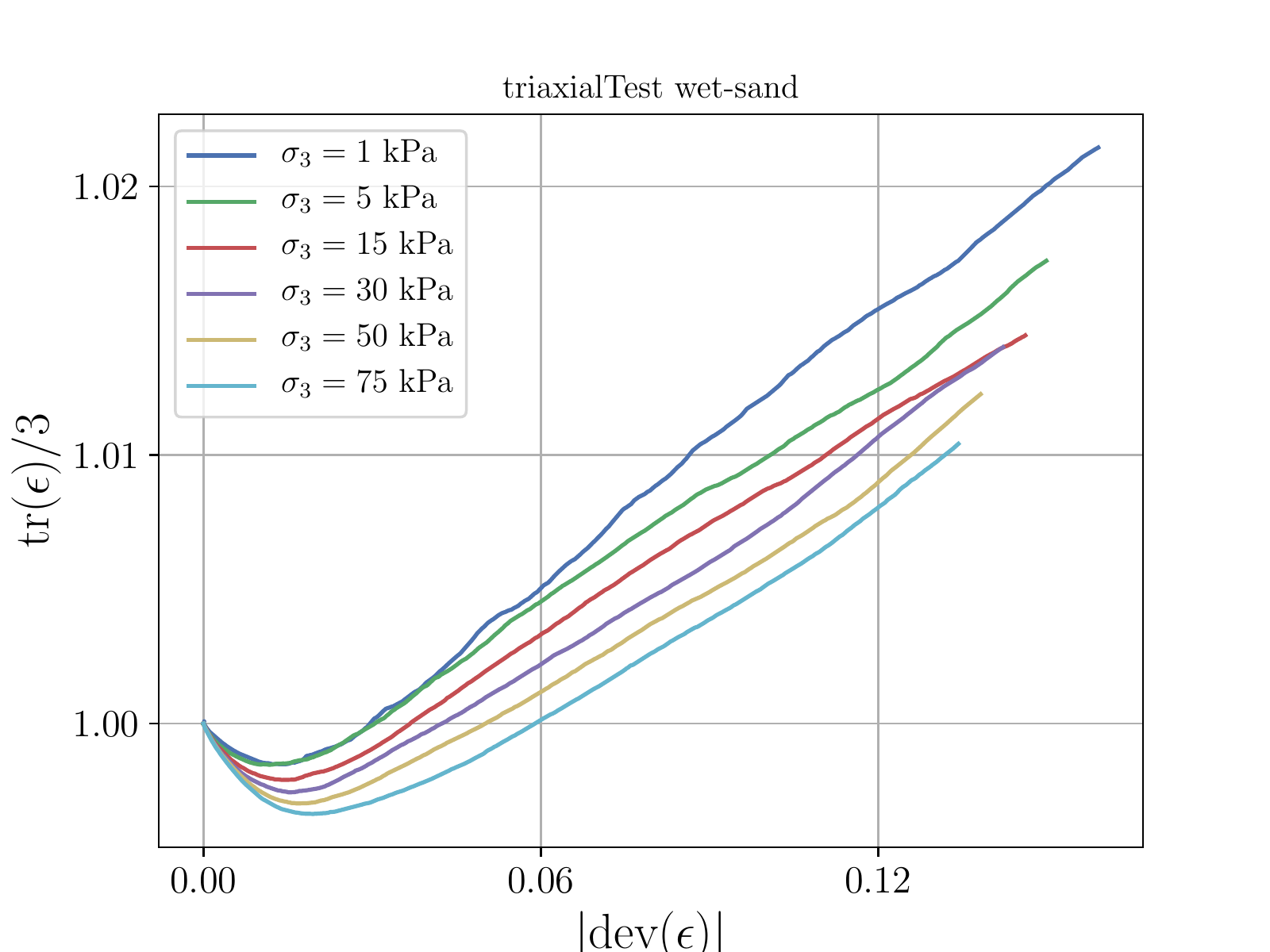}\\
    \caption{The volumetric strain as function of the deviatoric strain from triaxial test on \texttt{gravel-1} (left) and \texttt{wet-sand-1} (right).
    The dilatancy angle is measured during the shear failure phase, where the slope is positive and close to linear. It was found $\psi_{\texttt{gravel-1}} = 11^\circ$ and $\psi_{\texttt{wet-sand}} = 8^\circ$ for $\sigma_3 = 15$ kPa.}
    \label{fig:triaxial_strain}
\end{figure}

A (virtual) triaxial test was used for parameter calibration, following the procedure and setup in \cite{wiberg:2020:dem}, but with a lesser packing density. 
The model of the triaxial cell consists of frictionless rigid bodies for walls, 250 mm wide, driven by motors that are servo-controlled to maintain a given normal stress, $ \sigma_i $, $ i = 1, 2, 3 = xx, yy, zz $.
The particle samples have uniform size distribution with diameter ranging between 27 and 33 mm, and initial porosity $0.42$. 
The strength of each soil is tested by driving the horizontal walls at a vertical speed of 5 mm/s while controlling the vertical walls to maintain a given consolidation stress $\sigma_2 = \sigma_3$.
Tests are performed with different consolidation stresses in the range from 1 kPa and 75 kPa.
The motion of the walls and the stresses on them are registered during the simulation.
Example measurements of the stress deviator, $\sigma_1 - \sigma_3$, as a function of the lateral strain are shown in Fig.~\ref{fig:triaxial_stress} for the materials \texttt{gravel-1} and \texttt{wet-sand-1}.
The internal friction and bulk cohesion were determined from the Mohr's circles at failure stress, see Fig.~\ref{fig:triaxial_MC}.
The procedure was repeated, for each soil and consolidation stress, two or three times with different particle initial states.
The results are found in Table \ref{table:soil_library}.
Due to fluctuations and uncertainty in the Mohr method, the cohesion-free soils show an apparent cohesion of up to 2 kPa at most.
In these cases, the bulk cohesion is explicitly set to zero.
From the triaxial tests we also measure bulk elasticity as the secant modulus, $E_\text{b} = \Delta \sigma_{1} / \Delta \epsilon_{1}$ halfway before failure, and the dilatancy angle from $\psi = \arcsin\left[\Delta\text{tr}(\bm{\epsilon})/3\Delta\lVert\bar{\bm{\epsilon}}\rVert\right]$.
The dilatancy angles are computed from the strain curves in Fig.~\ref{fig:triaxial_strain}.
These bulk parameters are determined for a medium consolidation stress of $15$ kPa.
The procedure was repeated for all the 15 soils in Table \ref{table:soil_library}.

\begin{table}[ht] 
    \caption{A small library of soils with particle parameters pre-calibrated to desired bulk parameters using a triaxial test.}
     \footnotesize
      \centering
      \begin{tabular}{|l|rrrr|rrr|}
      \hline
            & \multicolumn{4}{|c|}{Bulk parameters} & \multicolumn{3}{|c|}{Particle parameters}   \\
      \hline
       Soil name & $\phi_\text{b}$ & $c_\text{b}$ & $\psi_\text{b}$ & $E_\text{b}$ & $\mu_\text{t}$ & $\mu_\text{r}$ & $c_\text{p}$  \\
      \hline

      \texttt{gravel-1} 	& $44^\circ$ 	&   $0$   	    & $11^\circ$ & $4.6$	& $0.4$ 	& $0.3$ 		& $0$ 			\\
      \texttt{sand-1} 	& $39^\circ$ 	&   $0$ 		& $9^\circ$ & $4.5$ 	&  $0.3$ 	& $0.1$ 		& $0$ 			\\
      \texttt{sand-2} 	& $32^\circ$ 	&   $0$ 		& $8^\circ$ & $3.5$ 	&  $0.3$	 & $0.02$ 		& $0$ 			\\
      \texttt{wet-sand-1} & $33^\circ$ 	&   $8.7$ 		& $8^\circ$  & $4.0$   &  $0.3$ 	& $0.1$ 		& $30$ 		    \\
      \texttt{dirt-1} 	& $40^\circ$ 	&   $2.1$ 		& $13^\circ$  & $6.5$   &  $0.4$ 	& $0.1$  		& $2.5$ 		\\
      \texttt{dirt-2} 	& $37^\circ$ 	&   $4.9$ 		& $10^\circ$ & $4.6$    &  $0.4$ 	& $0.1$  		& $12.7$ 		\\
      \texttt{dirt-3} 	& $35^\circ$ 	&   $21$ 		& $10^\circ$  & $5.8$    &  $0.4$ 	& $0.1$  		& $63.7$ 		\\
      \texttt{fs-strong} & $42^\circ$ 	&   $0$ 		& $13^\circ$ & $3.8$    &  $0.5$ 	& $0.1$ 		& $0$ 			\\
      \texttt{fs-weak}   & $35^\circ$ 	&   $0$ 		& $8^\circ$  & $2.8$   &  $0.3$ 	& $0.05$ 		& $0$ 			    \\
      \texttt{cfs-strong} & $32^\circ$ 	&   $7.1$ 		& $7^\circ$  & $4.6$   &  $0.3$ 	& $0.05$ 		& $23$ 		    \\
      \texttt{cfs-medium} & $32^\circ$ 	&   $4.3$ 		& $7^\circ$  & $4.2$   &  $0.3$ 	& $0.05$  		& $12$ 		    \\
      \texttt{cfs-weak}  & $21^\circ$ 	&   $4.8$ 		& $6^\circ$ & $3$    &  $0.15$ 	& $0.025$  		& $23$ 		    \\
      \texttt{cfs-weakest} & $12^\circ$ 	&  $2.4$ 		& $1^\circ$  & $1$   &  $0.06$ & $0.01$  		& $23$ 		    \\
      \texttt{cs-weak}   & $5^\circ$ 	&  $3.1$ 	    & $2^\circ$ & $0.7$    &  $0.0$ 	& $0.0$  		& $50$ 		        \\
      \texttt{frictionless} & $6^\circ$ 	&  $0$ 		    & $1^\circ$     & $0.25$  &  $0.0$ 	& $0.0$  		& $0$ 		\\
      \hline
      Units & deg & kPa & deg & MPa &  &   & kPa \\
      \hline
  \end{tabular}
  \label{table:soil_library}
\end{table}   

\section*{Simulation tests}
The multiscale model was tested and evaluated using two primary test systems, which are bulldozing and excavation in a flat soil bed.
For reference, the same siluations were performed using the microscale model. 
The digging resistance, resulting terrain surface heightfield and computational performance of the two models was then compared. 
All materials in Table \ref{table:soil_library} were subjected to tests, see the supplementary \href{https://www.algoryx.se/papers/terrain/}{Video 1} and \href{https://www.algoryx.se/papers/terrain/}{Video 2}.
The results for three frictional soils (\texttt{gravel-1}, \texttt{sand-1}, \texttt{frictionless}) and three cohesive soils (\texttt{dirt-1}, \texttt{wet-sand}, \texttt{cfs-weak}) are reported more detail.
In the microscale simulations the soil bed consisted of 50e3 particles, with size uniformly distributed between 55 and 45 mm
and specific mass density of $2200$ kg/m$^3$.
The voxel size in the multiscale model was $0.1$ m, the specific mass density also $2200$ kg/m$^3$ but the bulk mass density was set to $1474$ kg/m$^3$ which correspond to a void ratio of $0.33$.
Default values for the aggregate stiffness multiplier was set to $0.001$ and $1.0$ at the terrain and tool interface, respectively.

\begin{figure}[h] 
  \centering
  \includegraphics[width=0.45\textwidth]{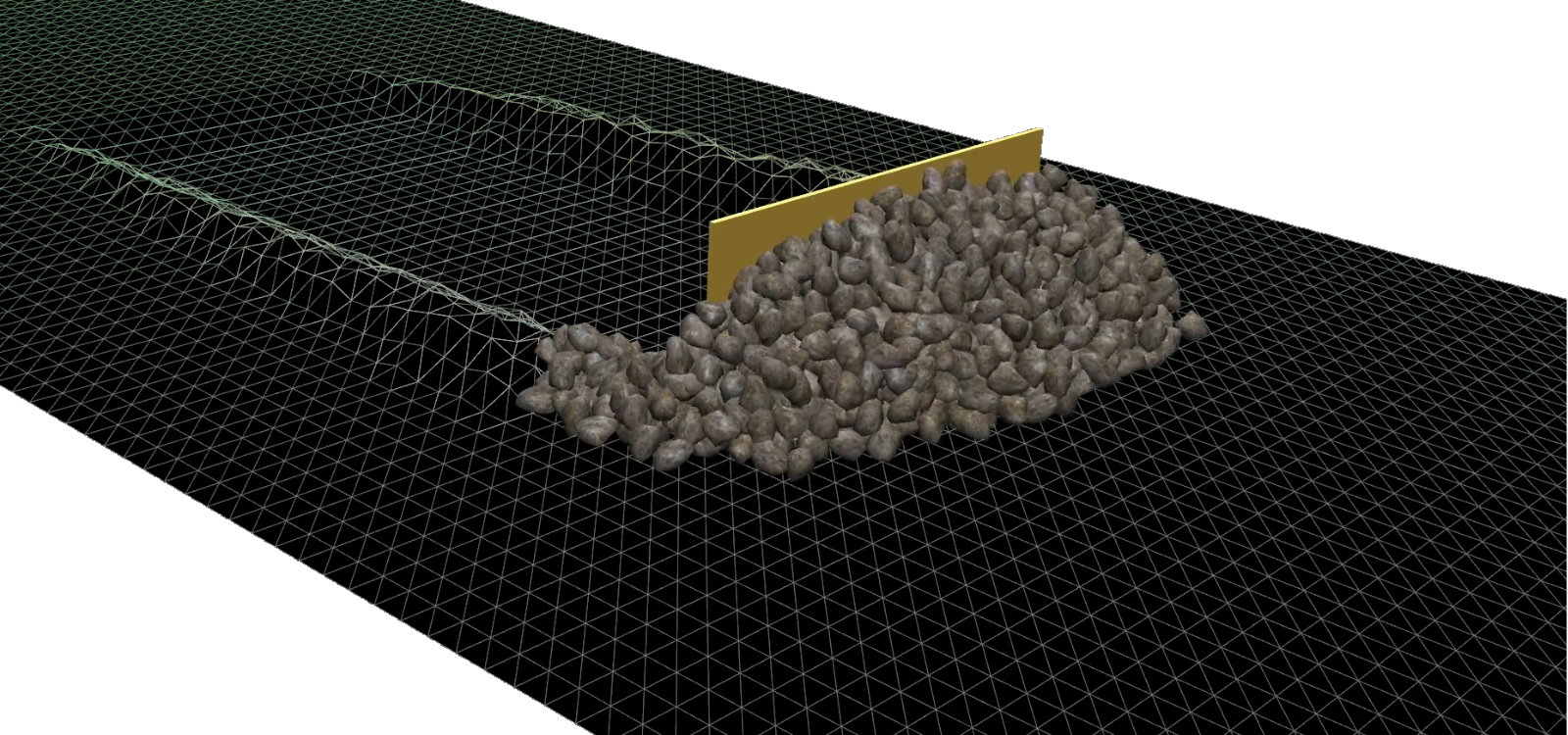}
  \includegraphics[width=0.45\textwidth]{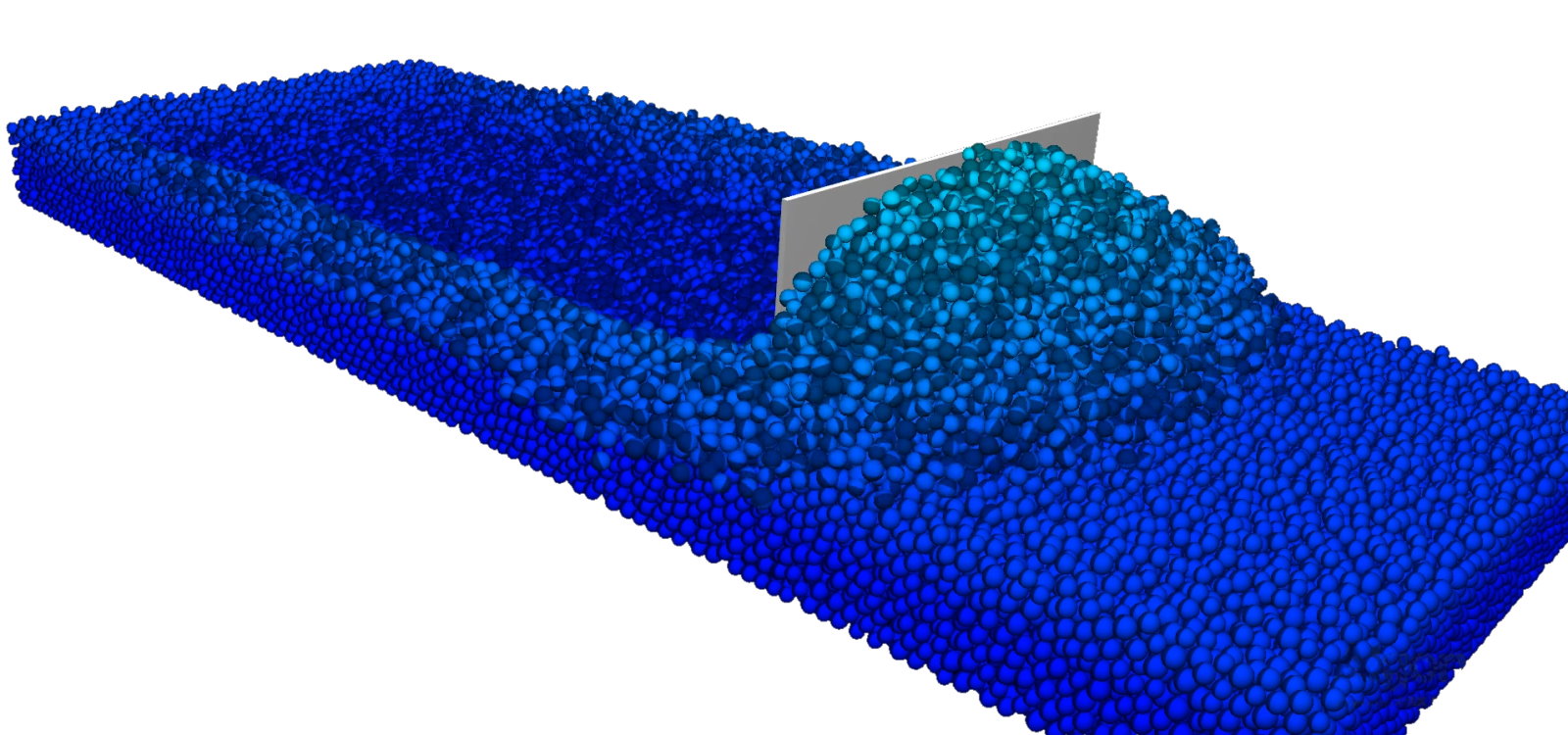}\\
  \includegraphics[width=0.12\textwidth]{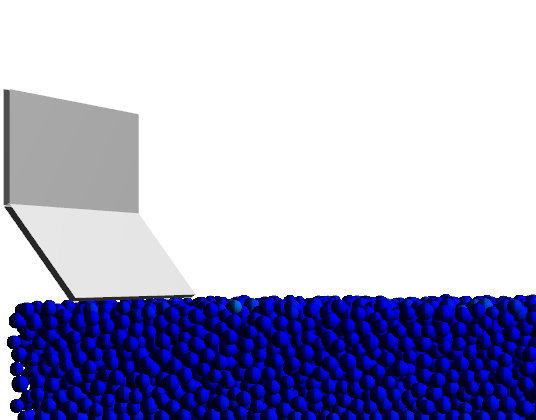} 
  \includegraphics[width=0.12\textwidth]{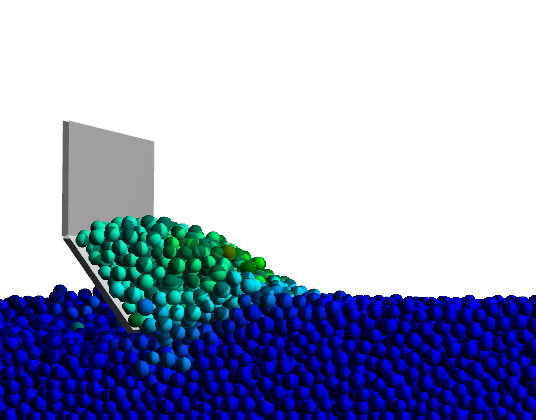} 
  \includegraphics[width=0.12\textwidth]{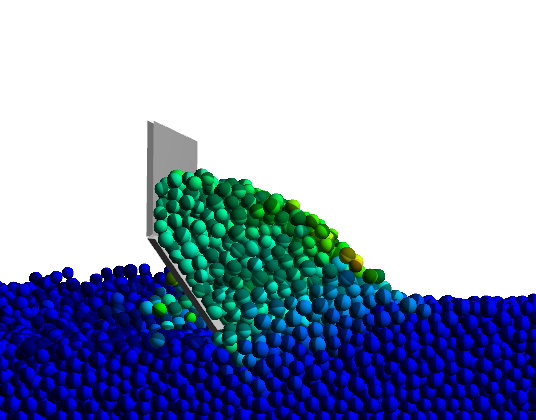} 
  \includegraphics[width=0.12\textwidth]{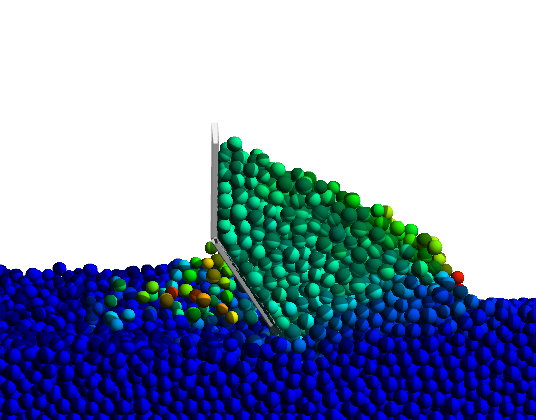} 
  \includegraphics[width=0.12\textwidth]{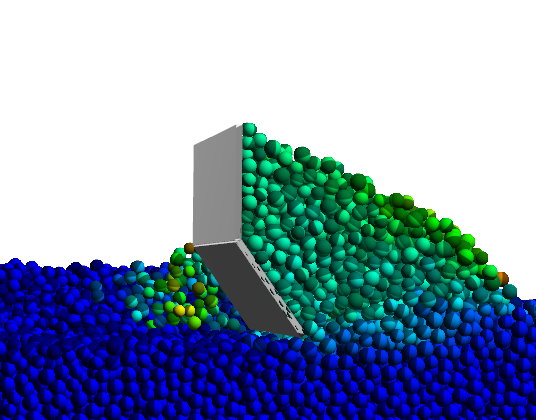} 
  \includegraphics[width=0.12\textwidth]{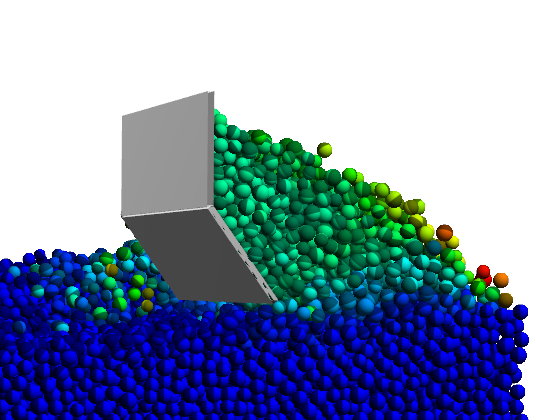} 
  \includegraphics[width=0.12\textwidth]{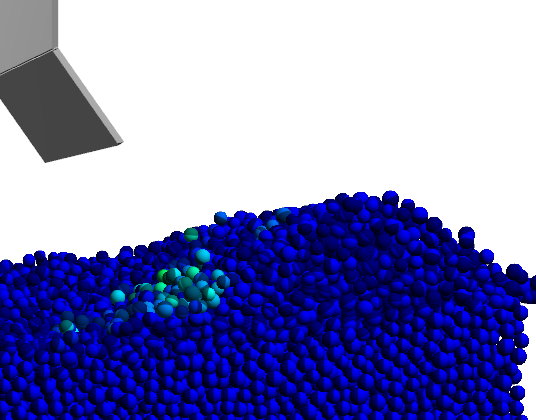}\\
  \includegraphics[width=0.12\textwidth]{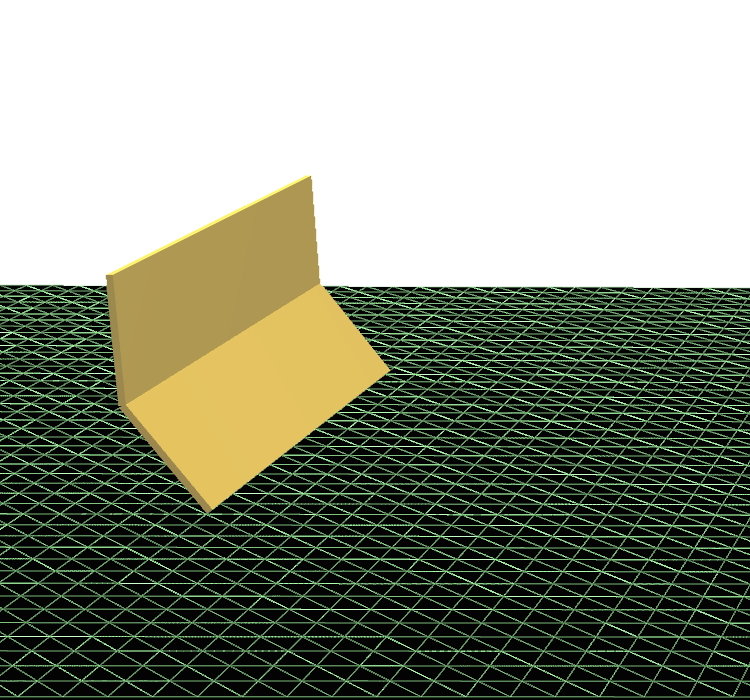}
  \includegraphics[width=0.12\textwidth]{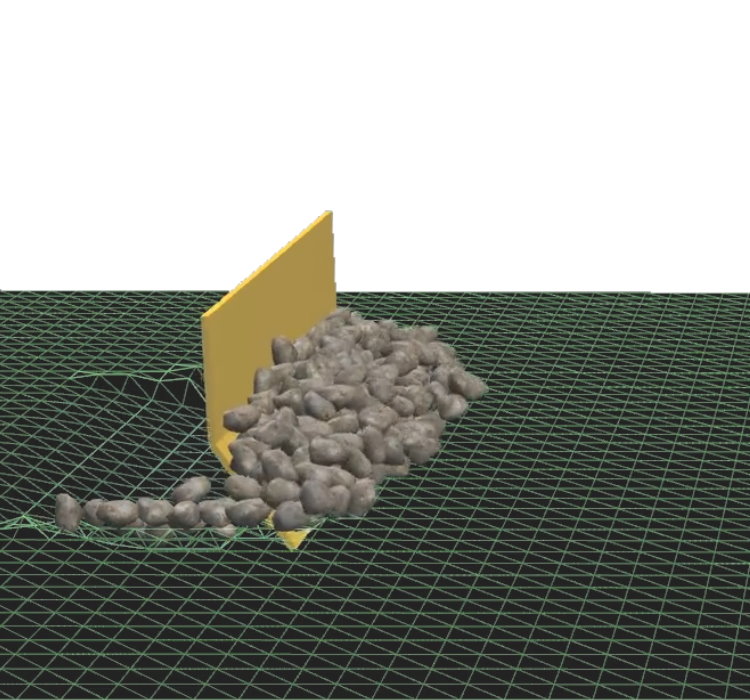}
  \includegraphics[width=0.12\textwidth]{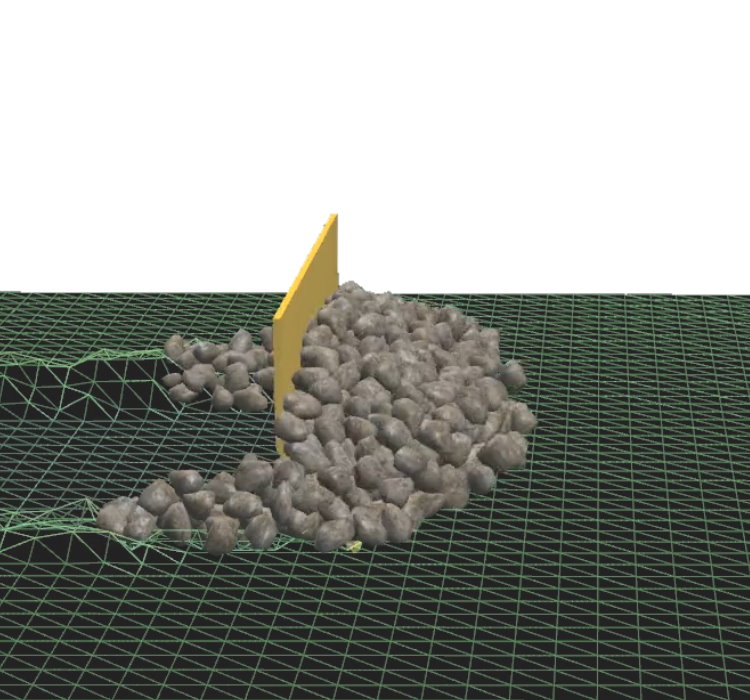}
  \includegraphics[width=0.12\textwidth]{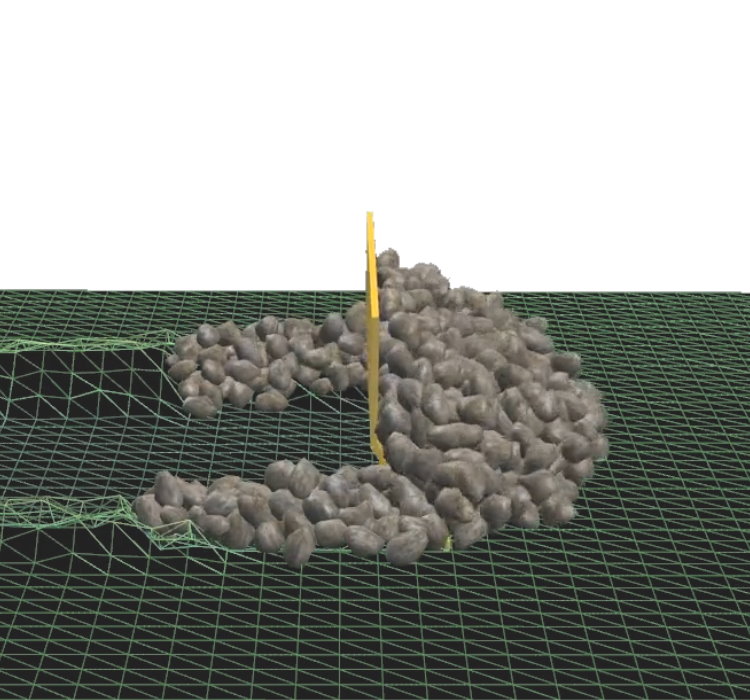}
  \includegraphics[width=0.12\textwidth]{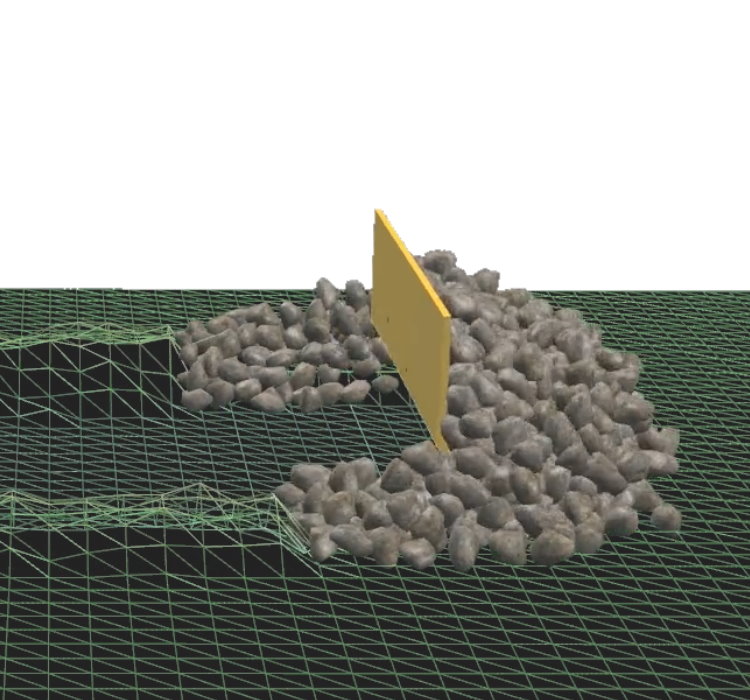}
  \includegraphics[width=0.12\textwidth]{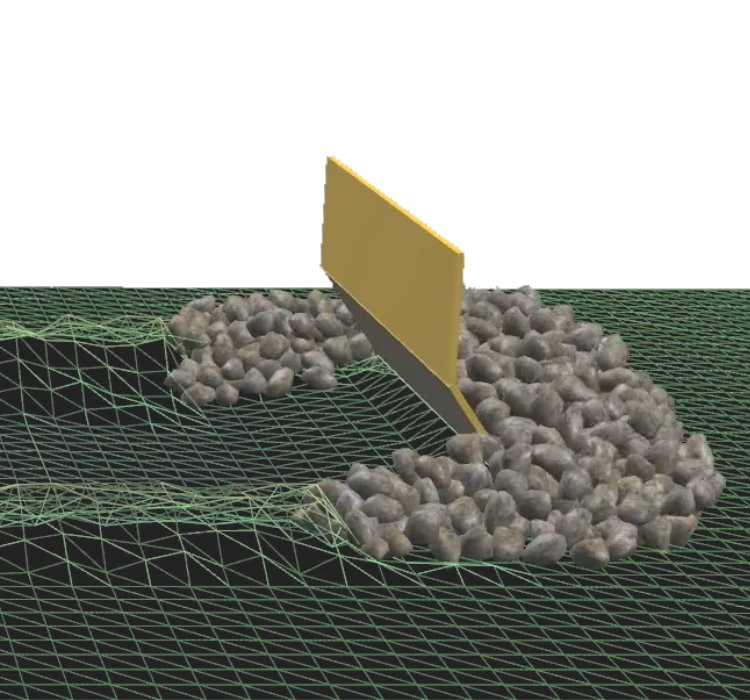}
  \includegraphics[width=0.12\textwidth]{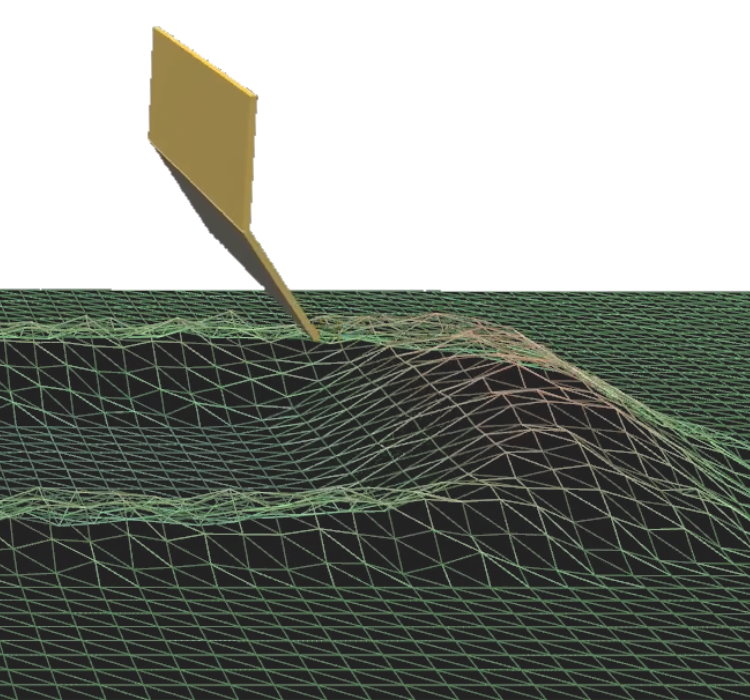}
  \caption{Simulated bulldozing for comparing the multiscale terrain model (left) and the microscale reference model (right, particles colour coded by particle height position) at time $8$ s and using material \texttt{sand-1}.
  Second and third row show sideview of the microscale model (particles colour coded by velocity, blue to red by 0 to 1.5 m/s) and the multiscale model, at time 0, 2, 4, 6, 8, 10 and 12 s. See supplementary \href{https://www.algoryx.se/papers/terrain/}{Video 3}.}
  \label{fig:NDEM_agxTerrain_bulldozing}
\end{figure}

In the bulldozing test, shown in Fig.~\ref{fig:NDEM_agxTerrain_bulldozing}, a blade cuts a bed of \texttt{sand-1} soil and pushes the material in front of it. 
After some distance, the material is dumped in a pile, as the blade is stopped, lifted, and reversed.
The cutting depth is $ 0.05 $ m and the length is $5$ m.
The blade has mass $100$ kg, sectioned vertically in two plates, $1.6$ m wide, $0.37$ m tall, $0.02$ m thick, and with a relative angle of $55^\circ$. 
It is attached with a lock constraint to a kinematic body that is driven with horizontal speed of $ 0.5 $ m/s.  See the supplementary \href{https://www.algoryx.se/papers/terrain/}{Video 3} for bulldozing in \texttt{sand-1} and \texttt{dirt-1}.
For the penetration resistance, the blade's edge is discretized by $80$ teeth with $10$ mm maximum radius and length, and with $2.5$ mm minimum radius.
The penetration force scaling is then calibrated to $10$ for all materials except for \texttt{gravel}, \texttt{dirt-1} and \texttt{wet-sand} which are given the value $20$.

\begin{figure}[h]
    \centering
    \hspace{1.5mm}
    \begin{picture}(102,45)
        \put(0,0){\includegraphics[width=0.2825\textwidth]{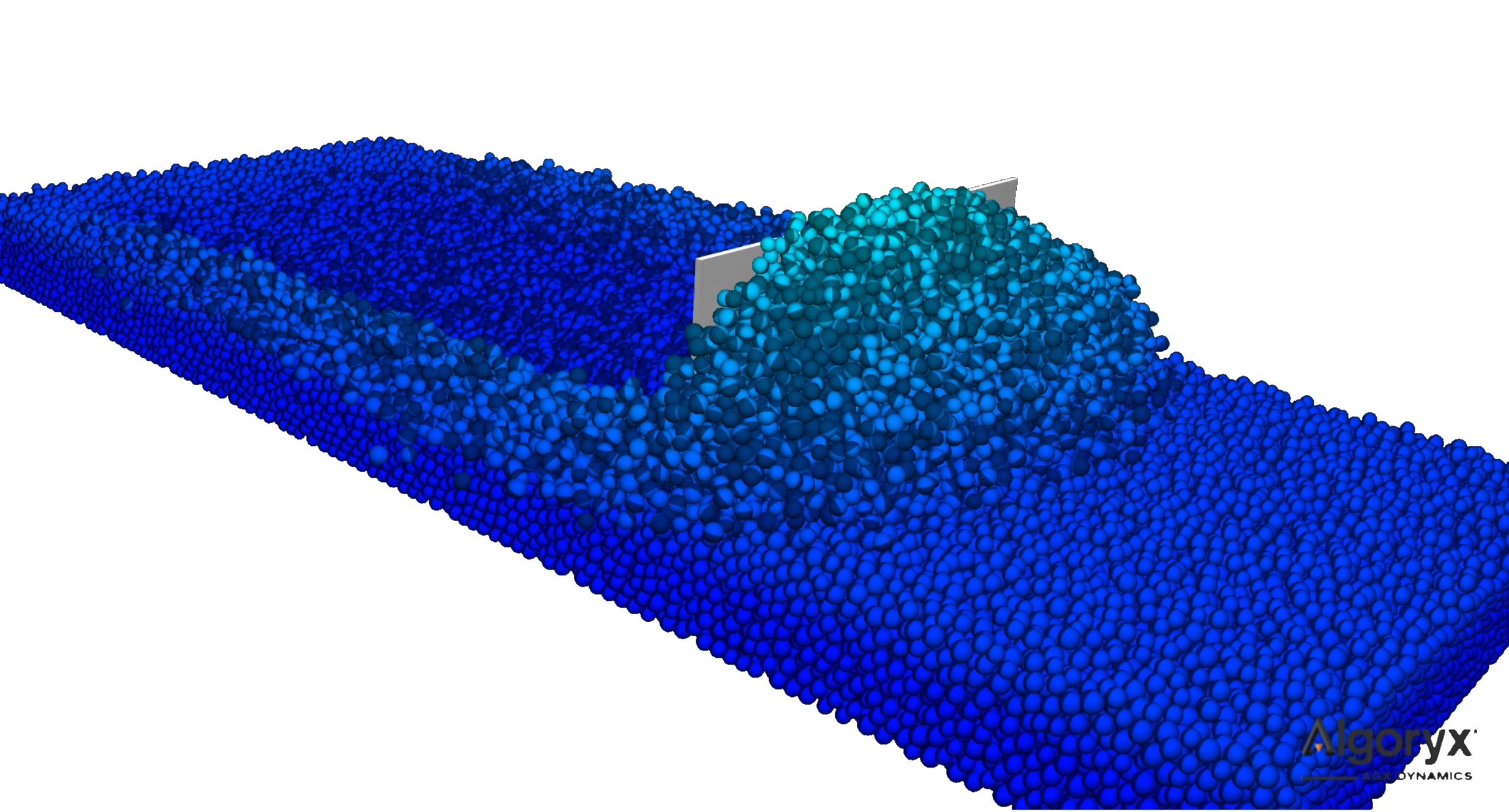}}
        \put(45,45){\tiny{gravel-1}}
    \end{picture}
    \begin{picture}(102,45)
        \put(0,0){\includegraphics[width=0.2825\textwidth]{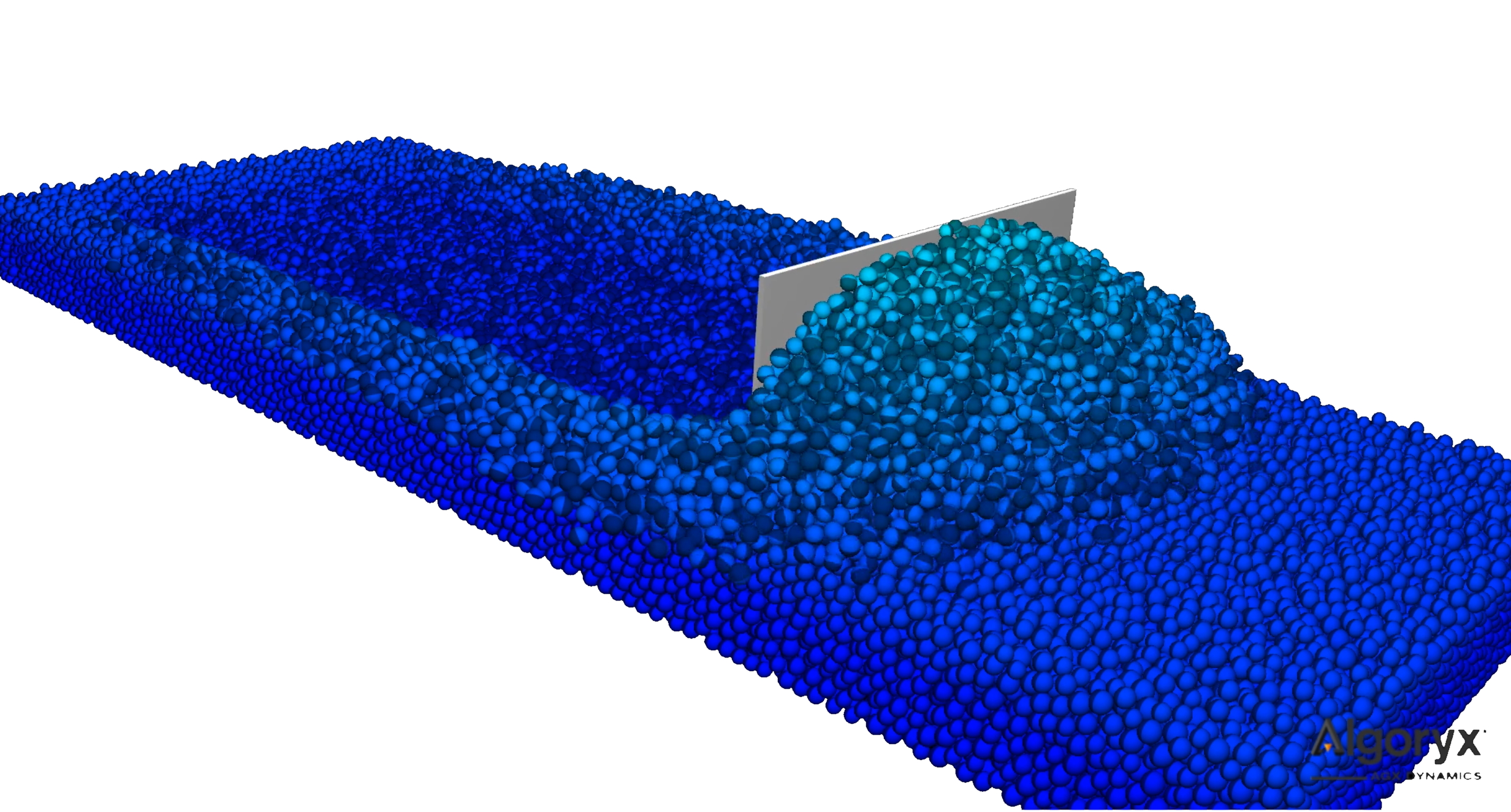}}
        \put(45,45){\tiny{sand-1}}
    \end{picture}
    \begin{picture}(102,45)
        \put(0,0){\includegraphics[width=0.2825\textwidth]{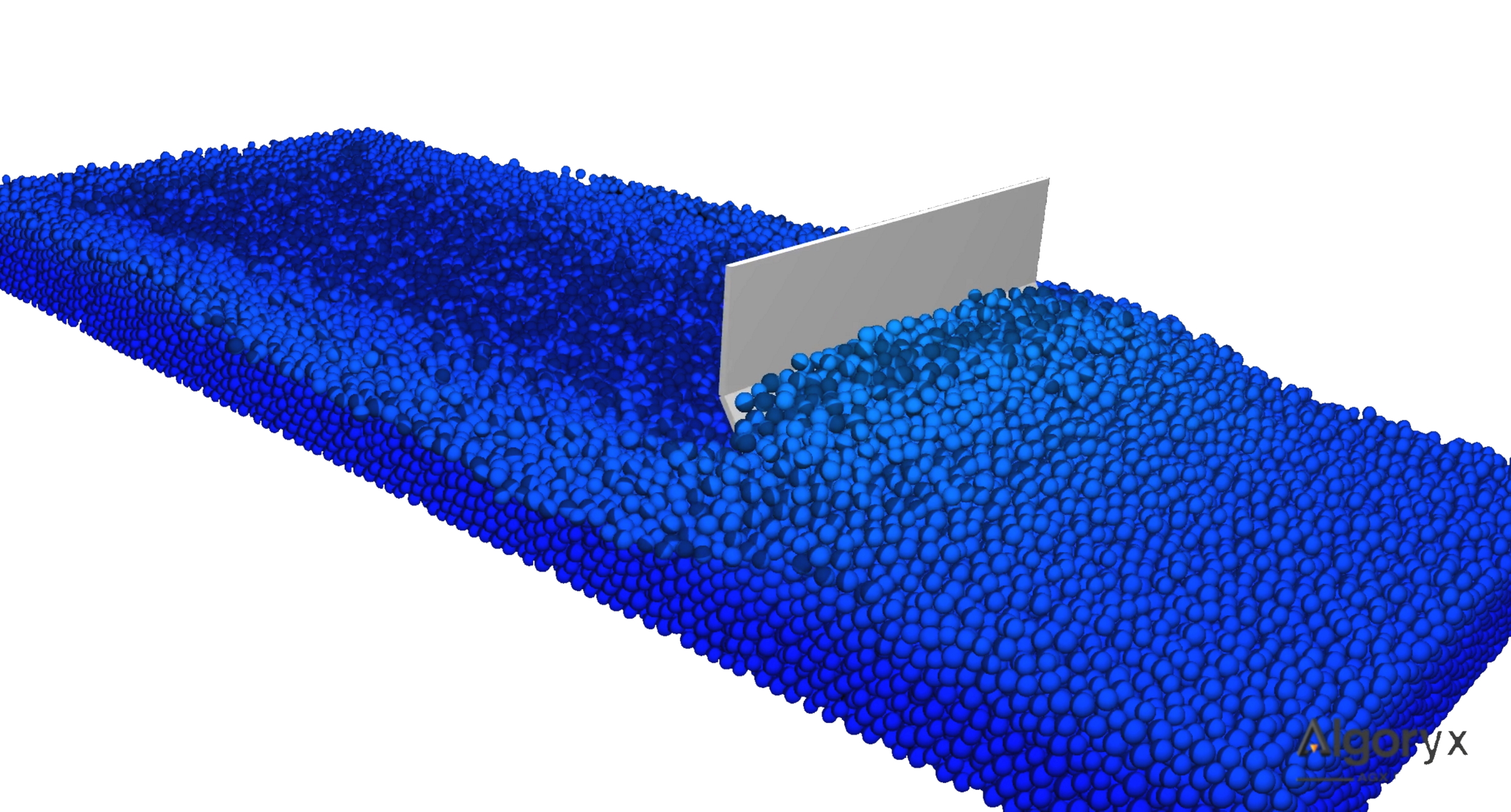}}
        \put(45,45){\tiny{frictionless}}
    \end{picture}
    \\
    \includegraphics[trim=0 0 40 35, clip, height=0.33\textwidth]{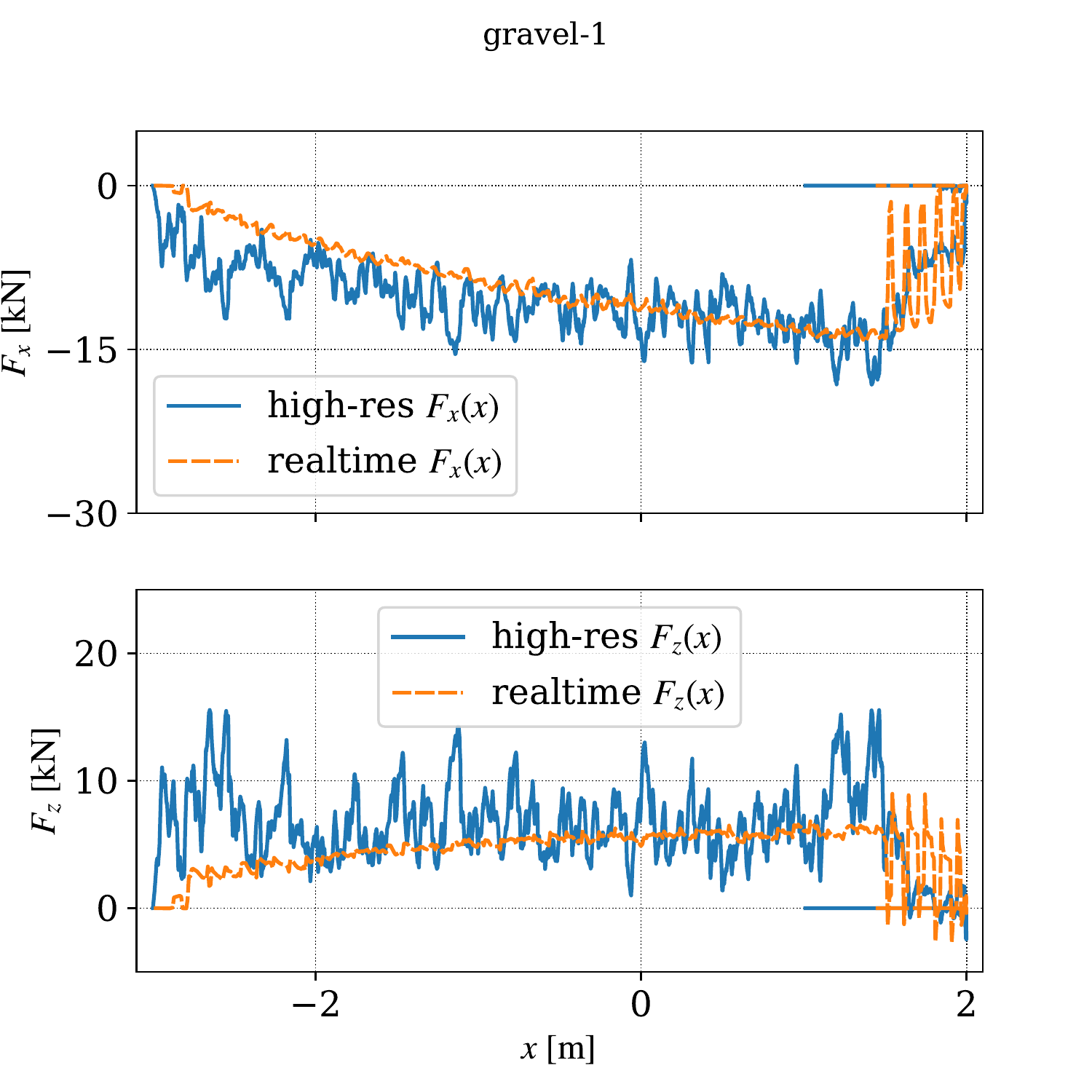}
    \includegraphics[trim=50 0 40 35, clip,height=0.33\textwidth]{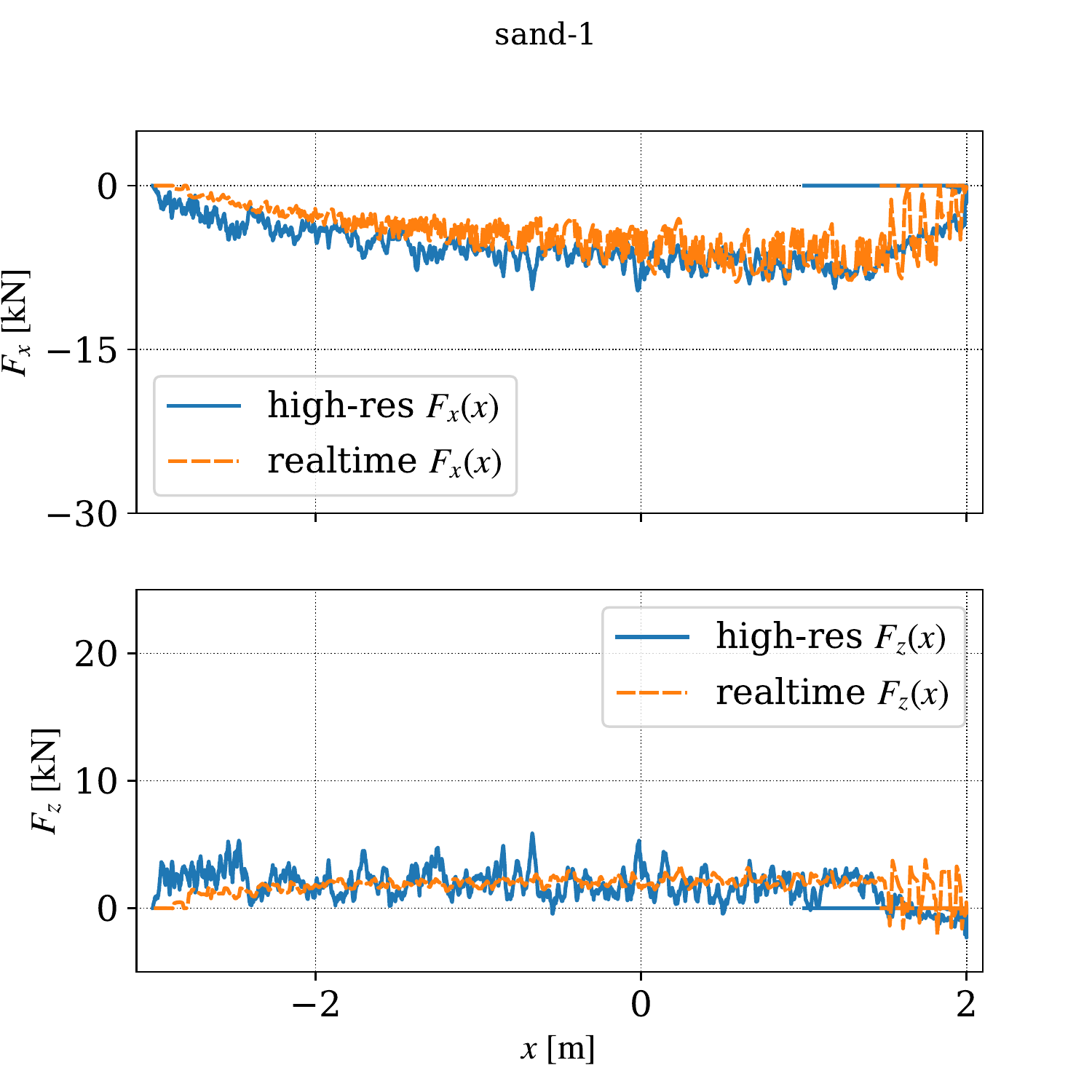}
    \includegraphics[trim=50 0 40 35, clip,height=0.33\textwidth]{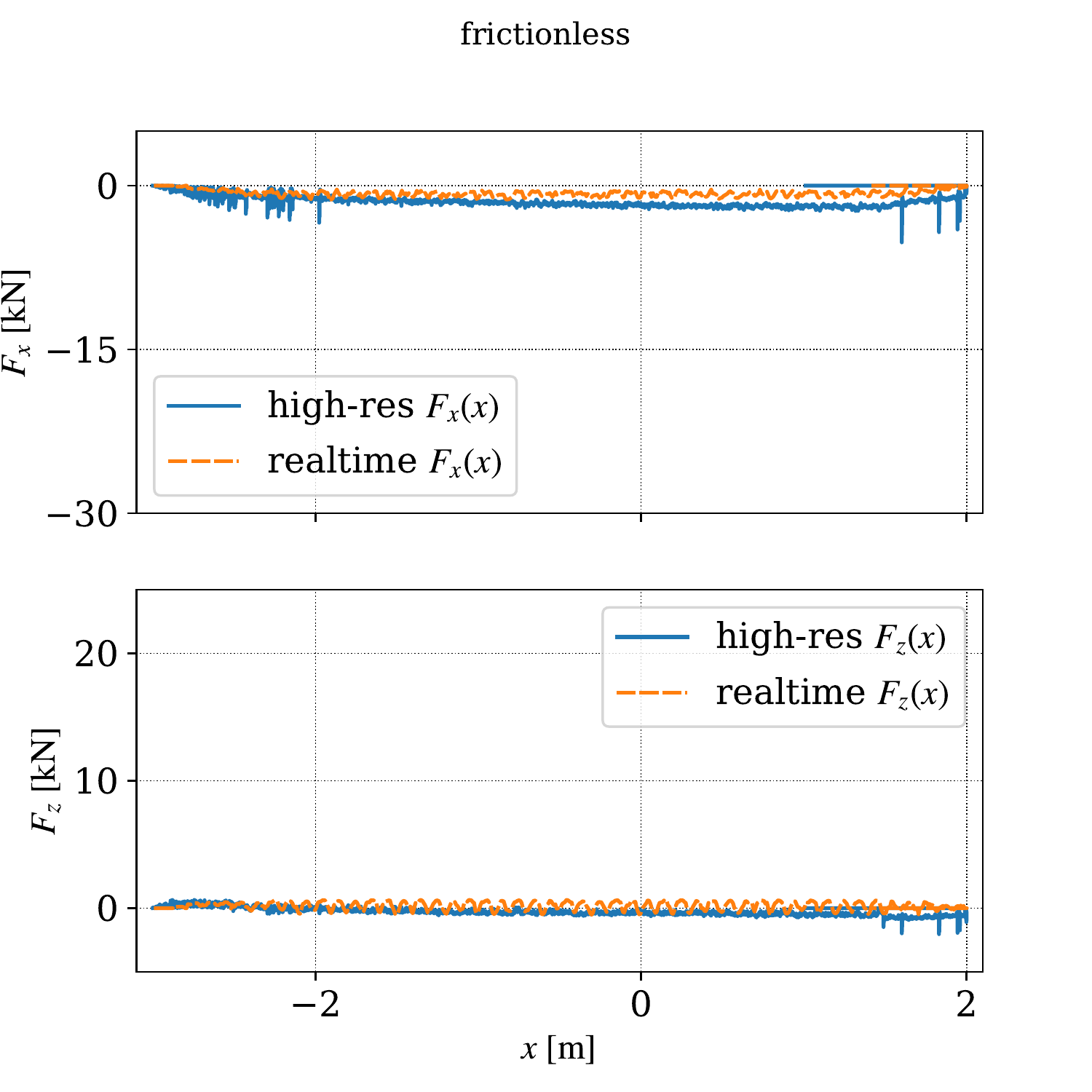}
    \caption{Force resistance from bulldozing using three frictional soils and the microscale model, with snapshot taken at 8 s.}
    \label{fig:NDEM_agxTerrain_bulldozer_frictional}
\end{figure}
\begin{figure}[h]
    \centering
    \hspace{1.5mm}
    \begin{picture}(102,45)
        \put(0,0){\includegraphics[width=0.2825\textwidth]{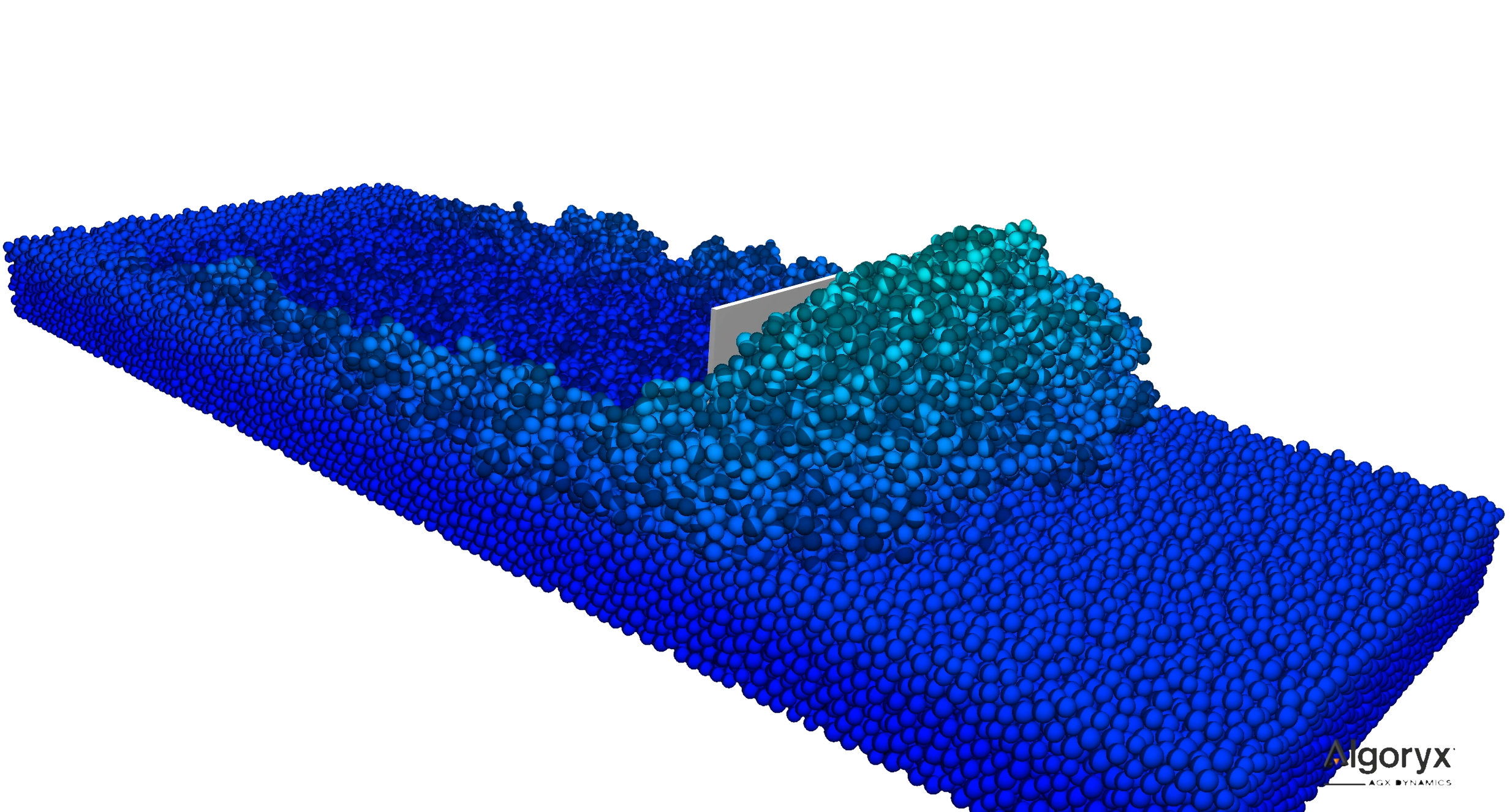}}
        \put(45,45){\tiny{dirt-1}}
    \end{picture}
    \begin{picture}(102,45)
        \put(0,0){\includegraphics[width=0.2825\textwidth]{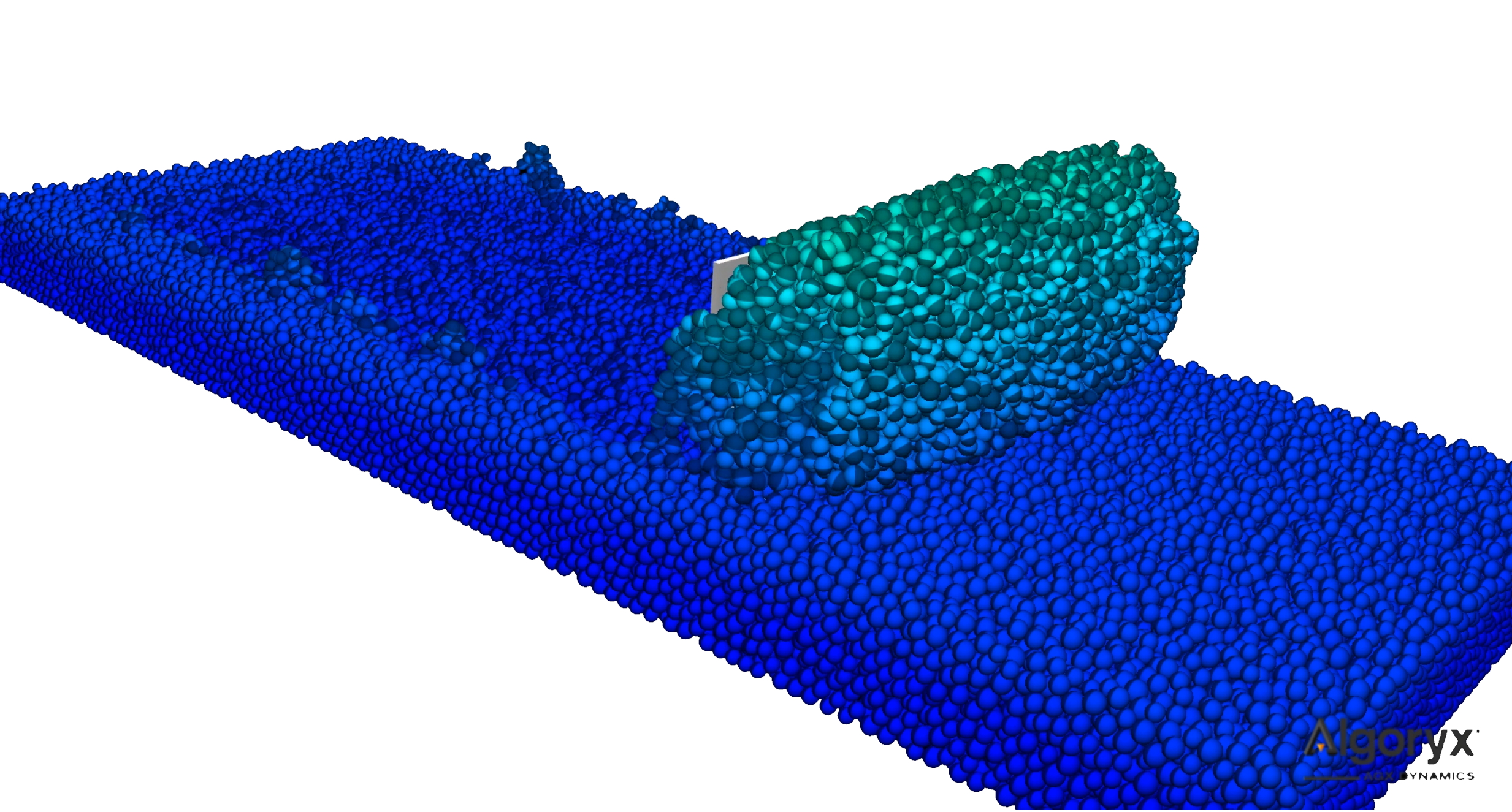}}
        \put(43,45){\tiny{wet-sand-1}}
    \end{picture}
    \begin{picture}(102,45)
        \put(0,0){\includegraphics[width=0.2825\textwidth]{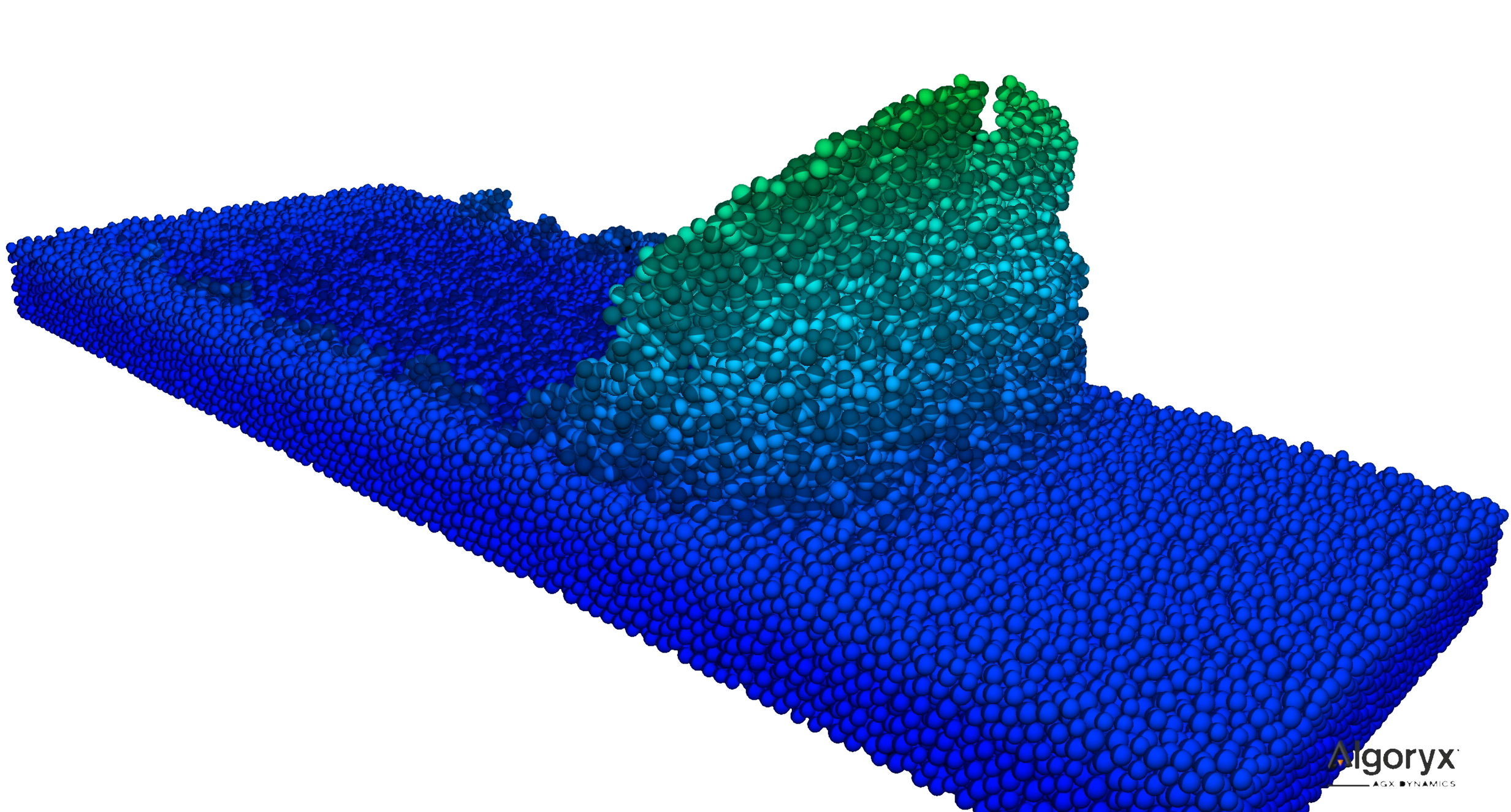}}
        \put(30,45){\tiny{cfs-weak}}
    \end{picture}
    \\
    \includegraphics[trim=0 0 40 35, clip, height=0.33\textwidth]{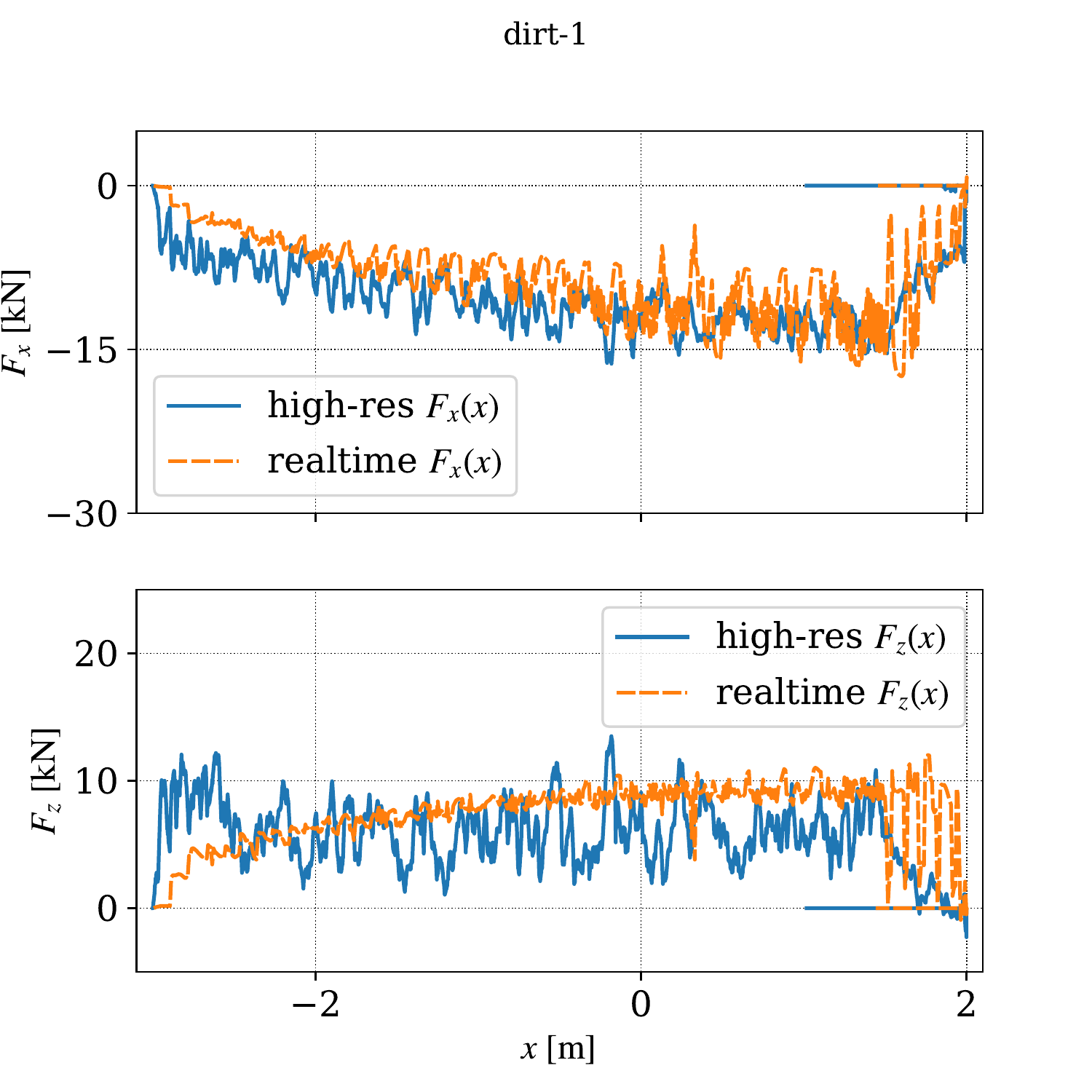}
    \includegraphics[trim=50 0 40 35, clip, height=0.33\textwidth]{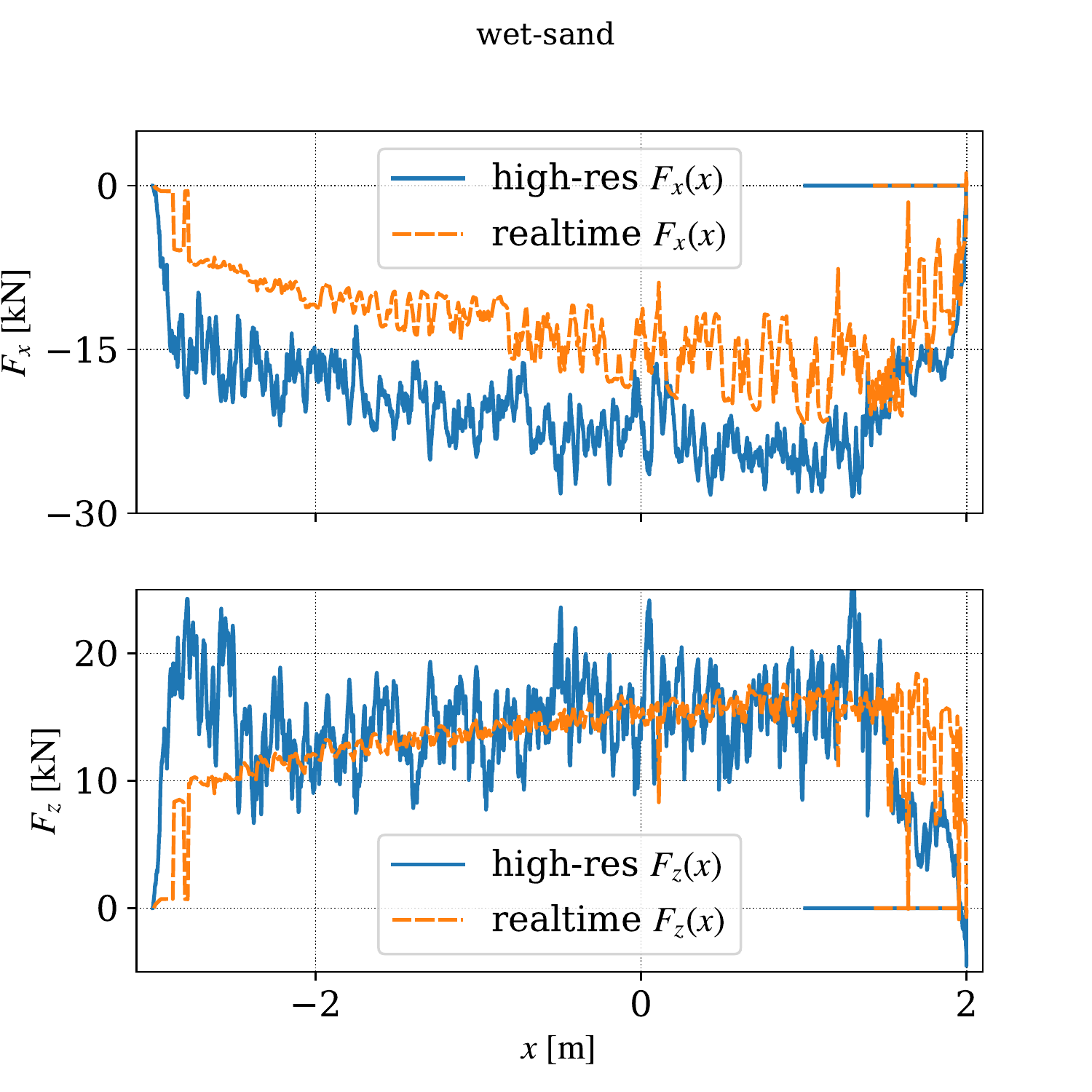}
    \includegraphics[trim=50 0 40 35, clip, height=0.33\textwidth]{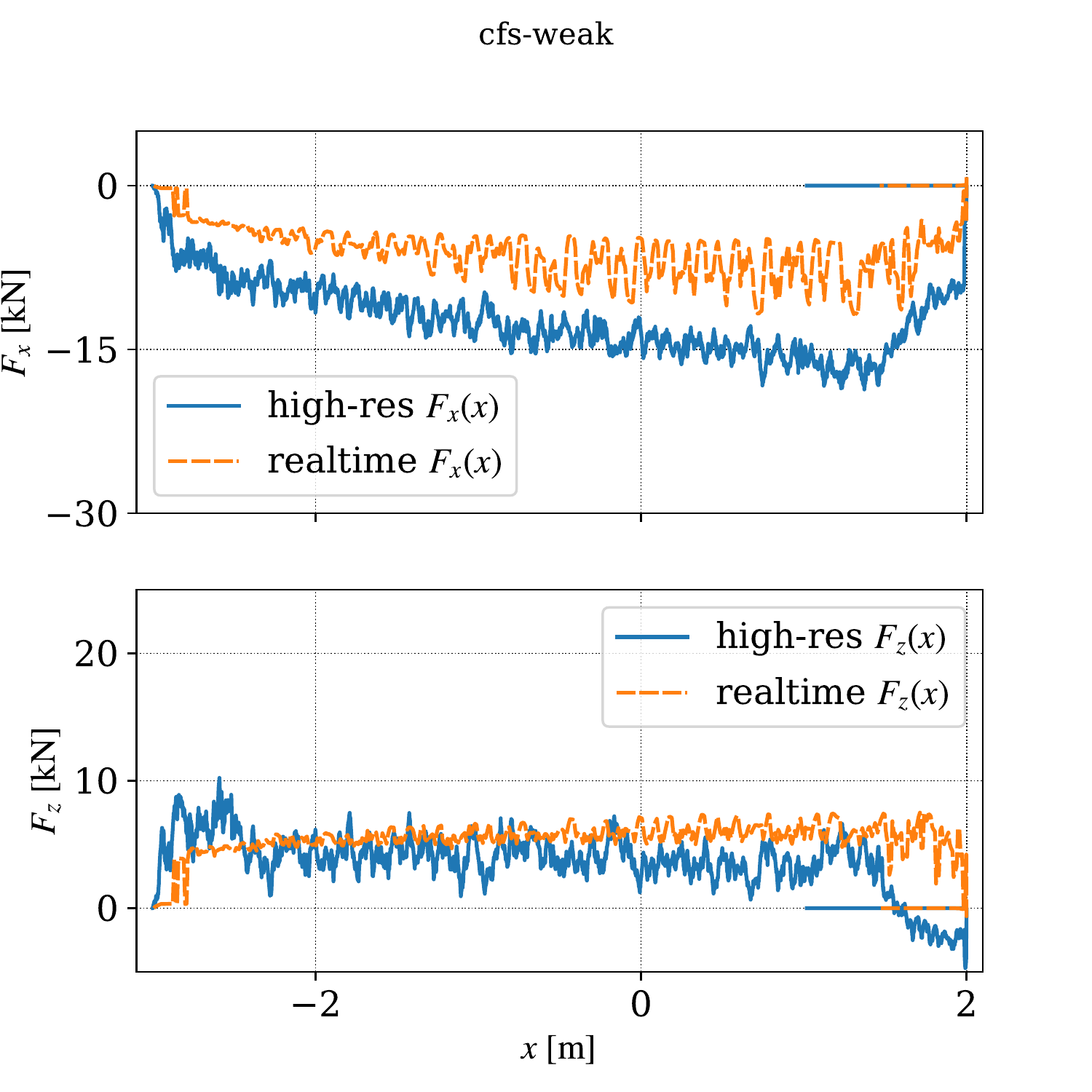}
    \caption{Force resistance from bulldozing using three cohesive soils and the microscale model, with snapshot taken at 8 s.}
    \label{fig:NDEM_agxTerrain_bulldozer_cohesive}
\end{figure}
\begin{figure}[h]
    \centering
    \includegraphics[trim=0 0 0 37, clip, width=0.45\textwidth]{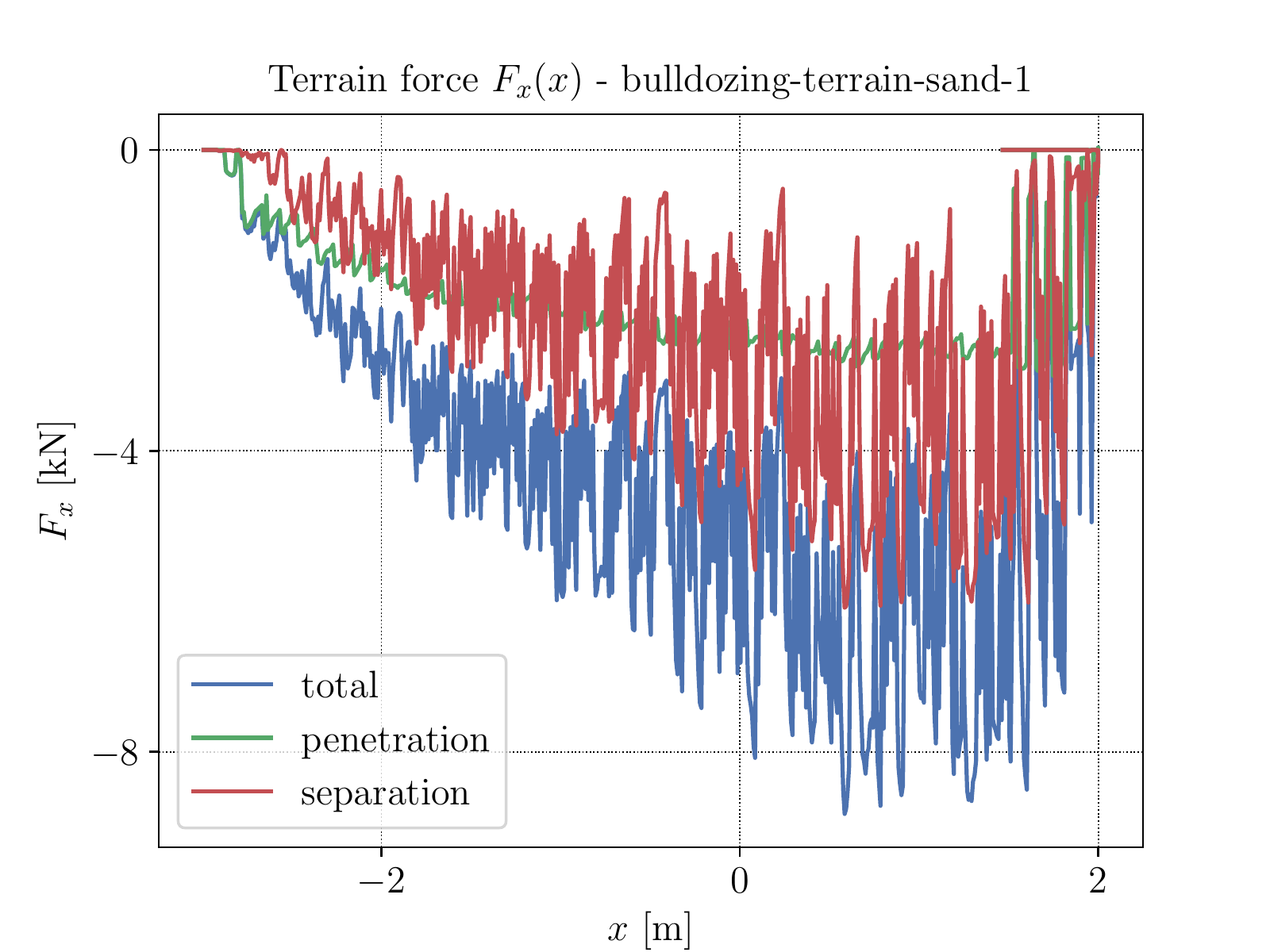}
    \includegraphics[trim=0 0 0 37, clip, width=0.45\textwidth]{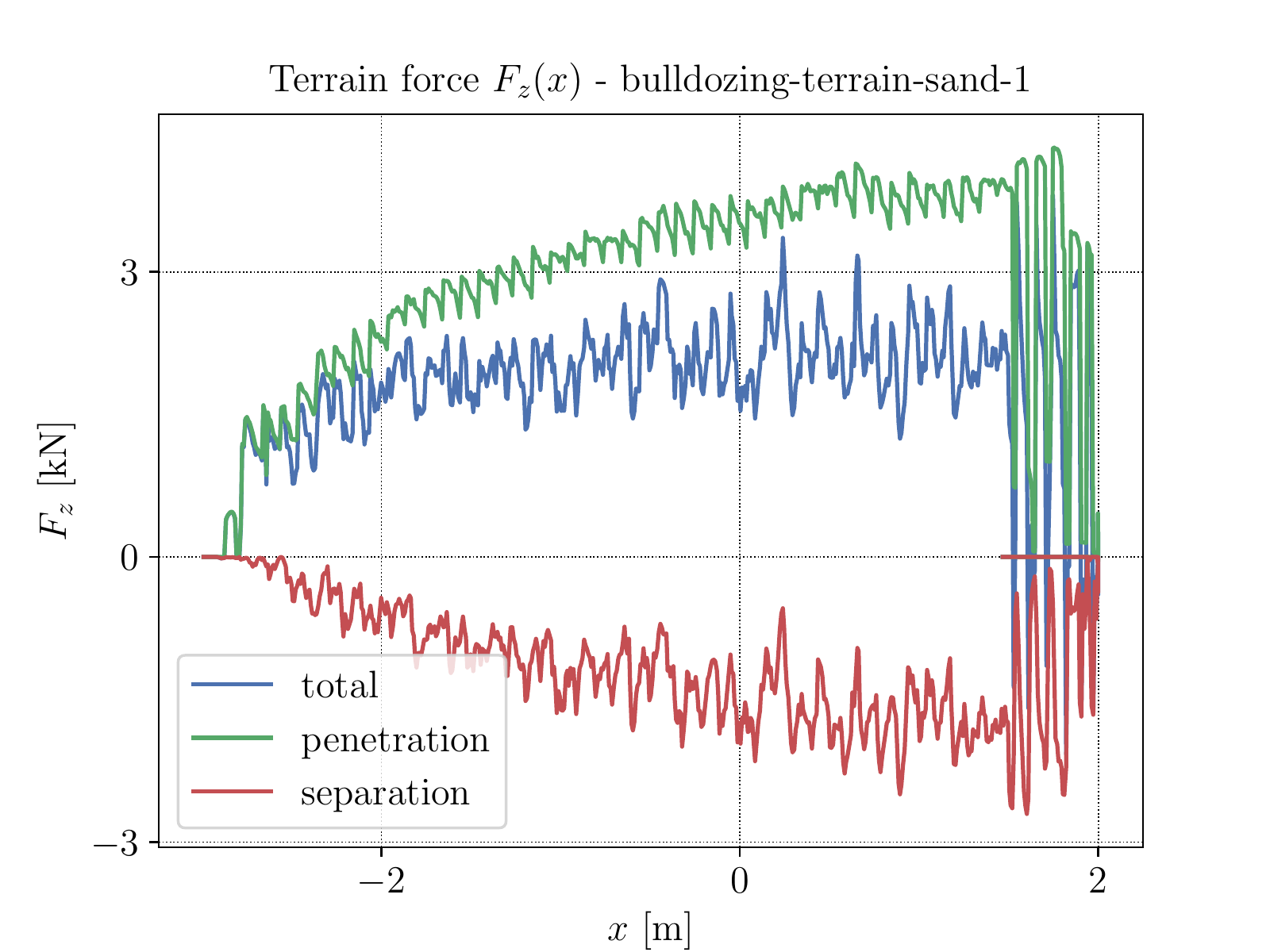}
    \caption{The digging resistance from bulldozing a bed of \texttt{sand-1} soil simulated using the multiscale model. The force is divided in the contributions from the penetration and separation models.}
    \label{fig:bulldozing_sand-1_force}
\end{figure}

Fig.~\ref{fig:NDEM_agxTerrain_bulldozer_frictional} and \ref{fig:NDEM_agxTerrain_bulldozer_cohesive} show the force from the terrain on the blade as a function of the horizontal position $x$.
The force from the multiscale and reference model largely follow each other, especially during the intermediate phase, starting after the blade is lowered into the ground ($ x = - 2.75$ m) and ending where the blade begins to rise ($ x = 1.5 $ m).
In the intermediate phase, the horizontal force grows gradually as soil accumulates in front of the blade, until material start spilling off at the same rate.
The vertical force is nearly constant in this phase.
The bulldozing resistance for gravel is larger than for sand and for the frictionless soil, as can be expected by the difference in internal friction.
The more fine-grained reference model has larger fluctuations than the coarse-grained multiscale model.
This may seem counter intuitive. 
Presumably this is due to the higher numerical precision (smaller time-step and larger PGS solver iteration count) of the reference model, that capture the granular nature of the soil better, i.e., formation of strong force chains, that grow and collapse in an irregular manner.
In the multiscale model, on the other hand, much of the fluctuations are lost due to the coarse-graining of particles and fluidized mass into a single aggregate body that fails more smoothly.
The time-averaged force of the models agrees generally within 25\%.
The largest deviations occur during the lowering and raising of the blade.
The contributions to the force on the blade from penetration and separation resistance in the multiscale model are shown in Fig.~\ref{fig:bulldozing_sand-1_force} for the case of \texttt{sand-1}.
It can be concluded that the penetration force is overestimated during the phase of raising the blade.
It dominates the vertical component of the force.
The resting and densely packed soil is forced to expand at the tip of the blade.
When the blade is raised, the cutting edge moves through the shear zone, which should be looser and offer lesser resistance to penetration than dense soil.
This is automatically captured by the resolved reference model but not in the multiscale model.
In all cases, the pile of accumulated material in front of the blade is higher and steeper with the reference model than with the multiscale model.
This is consistent with the larger time-step and lower PGS solver iteration count in the multiscale model, which imply larger numerical errors that manifest themselves as excessive slipping and rolling in the particle contacts.
However, since the soil's bulk properties have been calibrated in advance the errors affect the size of the aggregate but not on its fundamental shape or strength. 
In a sense, these mesoscale errors are filtered out in the aggregation process to the macroscale model and do not propagate into the digging resistance.
As expected, the digging resistance is much smaller for \texttt{frictionless} soil.
It is smaller for the multiscale model than for the reference model.
This can be understood by the fact that particles in the multiscale model slide over a frictionless plane while the motion of the particles in the refence model are damped by the dissipative contacts (zero restitution) with the irregular surface formed by the resting particles. 

\begin{figure}
    \centering
    \includegraphics[height=0.35\textwidth,trim={14mm 0mm 42mm 9mm},clip]{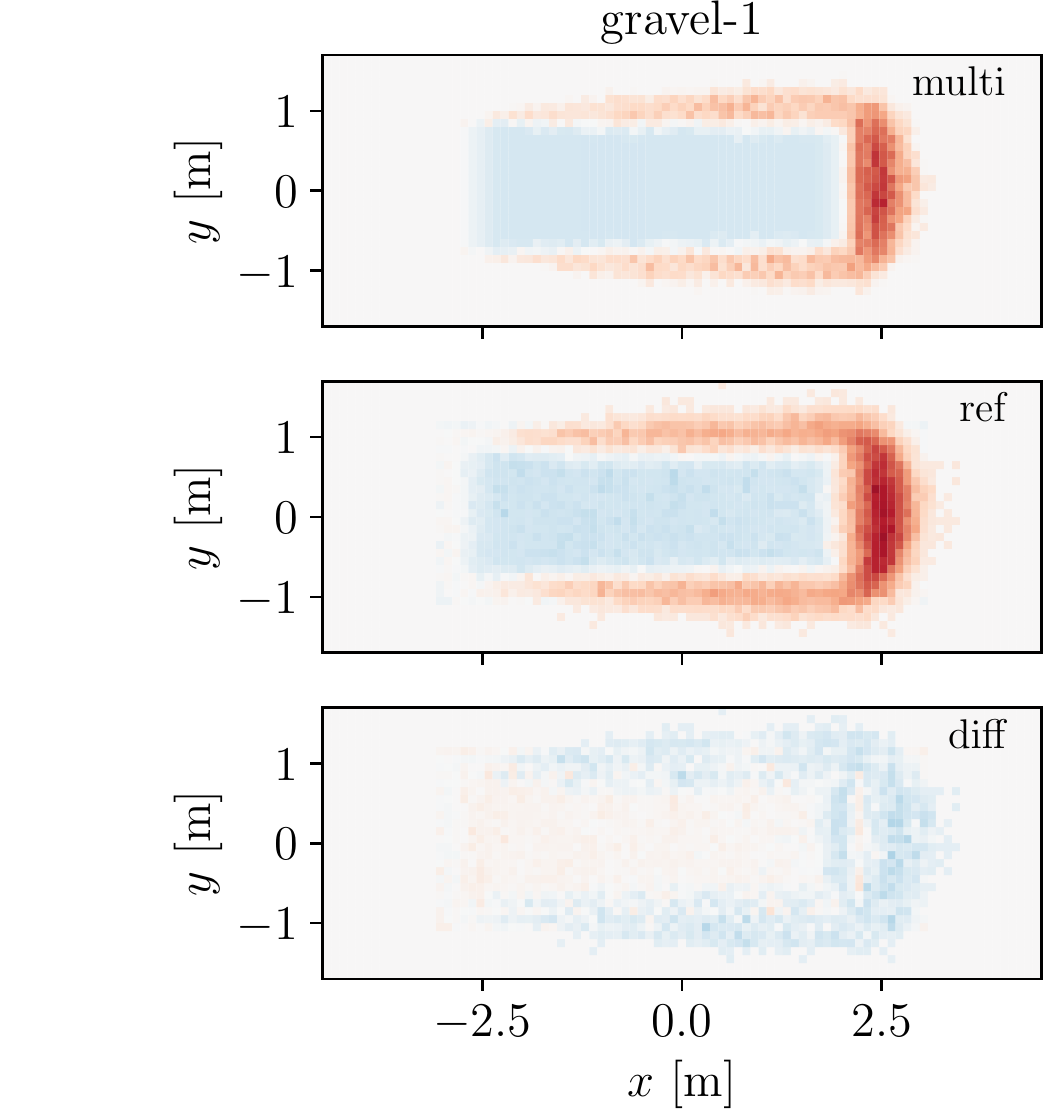}
    \includegraphics[height=0.35\textwidth,trim={45mm 0mm 42mm 9mm},clip]{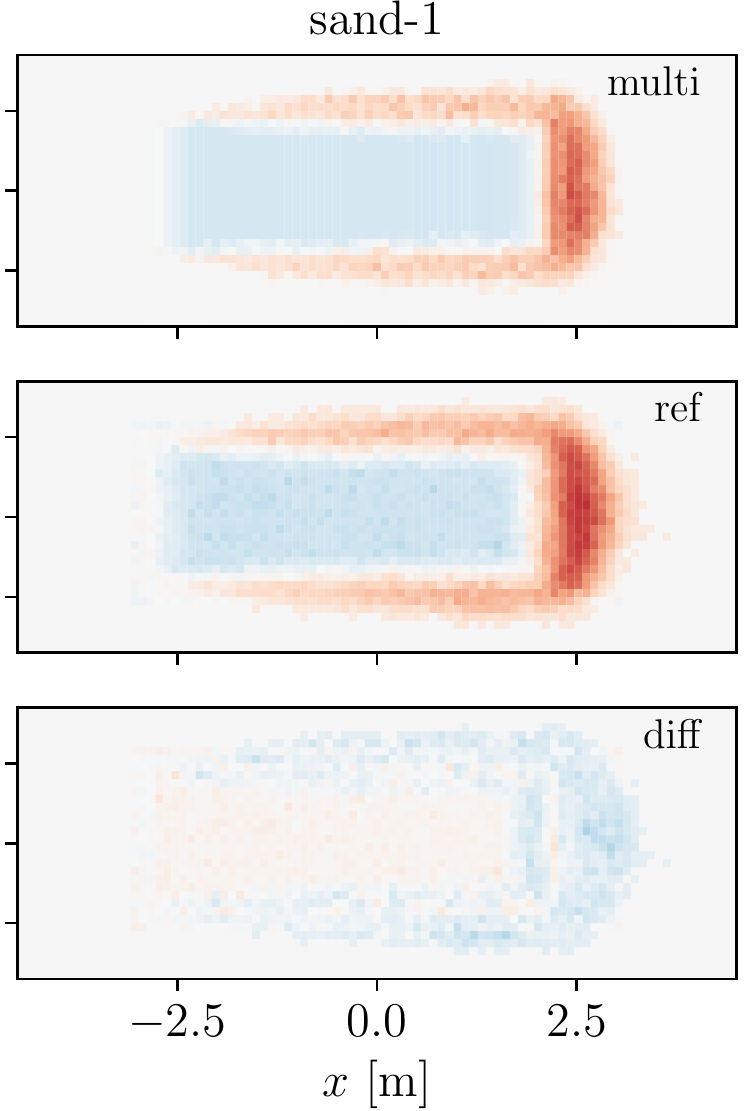}
    \includegraphics[height=0.35\textwidth,trim={45mm 0mm 42mm 9mm},clip]{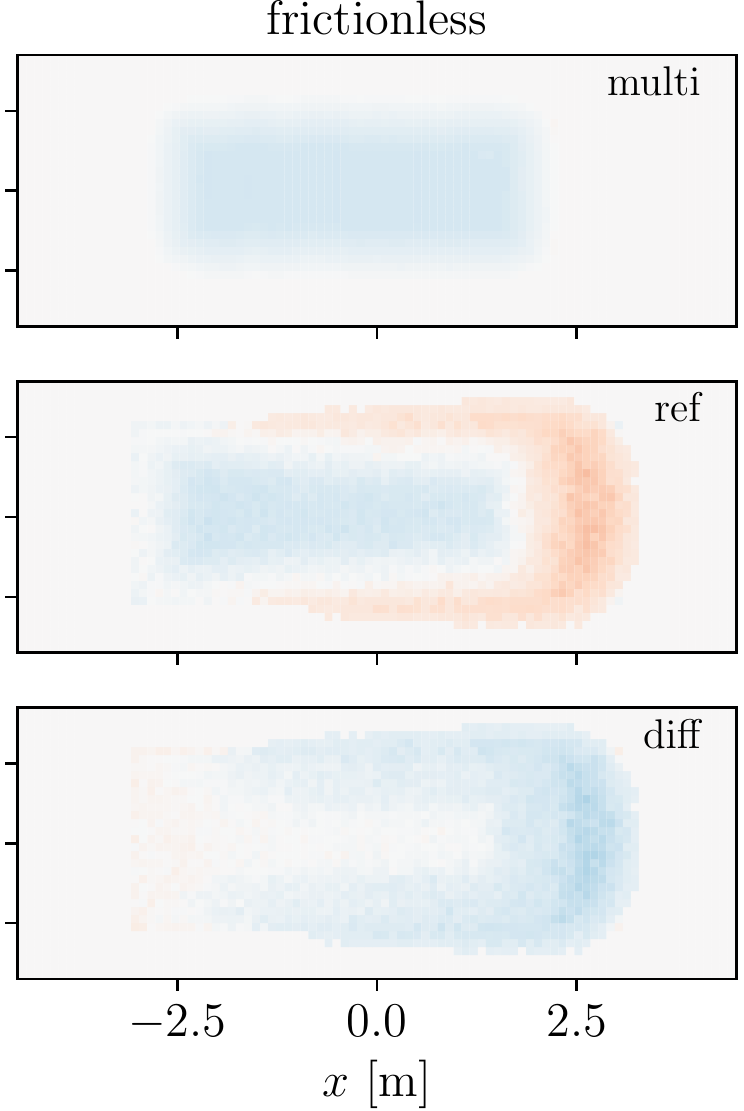}
    \includegraphics[height=0.35\textwidth,trim={0mm -13mm 0mm -5mm},clip]{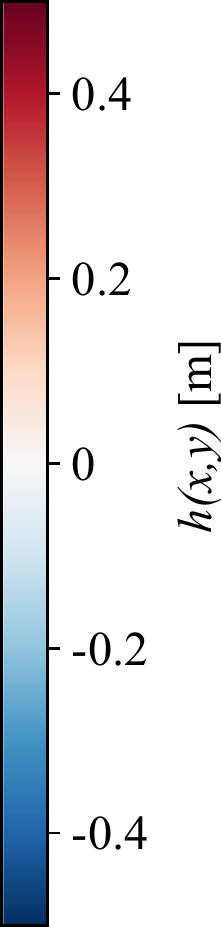}
    \caption{The resulting surfaces from bulldozing in three frictional soils using the multiscale model (top row) and the microscale reference model (middle row). The difference is shown at the bottom row.}
    \label{fig:surface_bull_frictional}
\end{figure}

\begin{figure}
    \centering
    \includegraphics[height=0.35\textwidth,trim={14mm 0mm 42mm 9mm},clip]{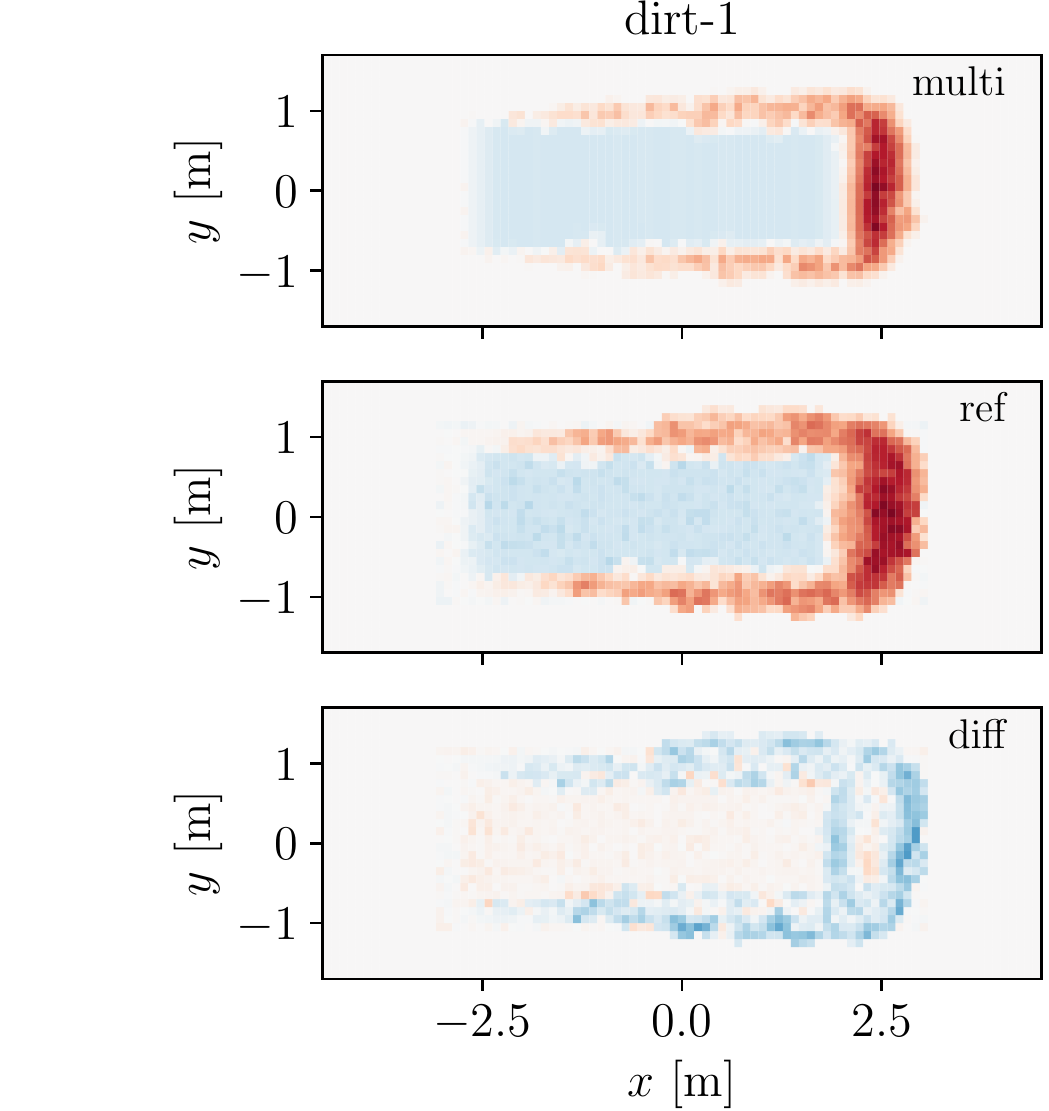}
    \includegraphics[height=0.35\textwidth,trim={45mm 0mm 42mm 9mm},clip]{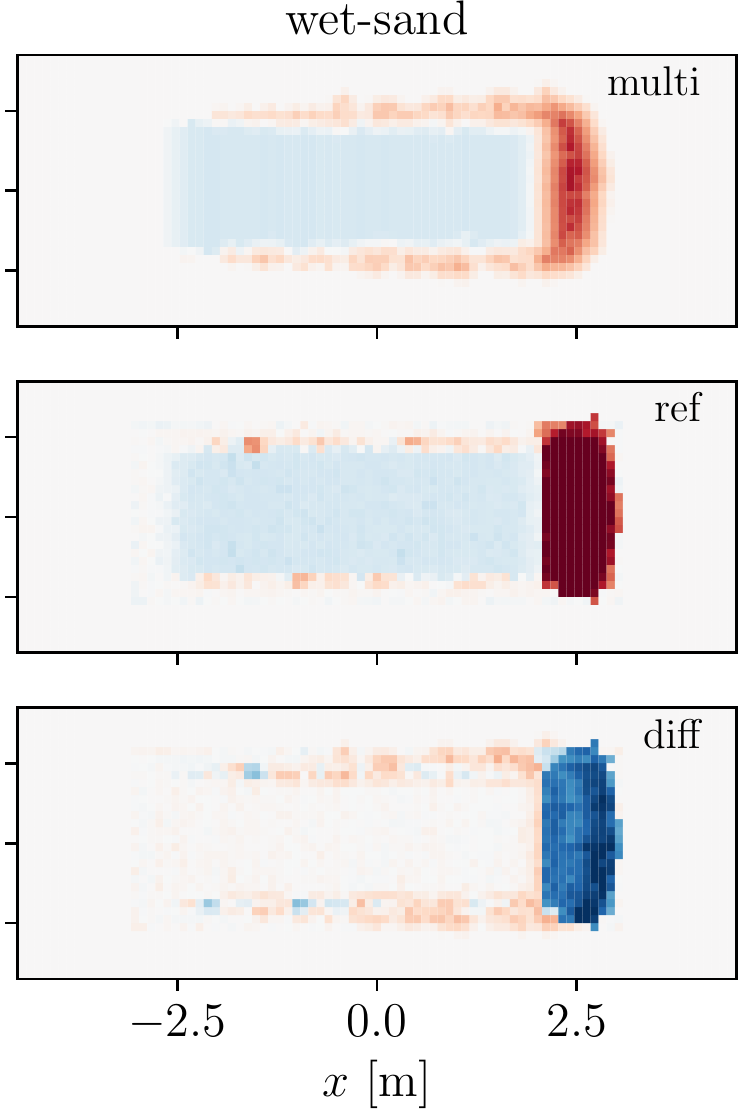}
    \includegraphics[height=0.35\textwidth,trim={45mm 0mm 42mm 9mm},clip]{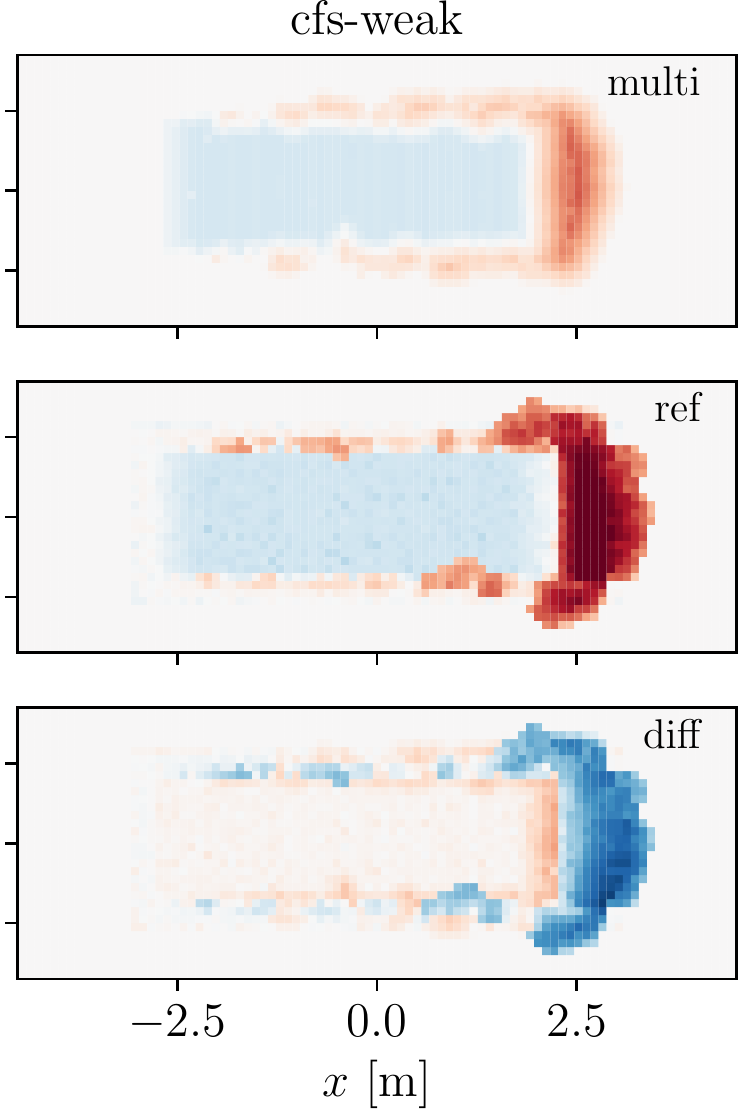}
    \includegraphics[height=0.35\textwidth,trim={0mm -13mm 0mm -5mm},clip]{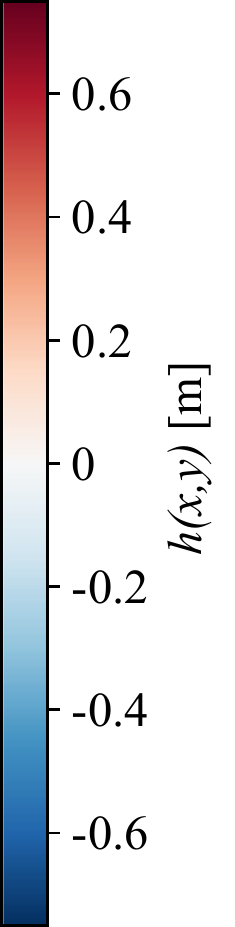}
    \caption{The resulting surfaces from bulldozing in three cohesive soils using the multiscale model (top row) and the microscale reference model (middle row). The difference is shown at the bottom row.}
    \label{fig:surface_bull_cohesive}
\end{figure}

The resulting surface height profiles, after a completed bulldozing cycle, are shown in Fig.~\ref{fig:surface_bull_frictional} and \ref{fig:surface_bull_cohesive}.
The multiscale and reference models are in good agreement for \texttt{gravel-1} and \texttt{sand-1}.
The depth of the cut strips agrees within 10 millimetres and the dimensions of the side berms are agree by roughly 10 \%.
The deviations in height profiles for the \texttt{frictionless} soil have the reason that is explained above.
For the cohesive soils the deviation between the models is larger. 
The reference model with \texttt{dirt-1} give wider side berms and a wider pile.  
The relative error is up to 50 \%.
For the strongly cohesive soils (\texttt{wet-sand-1} and \texttt{cfs-medium}) the difference is even larger, up to 100 \%. 
For the reference model almost all soil is accumulated in front of the blade.
The particle cohesion appear to be much stronger in the reference model than with the multiscale model.


\begin{figure}[h]
    \centering
    \hspace{4.0mm}
    \begin{picture}(102,40)
        \put(0,0){\includegraphics[width=0.2825\textwidth]{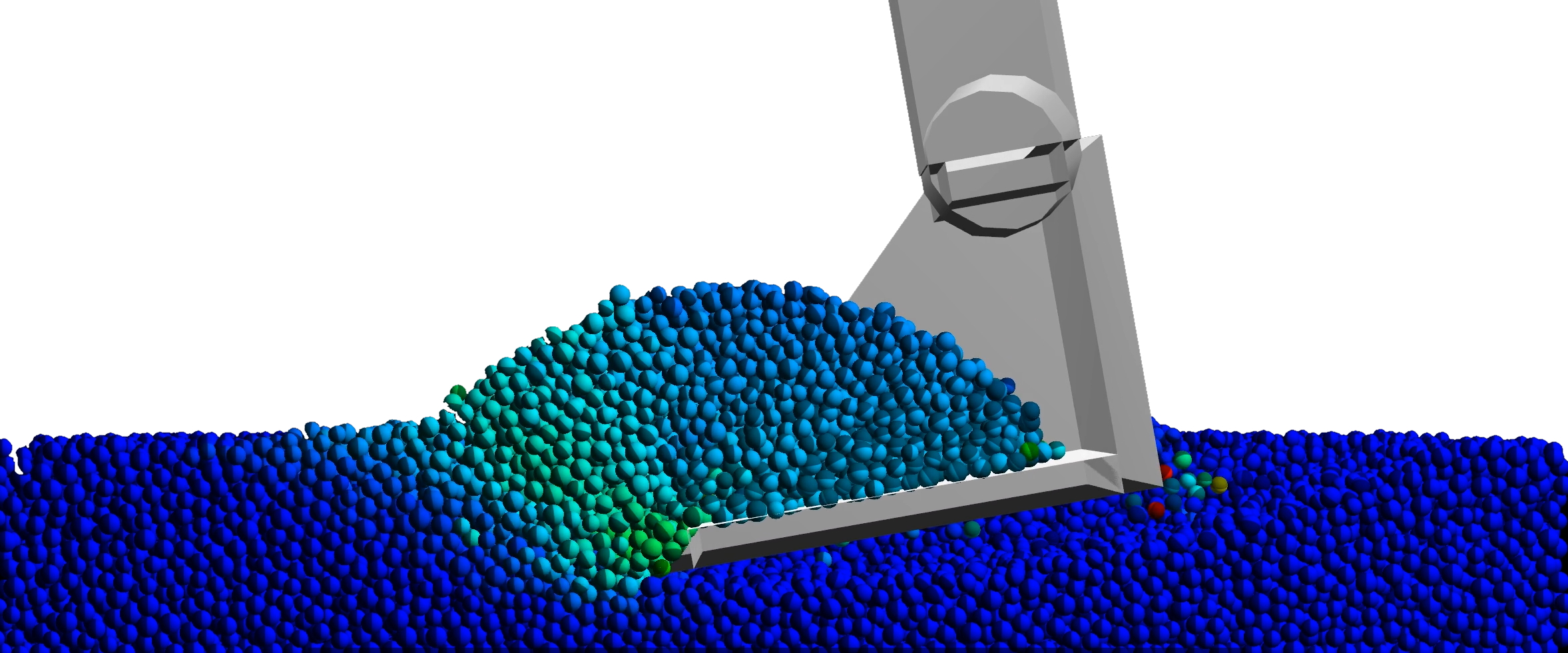}}
        \put(30,39){\tiny{gravel-1}}
    \end{picture}
    \begin{picture}(102,40)
        \put(0,0){\includegraphics[width=0.2825\textwidth]{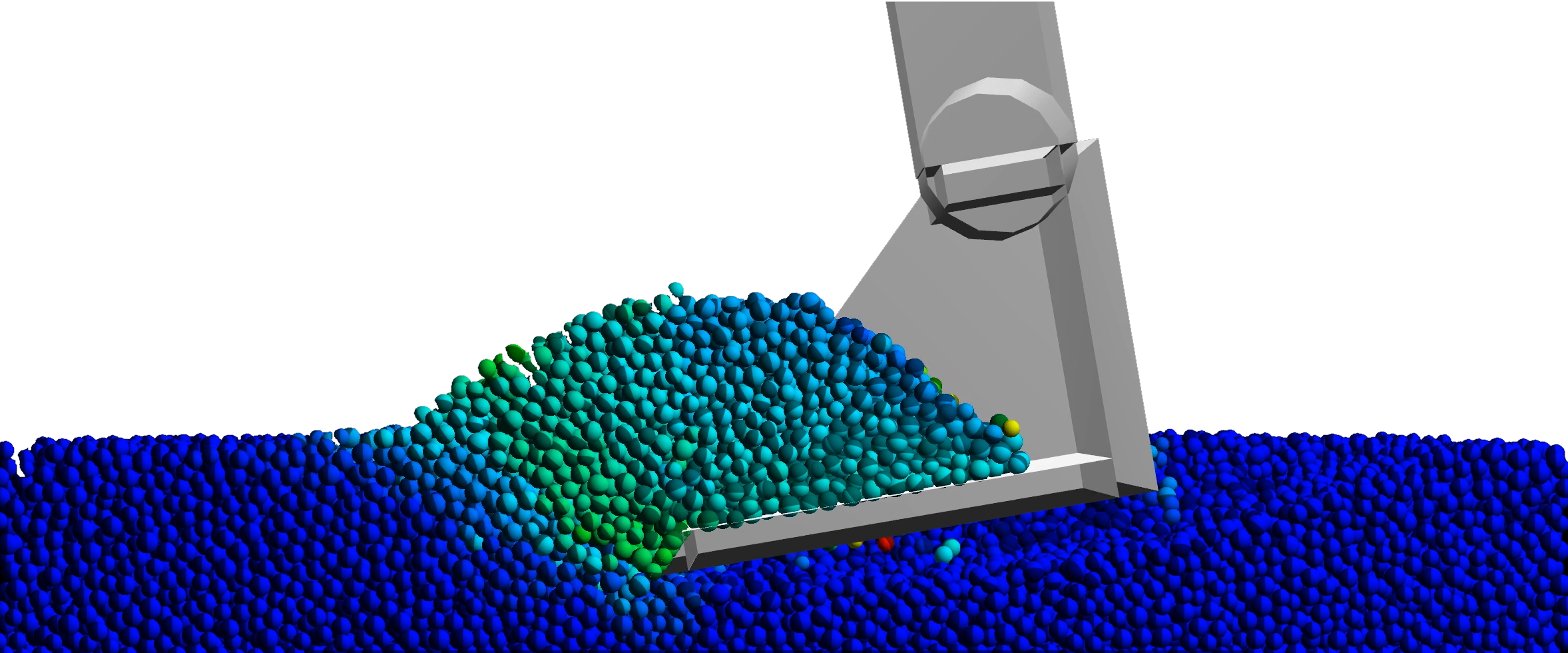}}
        \put(30,39){\tiny{sand-1}}
    \end{picture}    
    \begin{picture}(102,40)
        \put(0,0){\includegraphics[width=0.2825\textwidth]{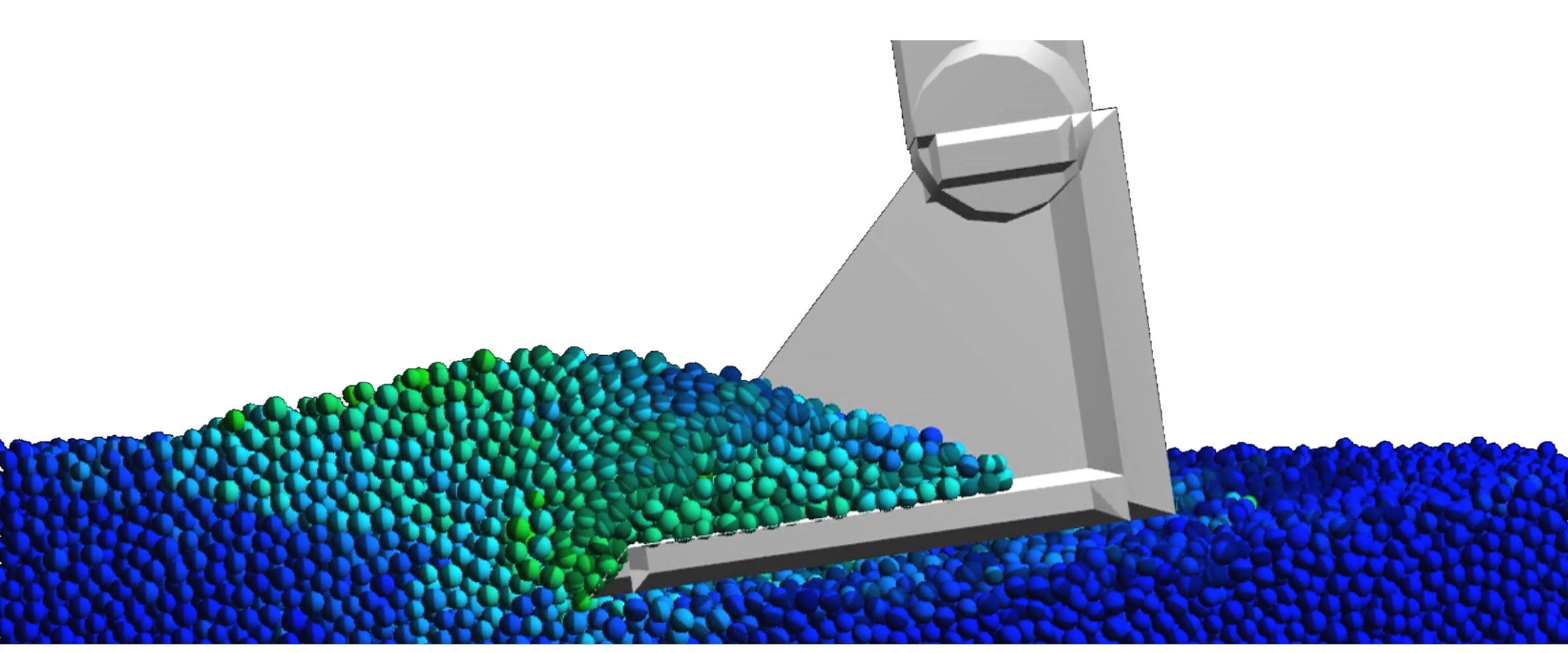}}
        \put(20,38){\tiny{frictionless}}
    \end{picture}
    \\
    \includegraphics[trim=0 0 40 35, clip, height=0.33\textwidth]{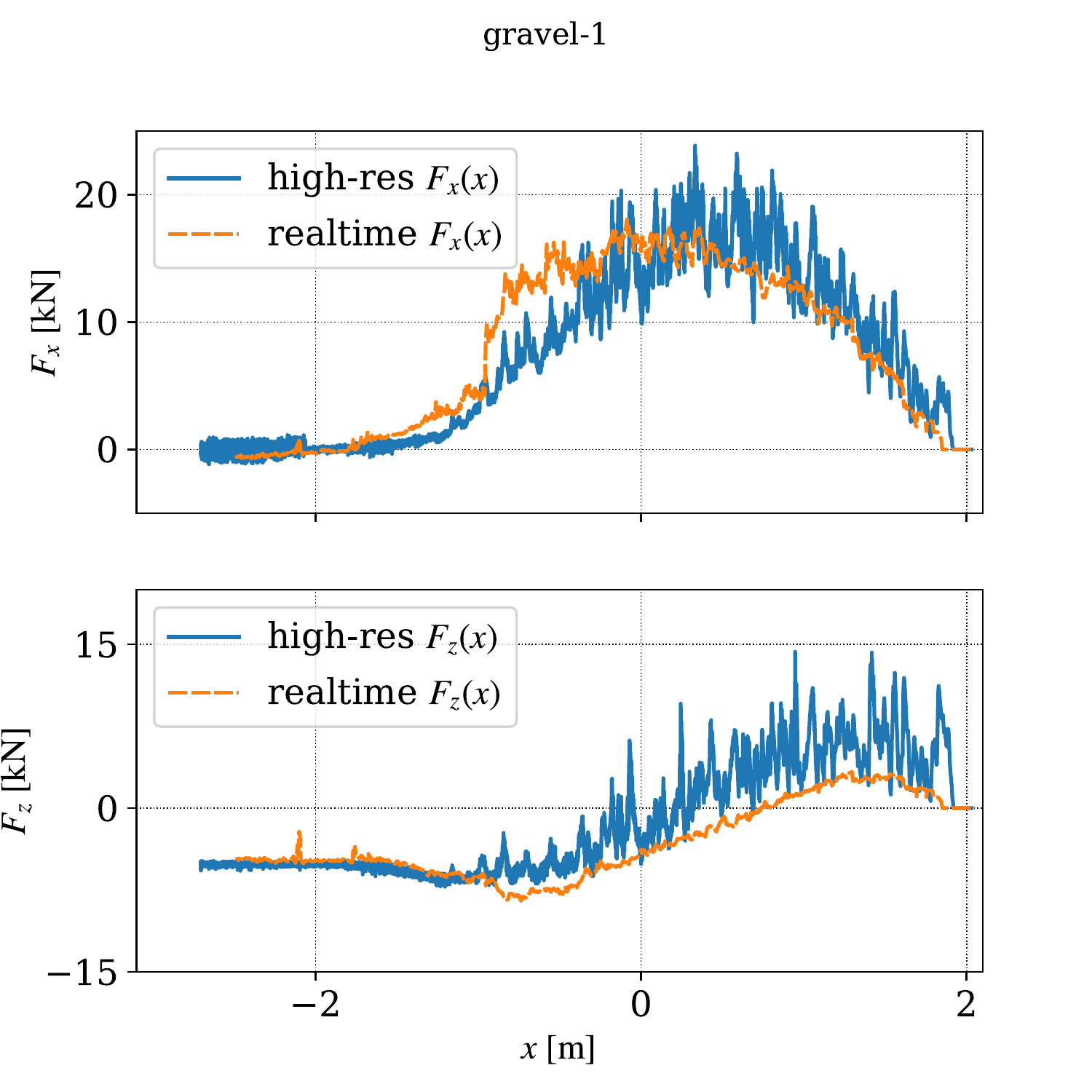}
    \includegraphics[trim=50 0 40 35, clip,height=0.33\textwidth]{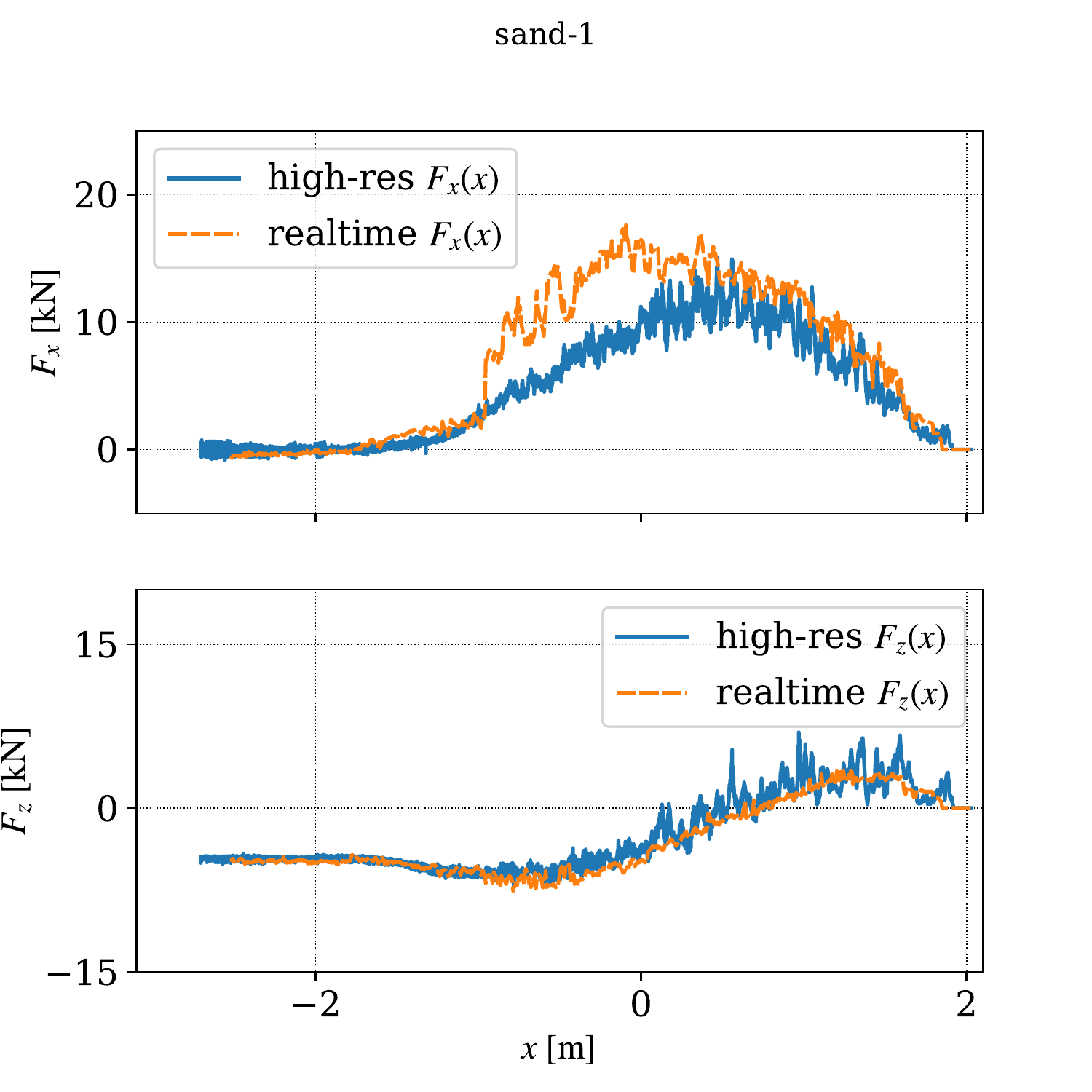}
    \includegraphics[trim=50 0 40 35, clip,height=0.33\textwidth]{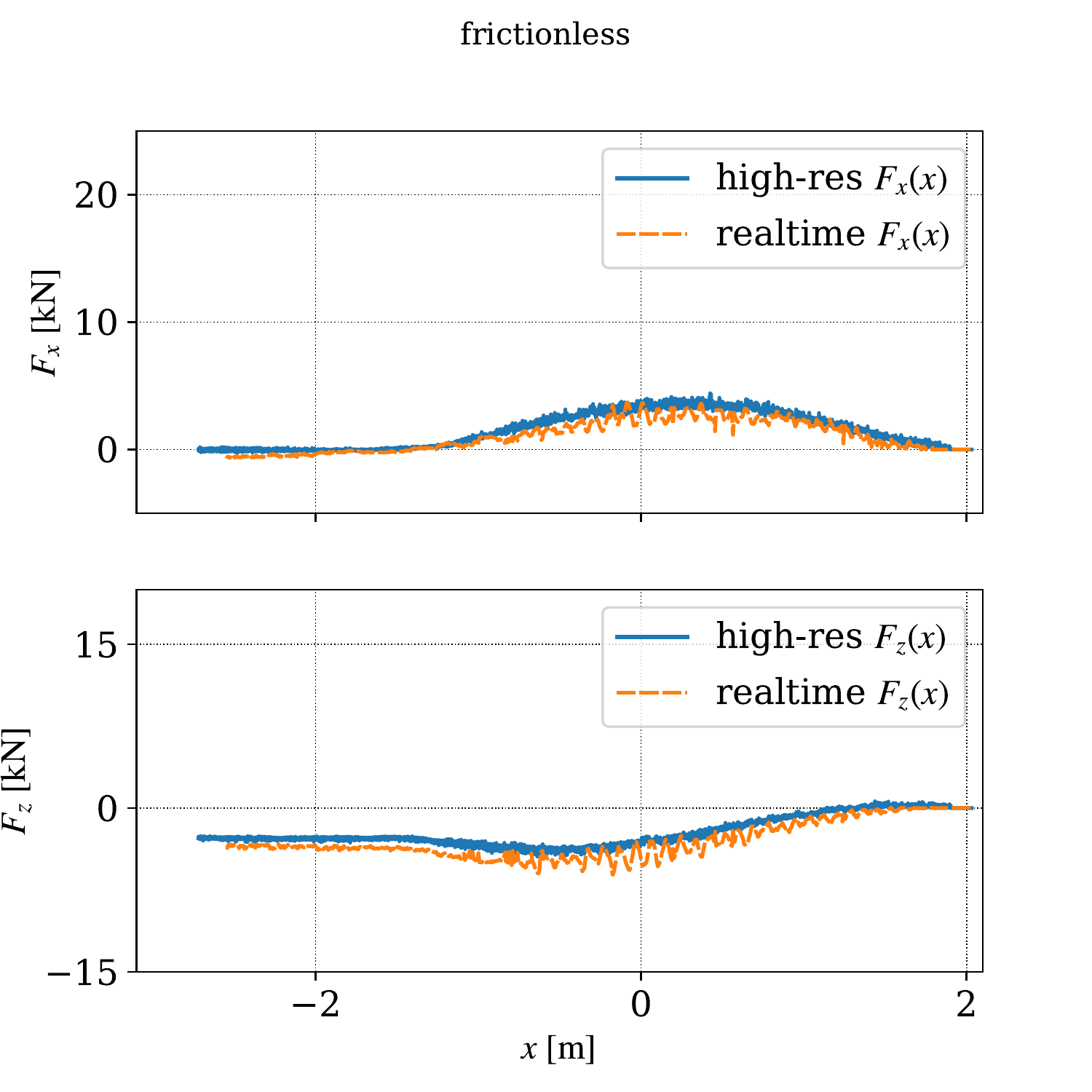}
    \caption{Force resistance from excavation using three frictional soils and the microscale model, with snapshot taken at 5 s.}
    \label{fig:NDEM_agxTerrain_excavator_frictional}
\end{figure}

\begin{figure}[h]
    \centering
    \hspace{4.0mm}
    \begin{picture}(102,40)
        \put(0,0){\includegraphics[width=0.2825\textwidth]{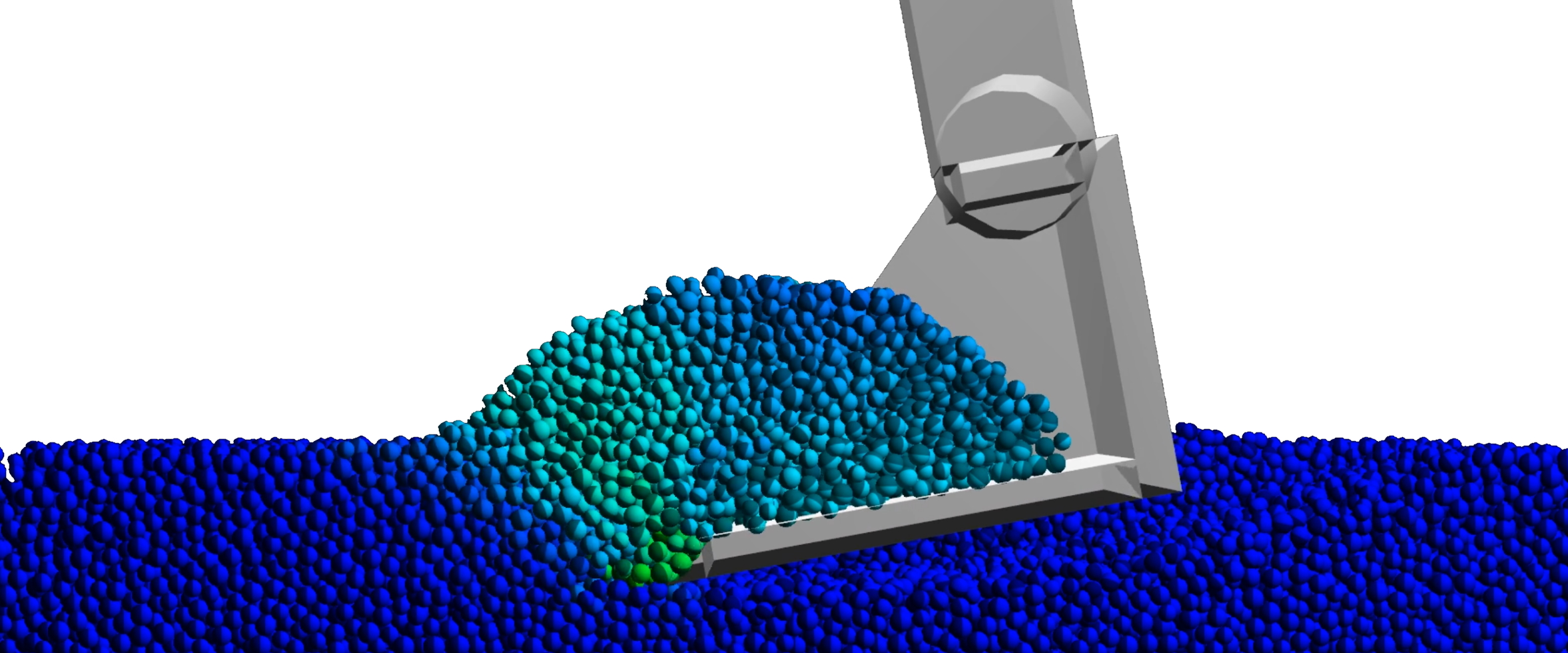}}
        \put(30,40){\tiny{dirt-1}}
    \end{picture} 
    \begin{picture}(102,40)
        \put(0,0){\includegraphics[width=0.2825\textwidth]{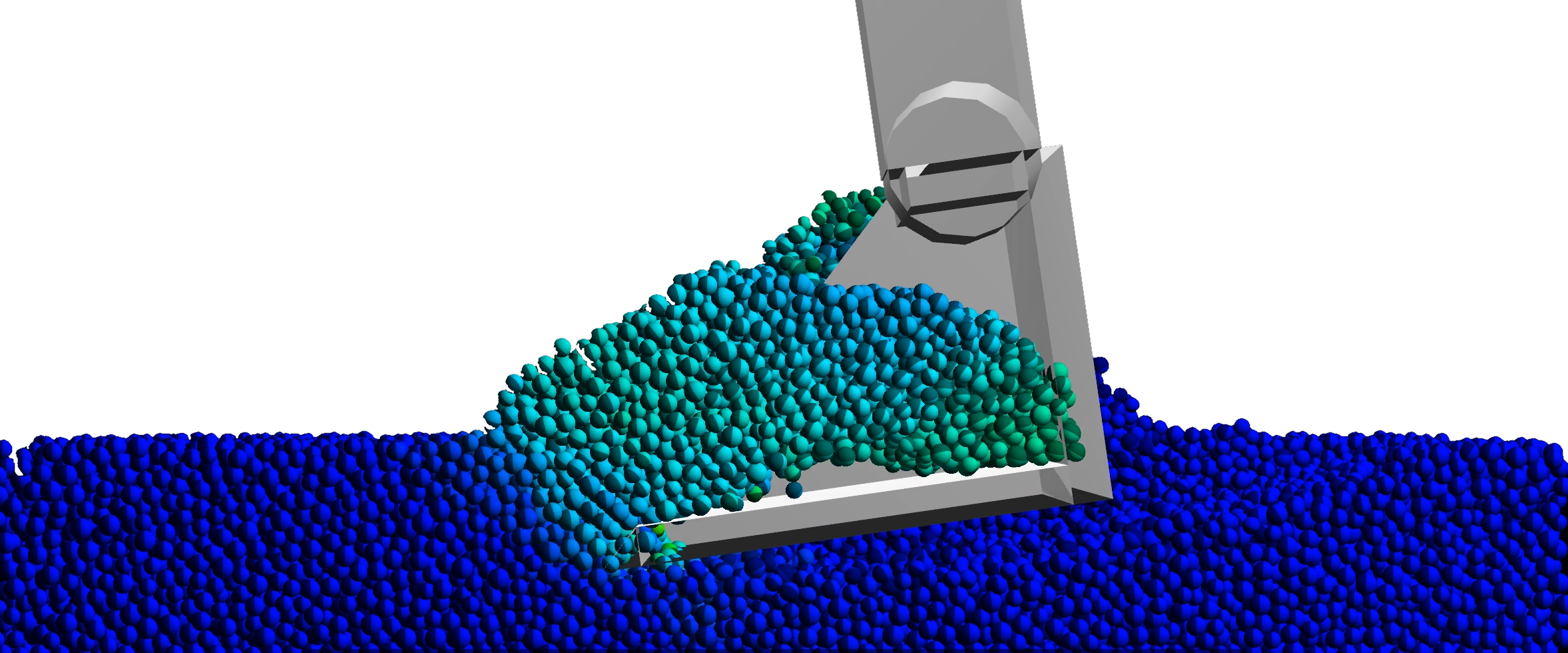}}
        \put(25,40){\tiny{wet-sand-1}}
    \end{picture} 
    \begin{picture}(102,40)
        \put(0,0){\includegraphics[width=0.2825\textwidth]{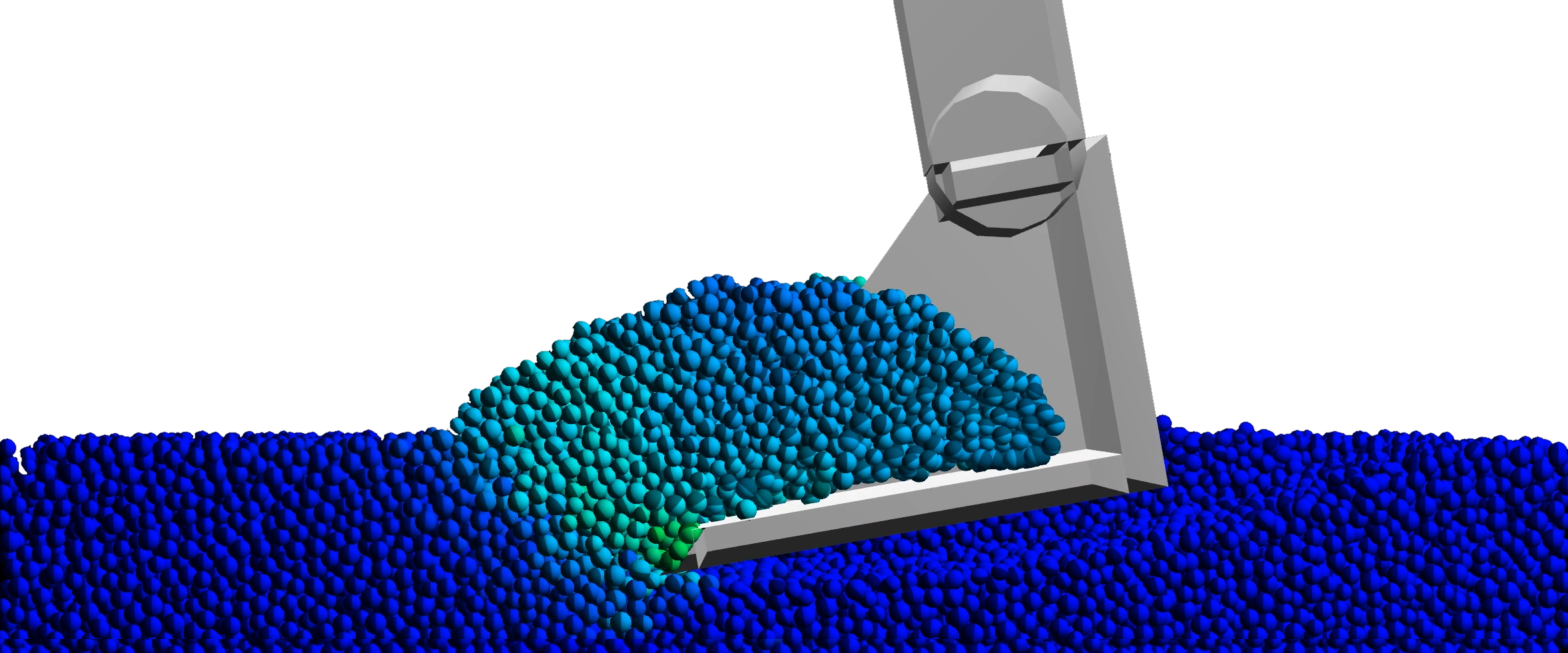}}
        \put(30,39){\tiny{cfs-weak}}
    \end{picture} 
    \\
    \includegraphics[trim=0 0 40 35, clip, height=0.33\textwidth]{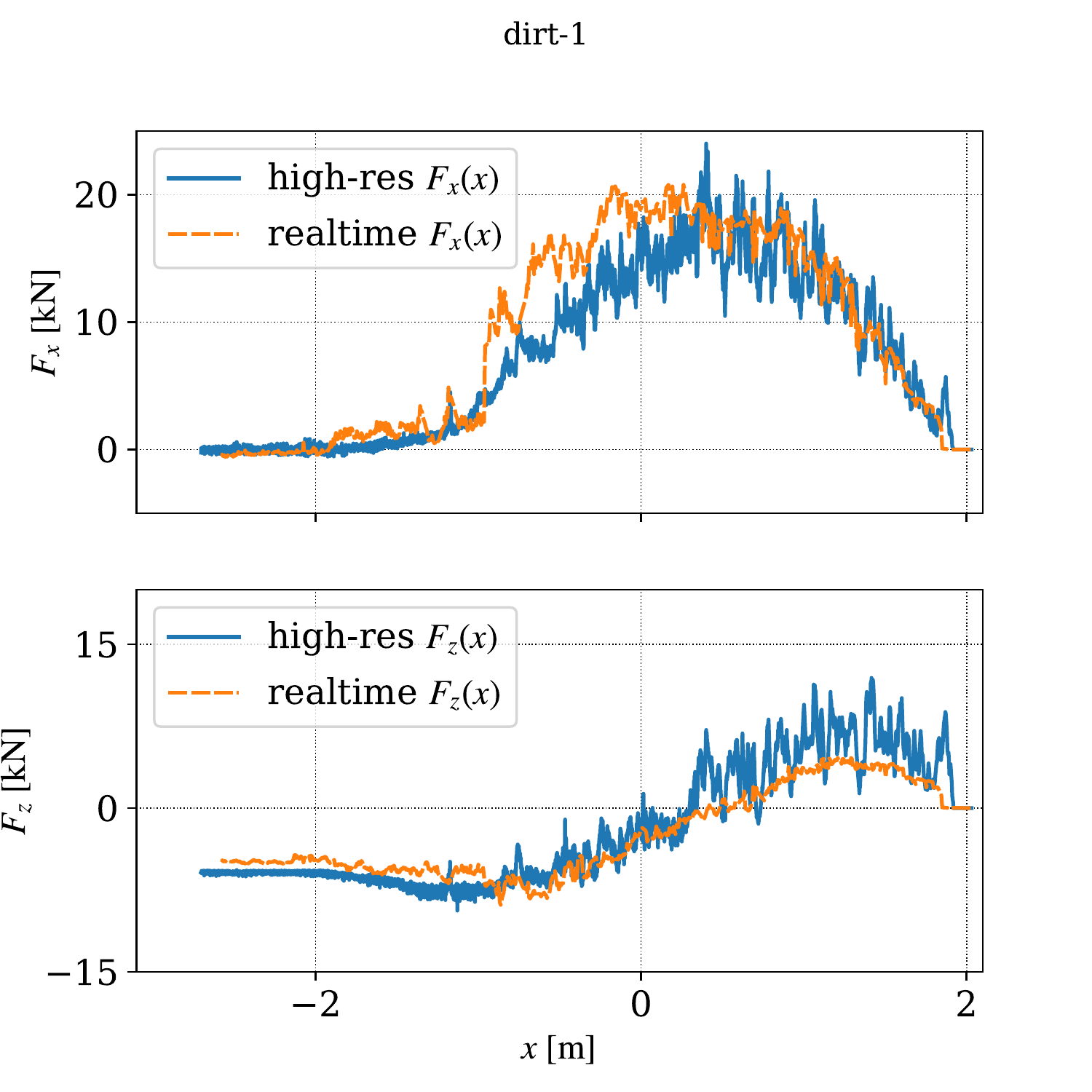}
    \includegraphics[trim=50 0 40 35, clip, height=0.33\textwidth]{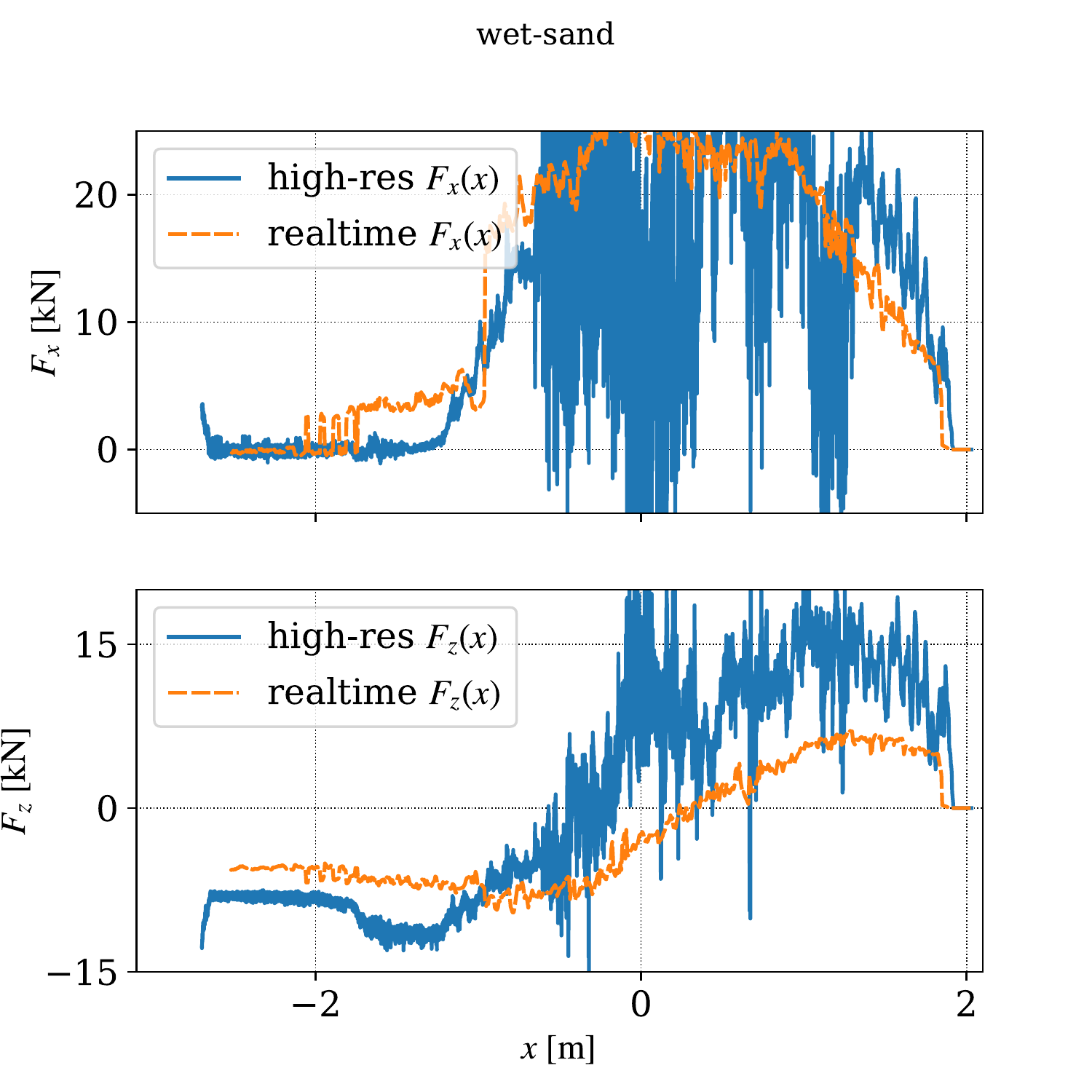}
    \includegraphics[trim=50 0 40 35, clip, height=0.33\textwidth]{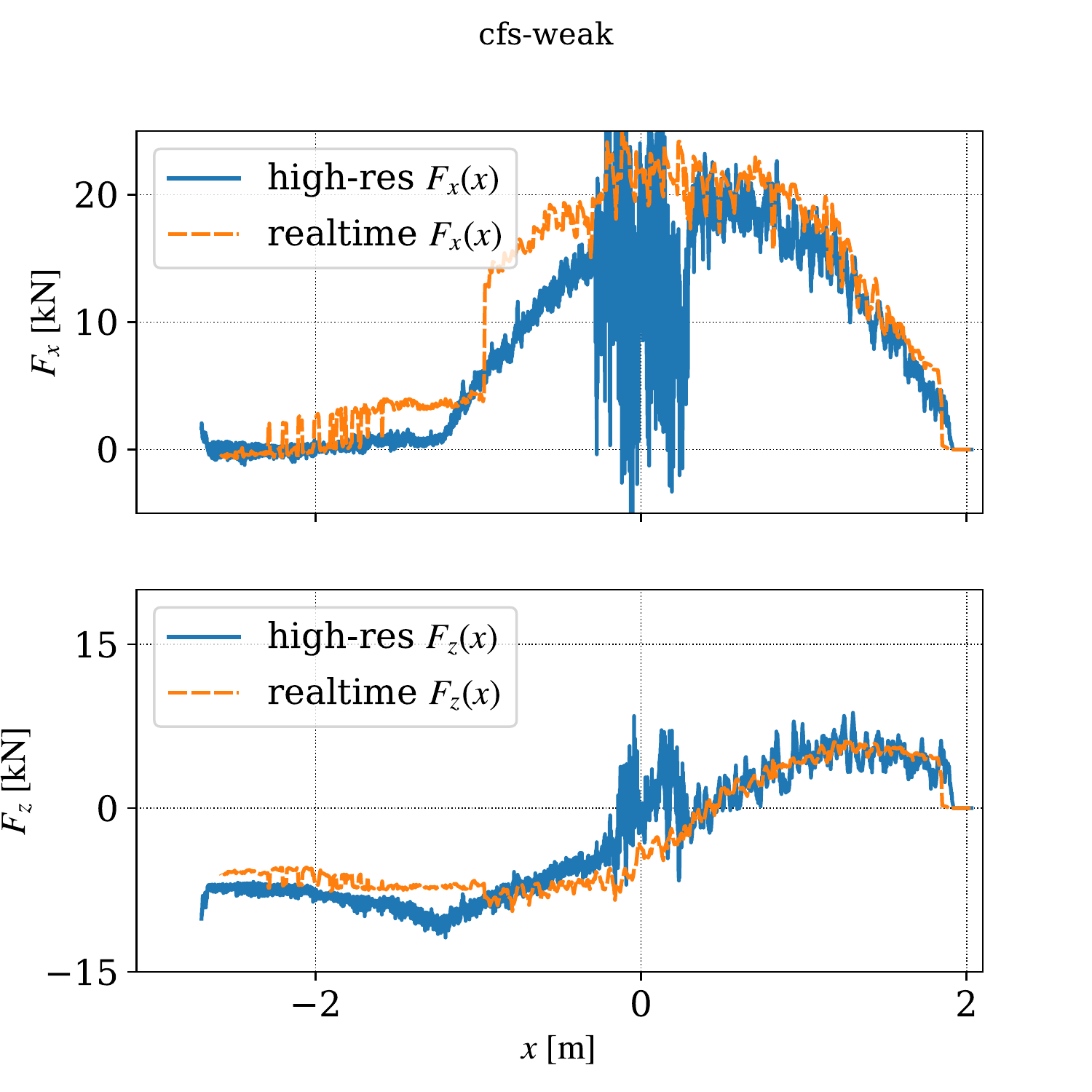}
    \caption{Force resistance from excavation using three cohesive soils and the microscale model, with snapshot taken at 5 s.}
    \label{fig:NDEM_agxTerrain_excavator_cohesive}
\end{figure}

\begin{figure}[h]
    \centering
    \includegraphics[trim=0 0 0 37, clip, width=0.45\textwidth]{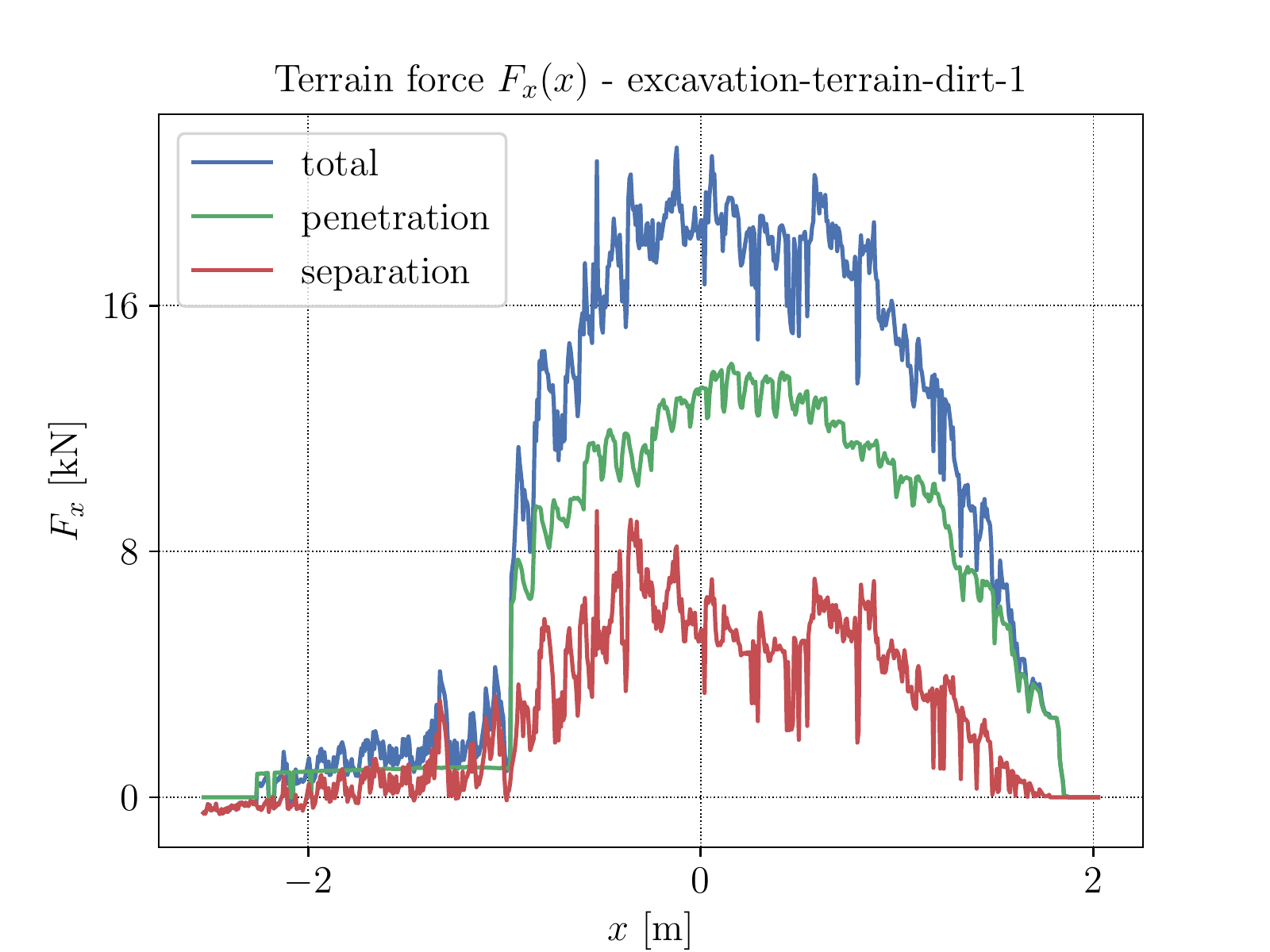}
    \includegraphics[trim=0 0 0 37, clip, width=0.45\textwidth]{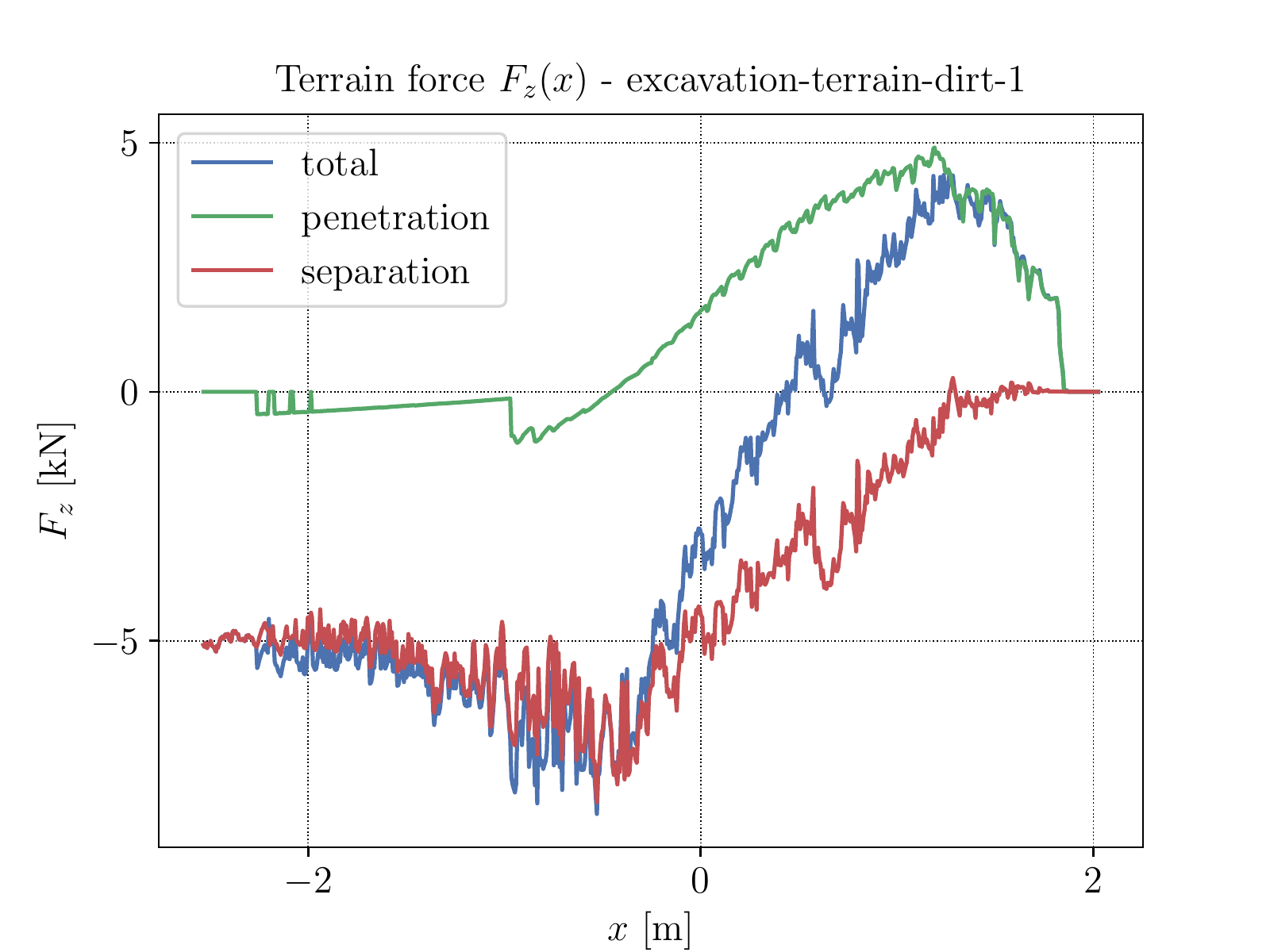}
    \caption{The digging resistance from excavation in a bed of \texttt{dirt-1} soil simulated using the multiscale model. The force is divided in the contributions from the penetration and separation models.}
    \label{fig:NDEM_agxTerrain_excavator_force}
\end{figure}

\begin{figure}
    \centering
    \includegraphics[height=0.35\textwidth,trim={14mm 0mm 46mm 9mm},clip]{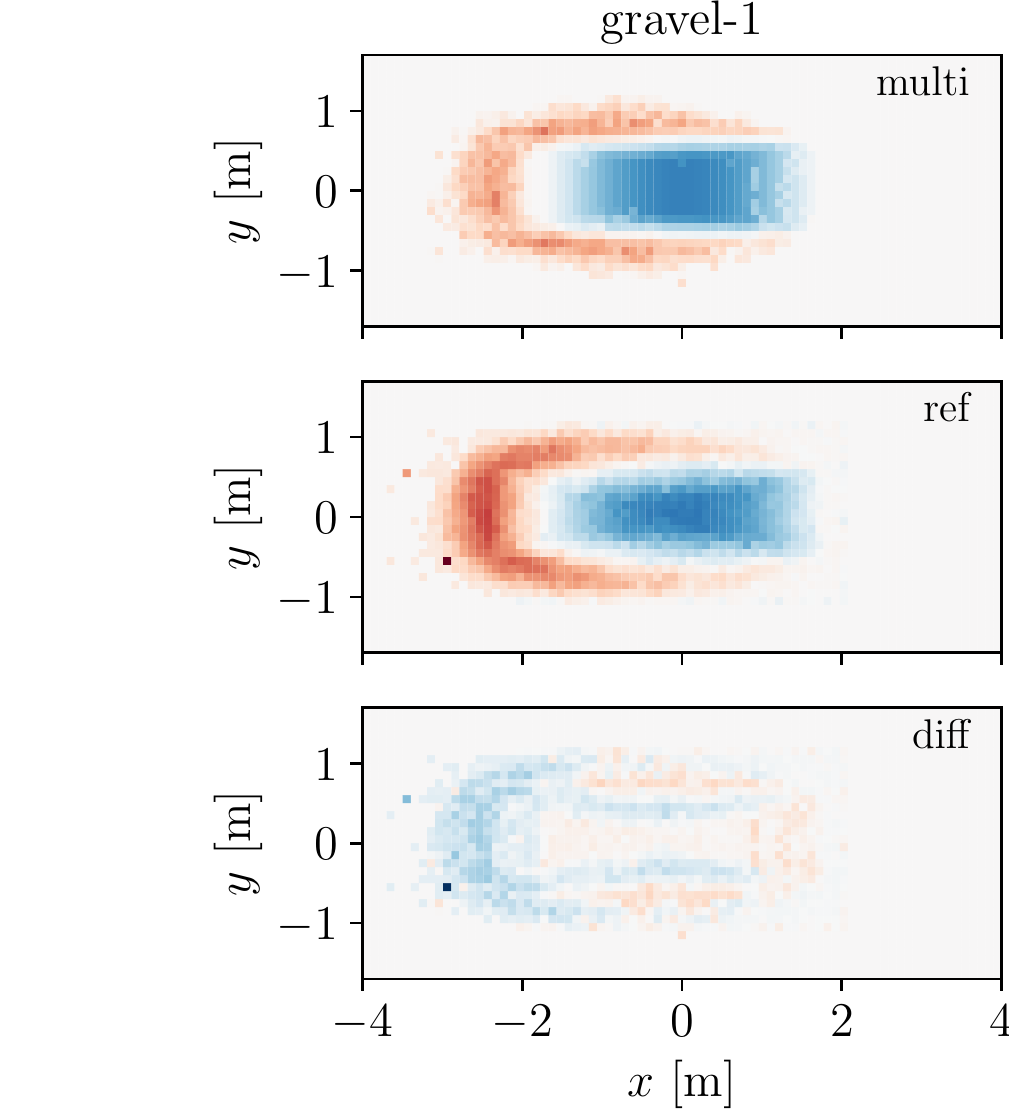}
    \includegraphics[height=0.35\textwidth,trim={49mm 0mm 46mm 9mm},clip]{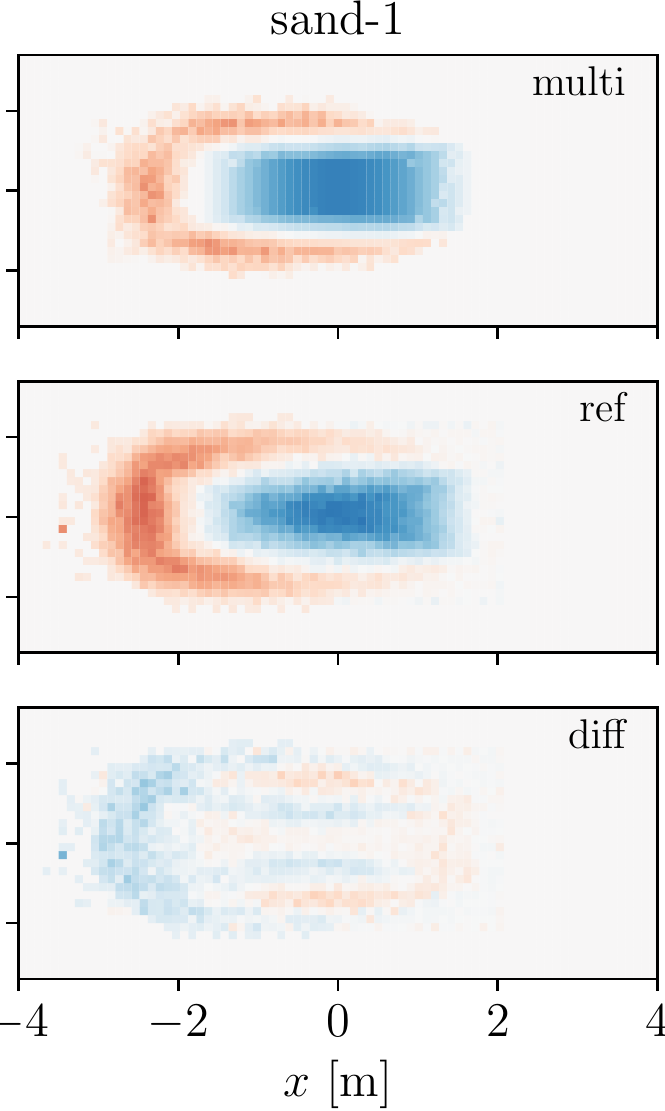}
    \includegraphics[height=0.35\textwidth,trim={49mm 0mm 46mm 9mm},clip]{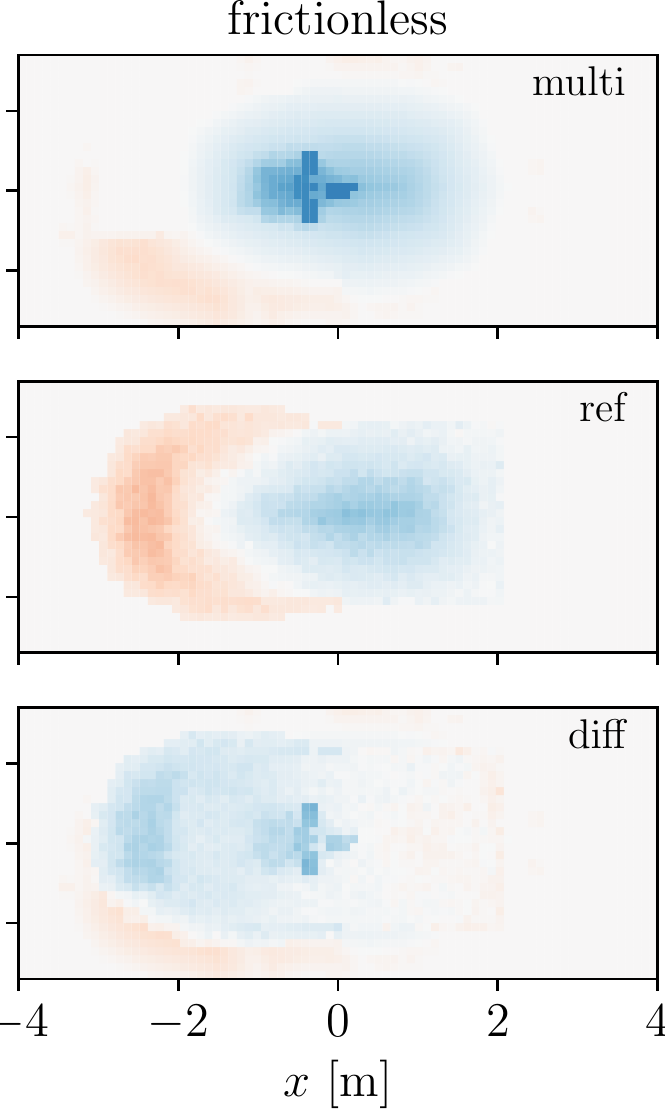}
    \includegraphics[height=0.35\textwidth,trim={0mm -13mm 0mm -5mm},clip]{colorbar_frictional_exc.pdf}
    \caption{The resulting surfaces from excavation in three frictional soils using the multiscale model (top row) and the microscale reference model (middle row). The difference is shown at the bottom row.}
    \label{fig:surface_exc_frictional}
\end{figure}

\begin{figure}
    \centering
    \includegraphics[height=0.35\textwidth,trim={14mm 0mm 46mm 9mm},clip]{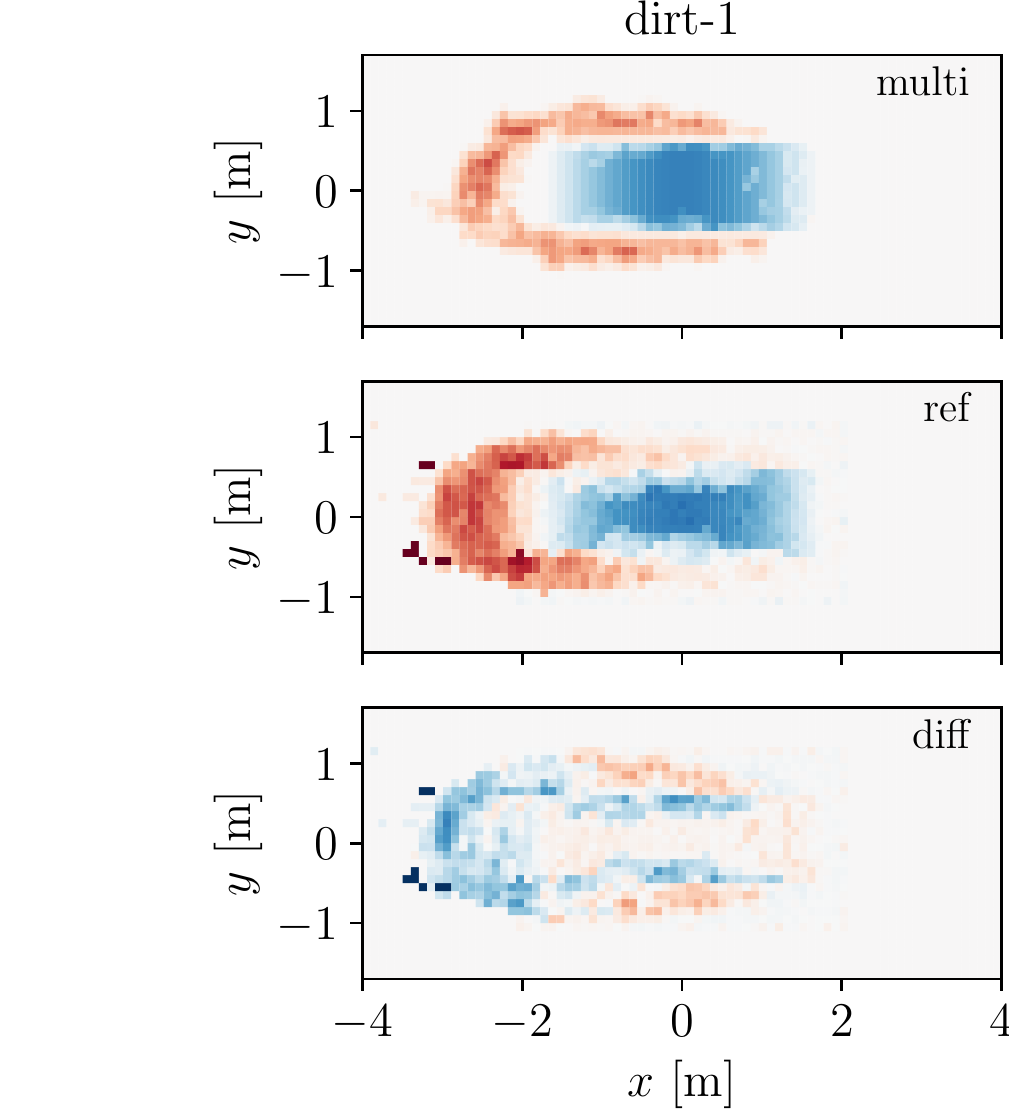}
    \includegraphics[height=0.35\textwidth,trim={49mm 0mm 46mm 9mm},clip]{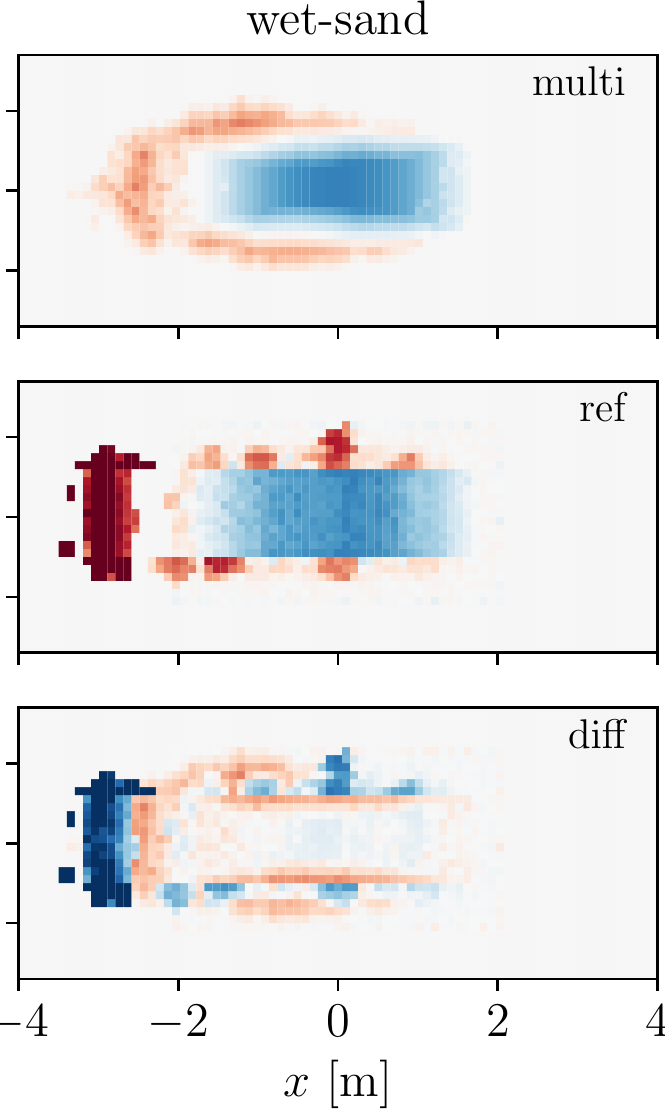}
    \includegraphics[height=0.35\textwidth,trim={49mm 0mm 46mm 9mm},clip]{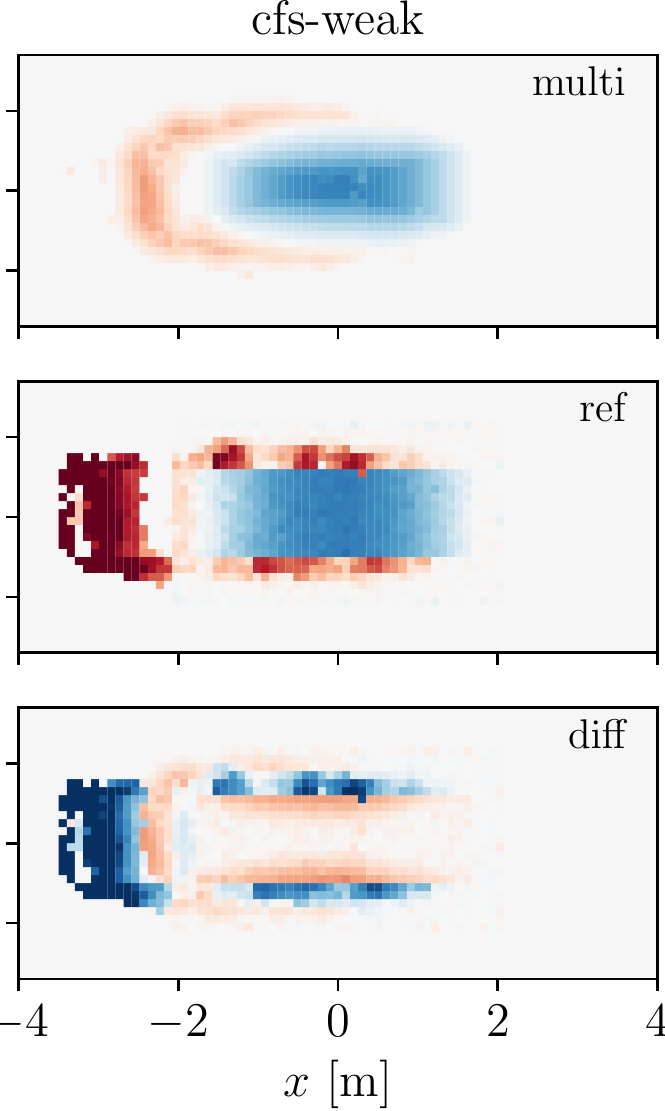}
    \includegraphics[height=0.35\textwidth,trim={0mm -13mm 0mm -5mm},clip]{colorbar_cohesive_exc.pdf}
    \caption{The resulting surfaces from excavation in three cohesive soils using the multiscale model (top row) and the microscale reference model (middle row). The difference is shown at the bottom row.}
    \label{fig:surface_exc_cohesive}
\end{figure}

The excavation test with \texttt{dirt-1} can be seen in Fig.~\ref{fig:NDEM_agxTerrain_excavator} and in the supplementary \href{https://www.algoryx.se/papers/terrain/}{Video 4}.
\begin{figure}[h]
    \centering
    \includegraphics[width=0.45\textwidth]{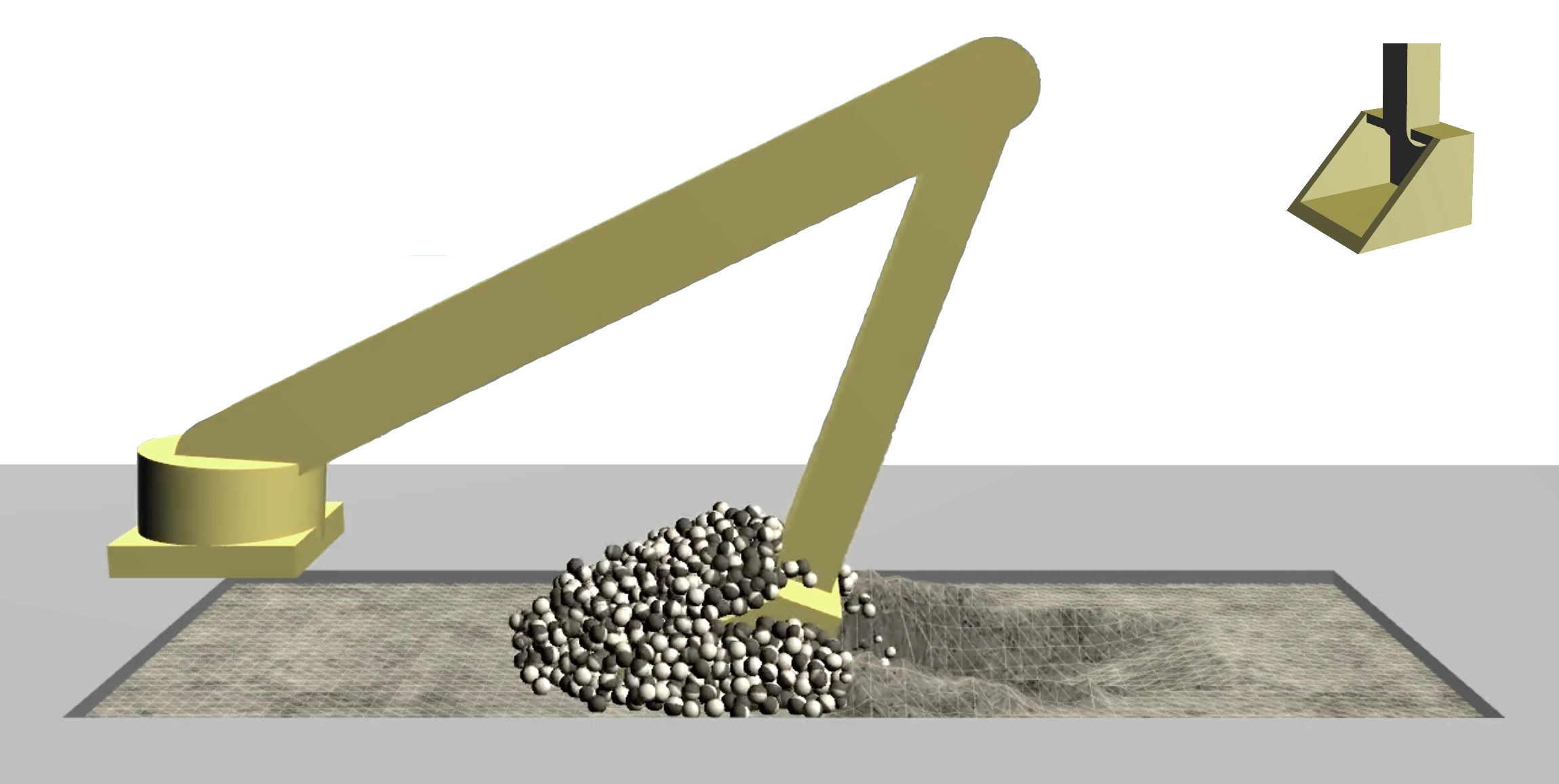}
    \includegraphics[width=0.45\textwidth]{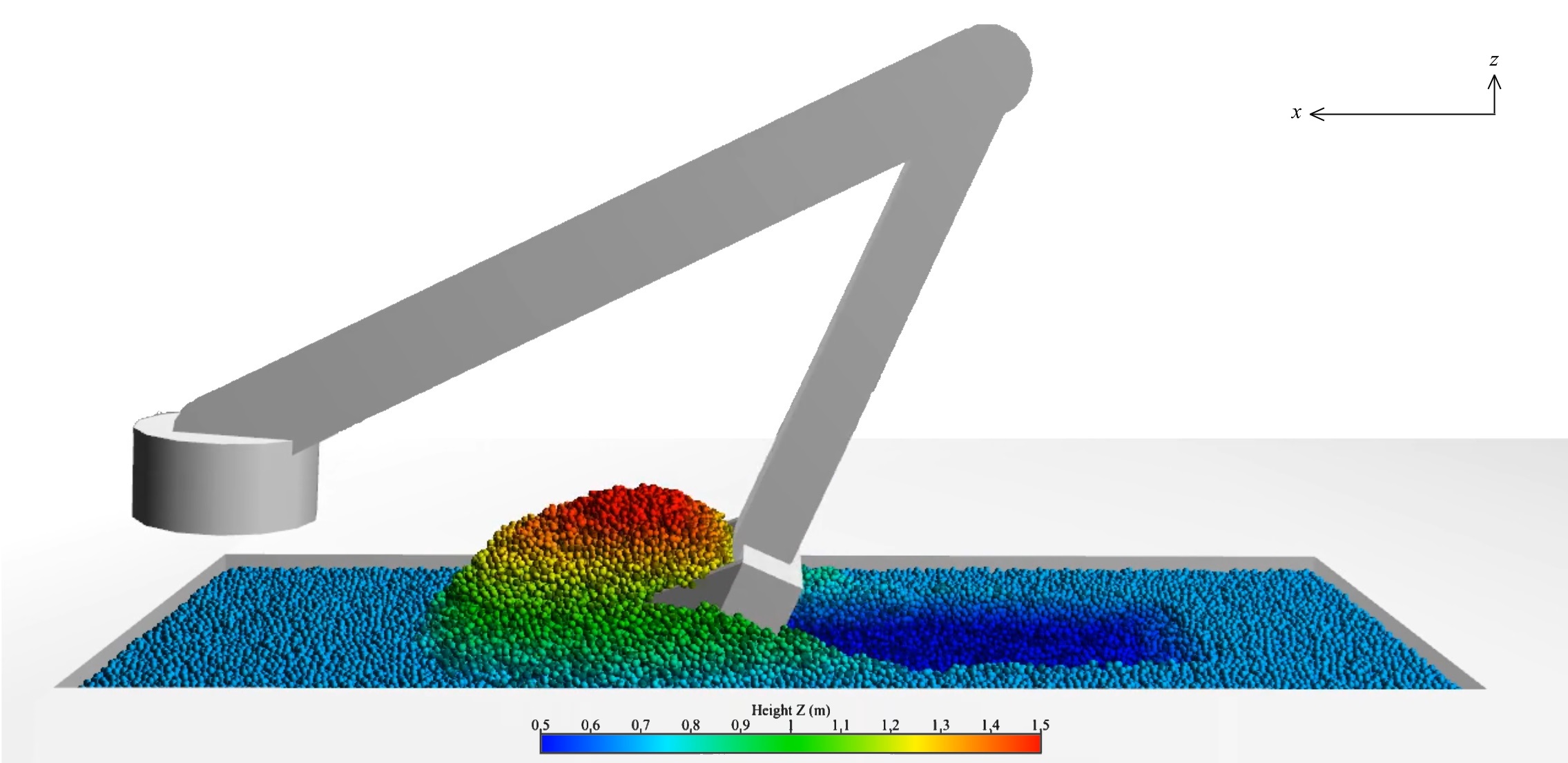}\\
    \vspace{2mm}
    \includegraphics[width=0.12\textwidth]{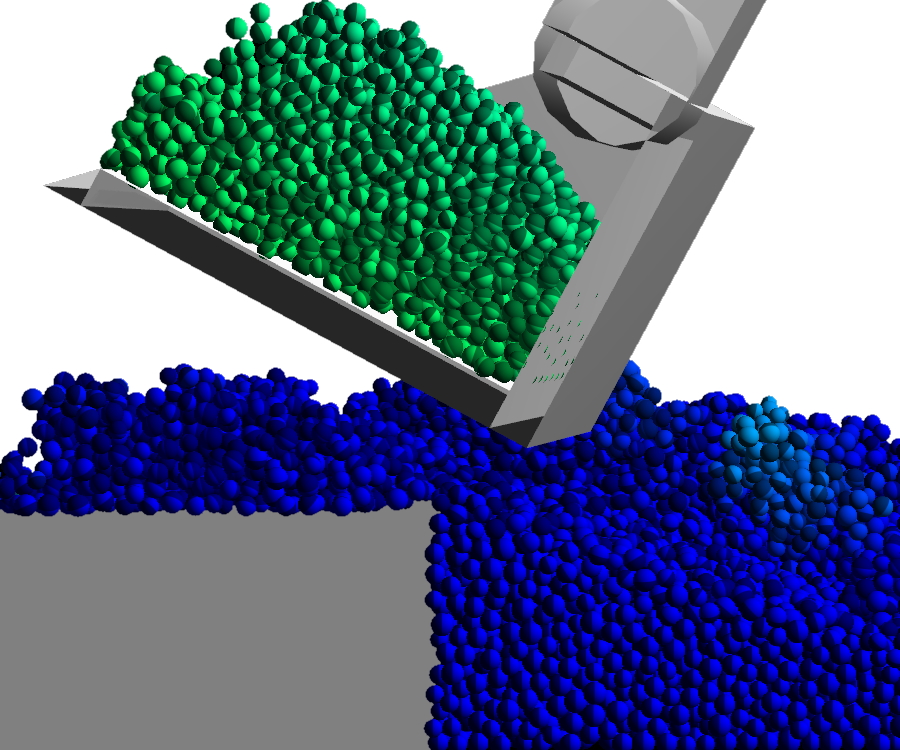}
    \includegraphics[width=0.12\textwidth]{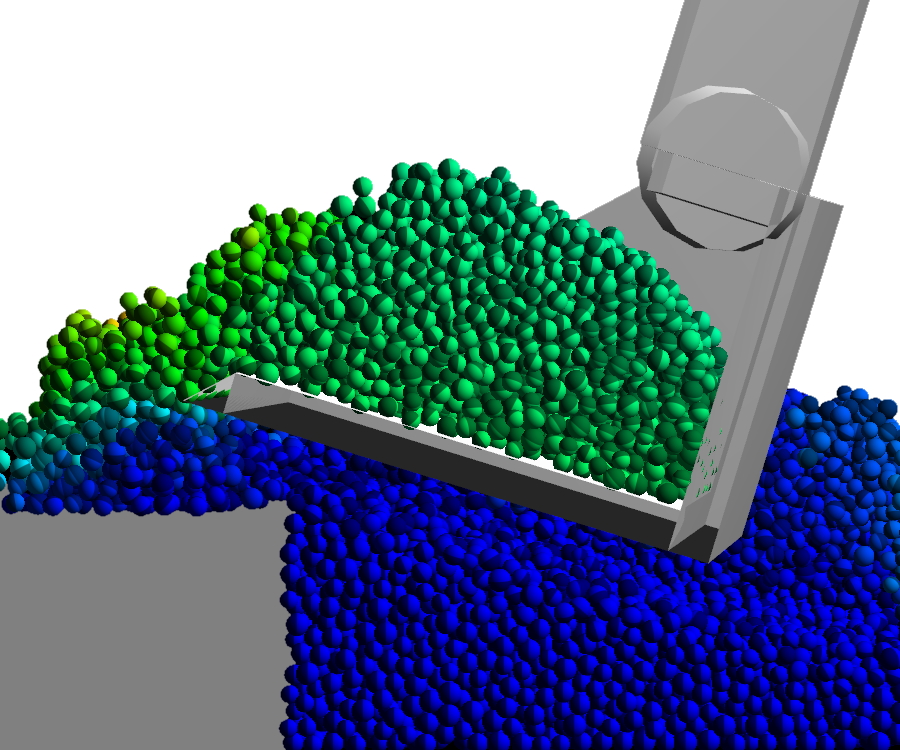}
    \includegraphics[width=0.12\textwidth]{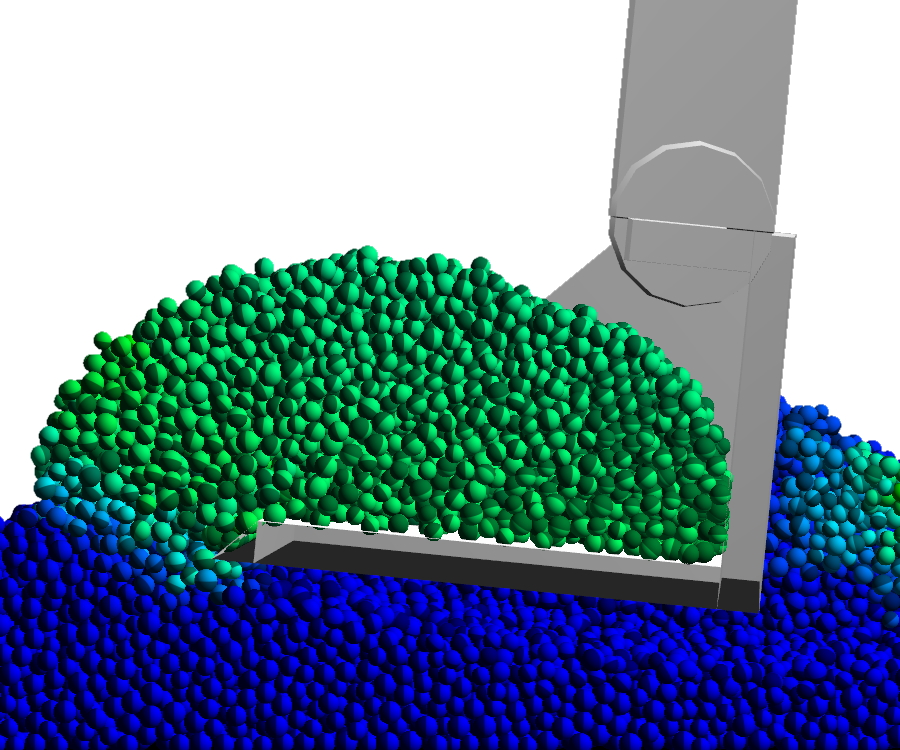}
    \includegraphics[width=0.12\textwidth]{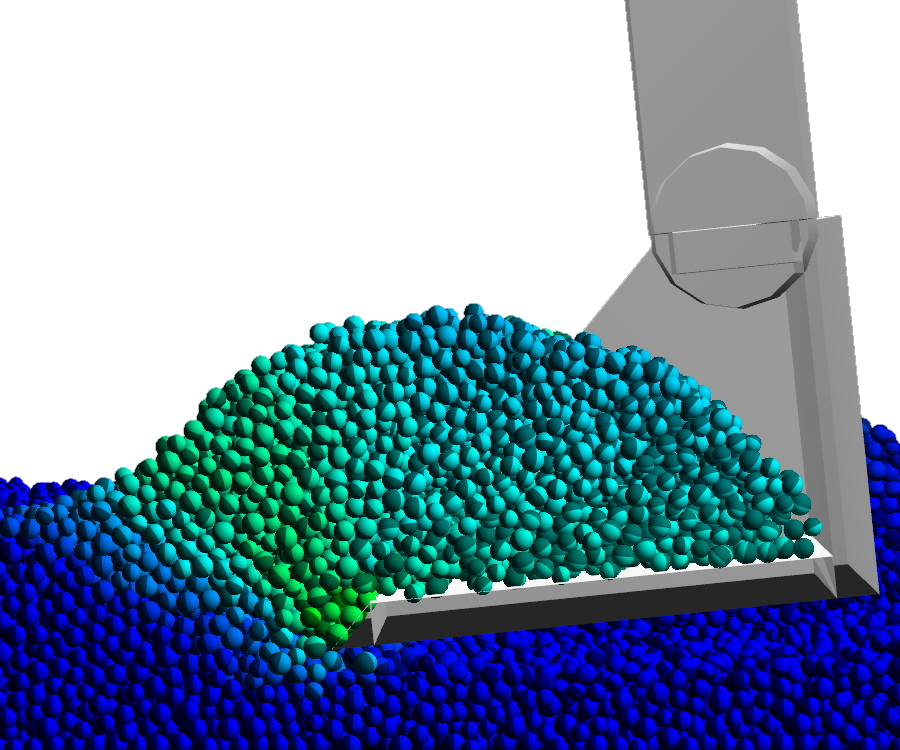}
    \includegraphics[width=0.12\textwidth]{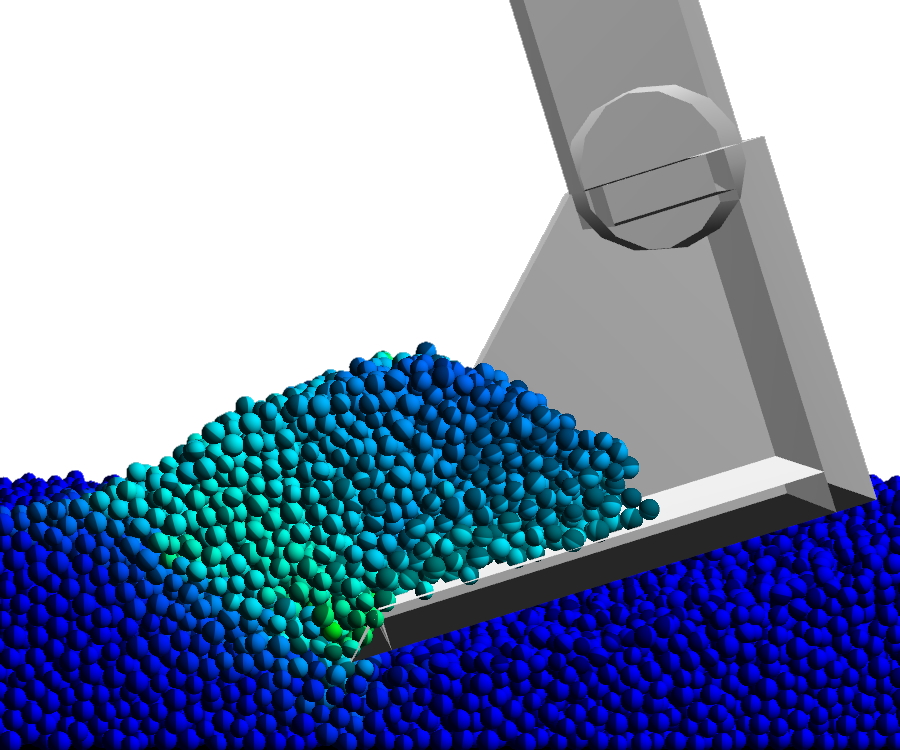}
    \includegraphics[width=0.12\textwidth]{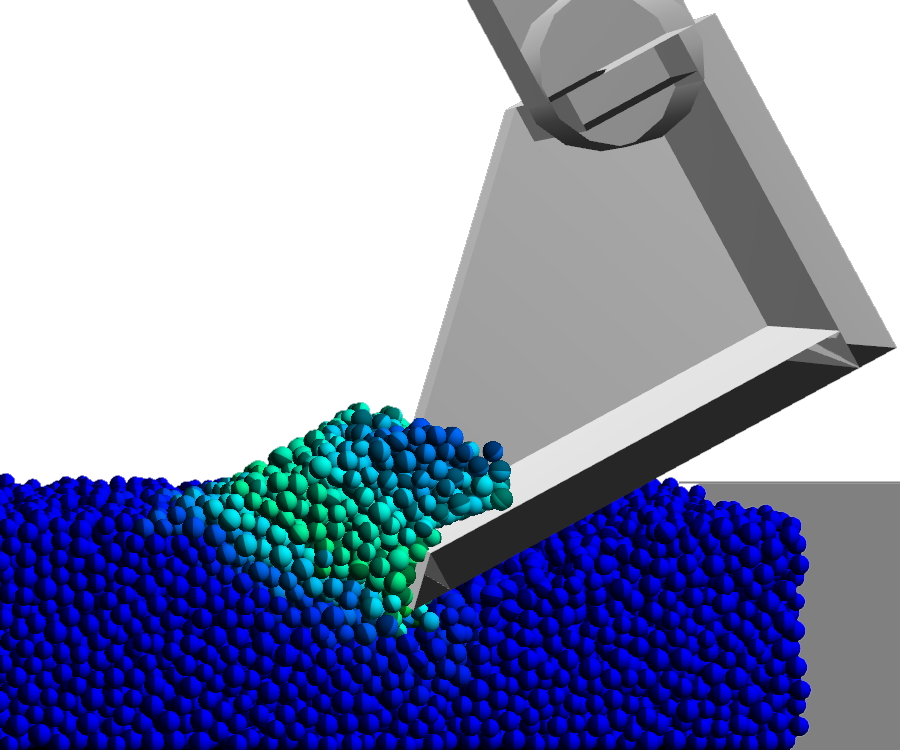}
    \includegraphics[width=0.12\textwidth]{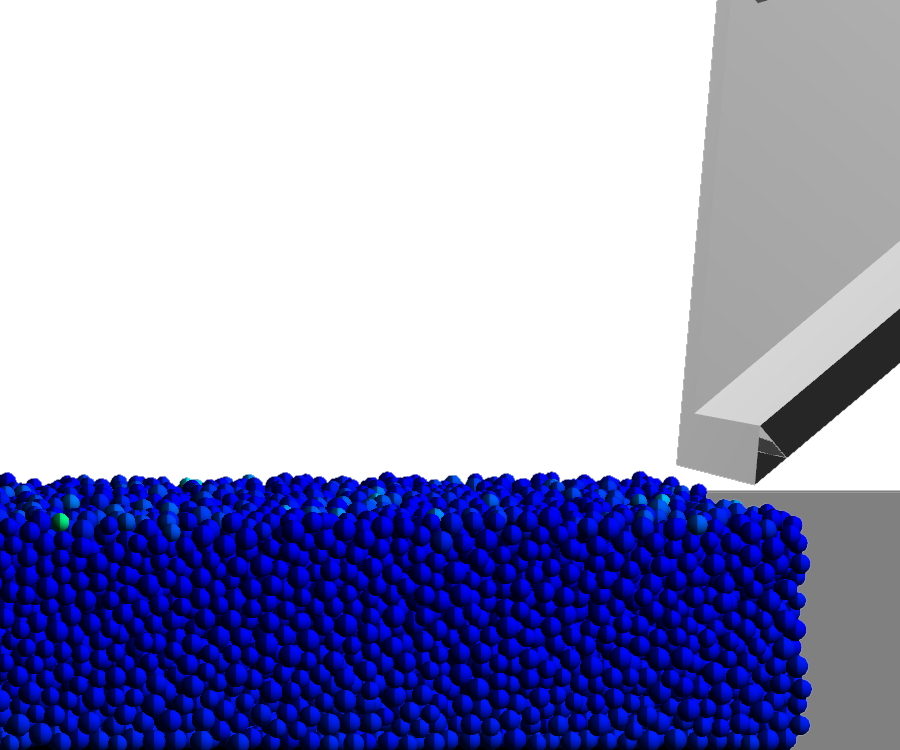}\\
    \includegraphics[width=0.12\textwidth]{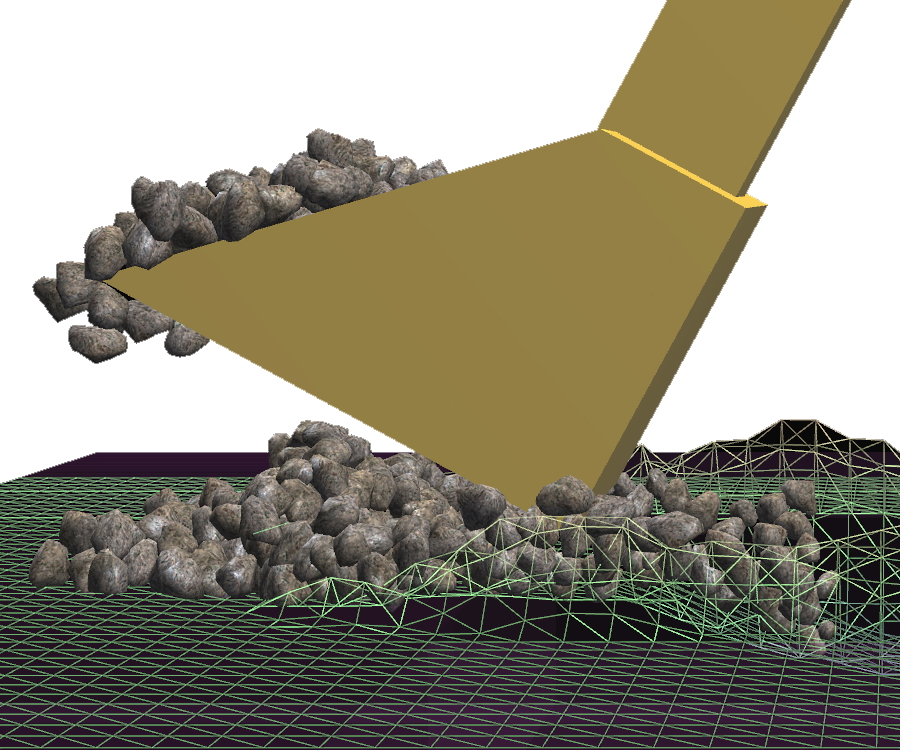}
    \includegraphics[width=0.12\textwidth]{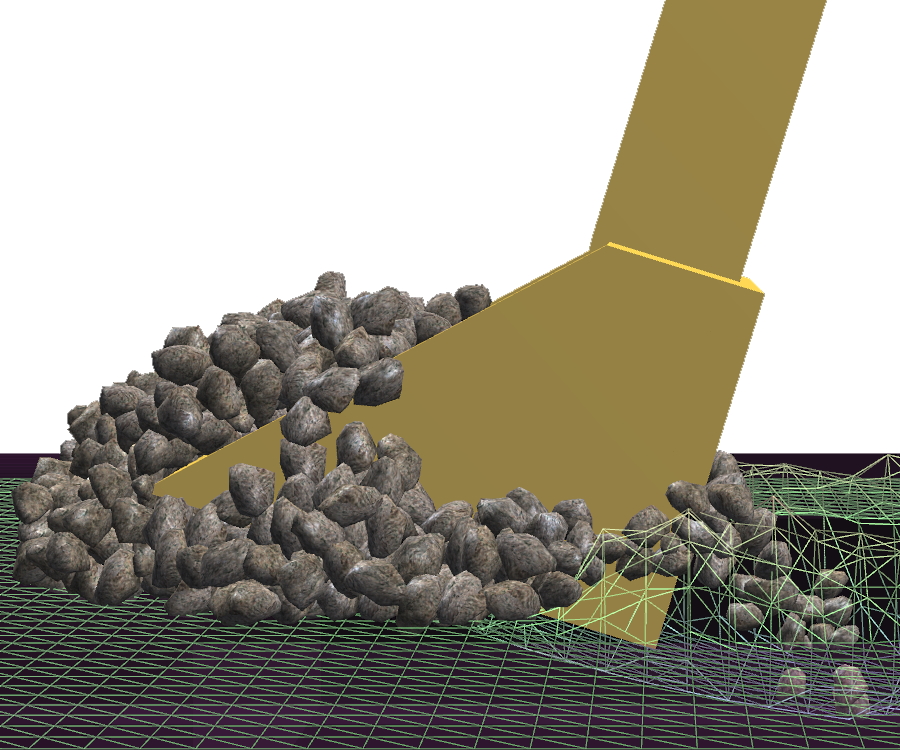}
    \includegraphics[width=0.12\textwidth]{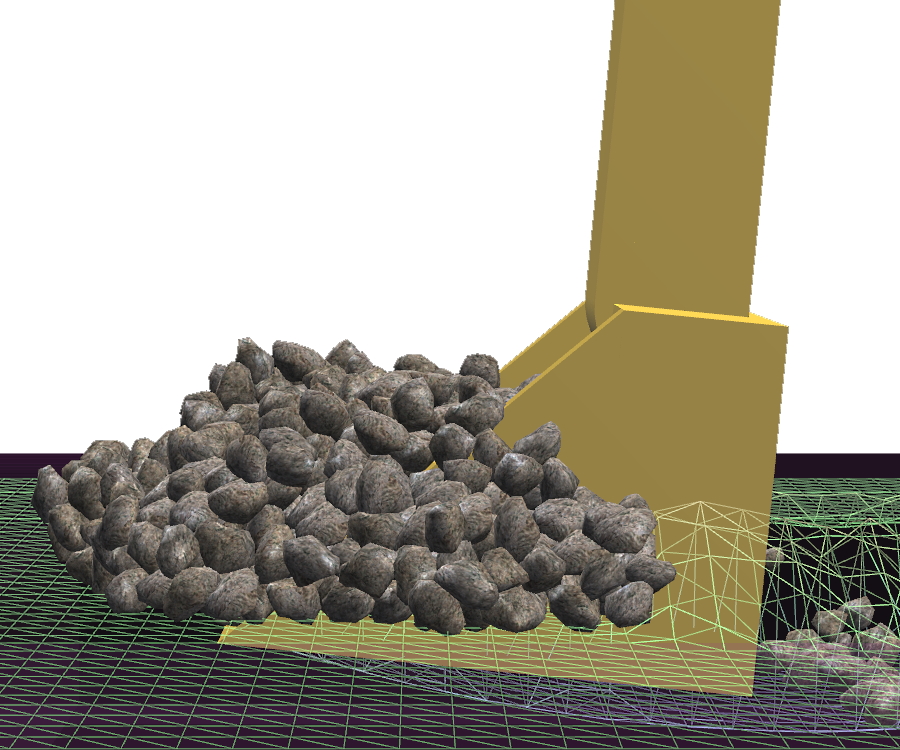}
    \includegraphics[width=0.12\textwidth]{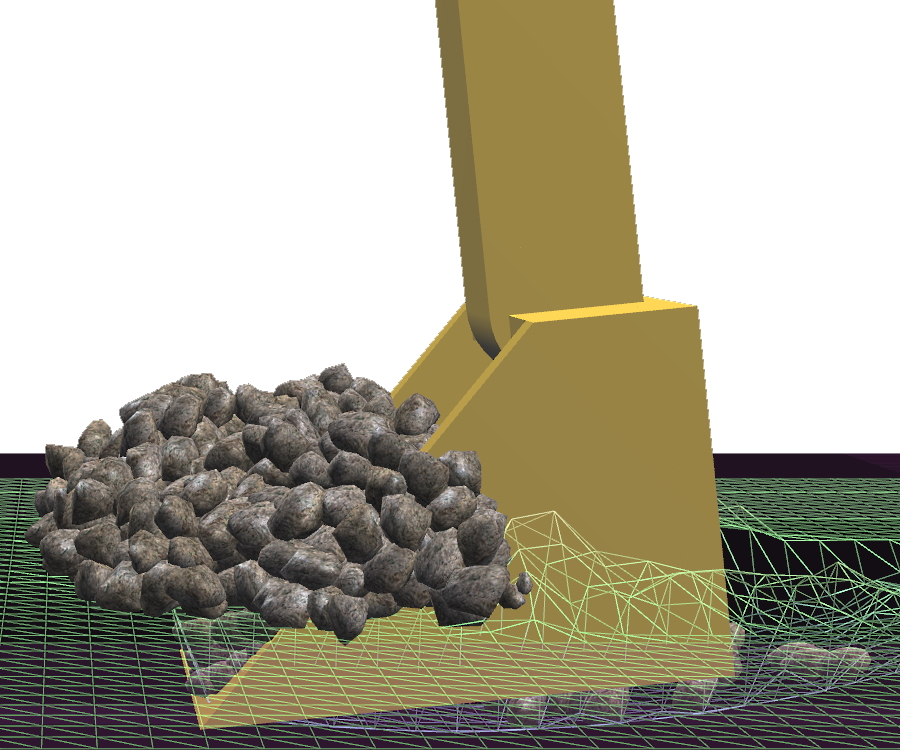}
    \includegraphics[width=0.12\textwidth]{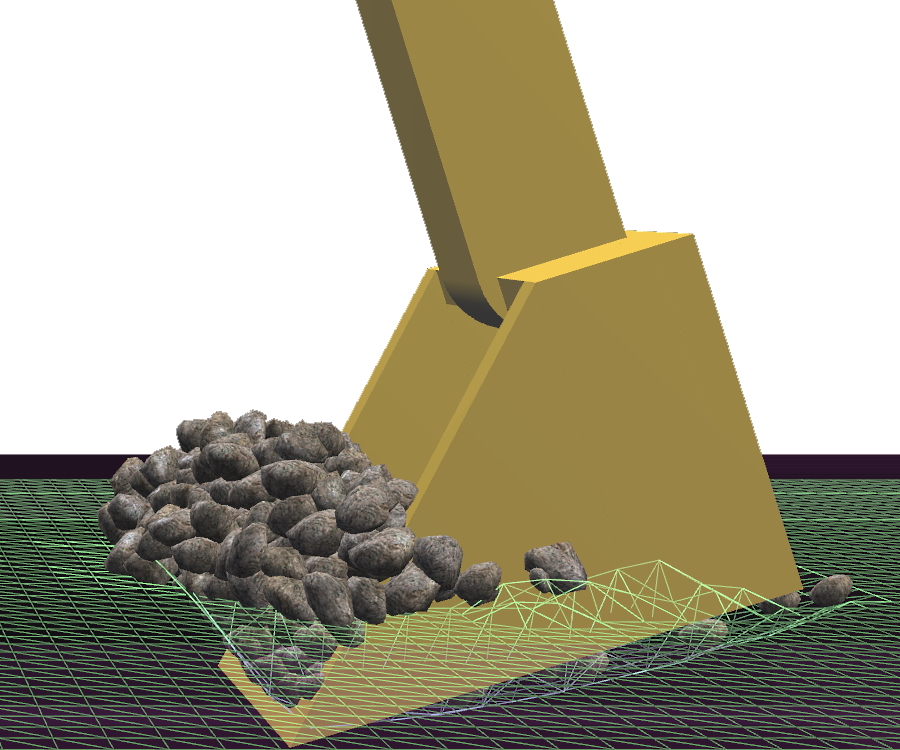}
    \includegraphics[width=0.12\textwidth]{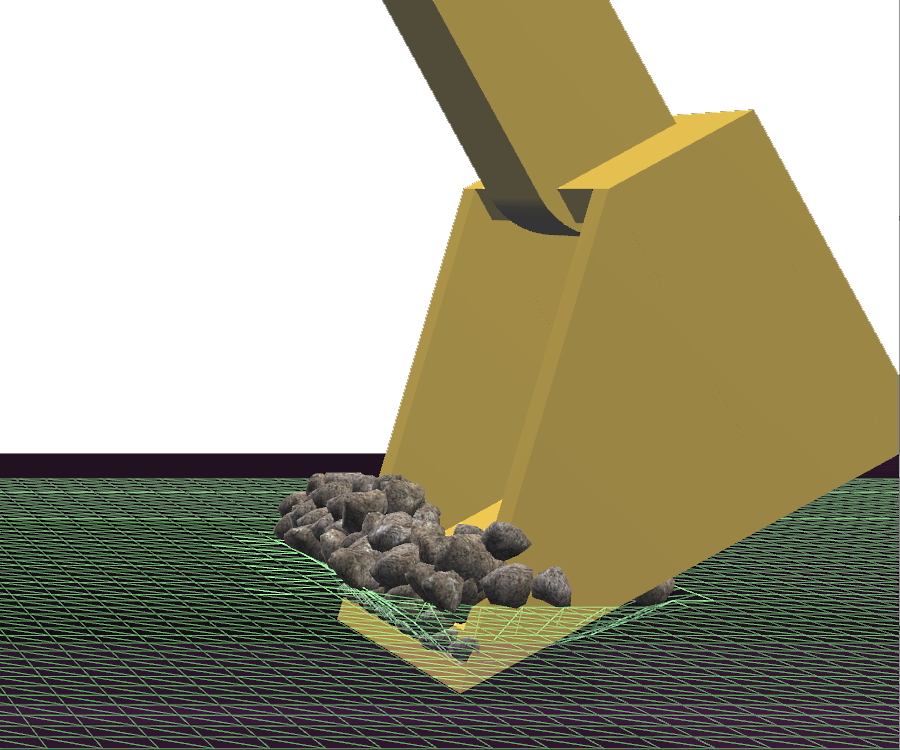}
    \includegraphics[width=0.12\textwidth]{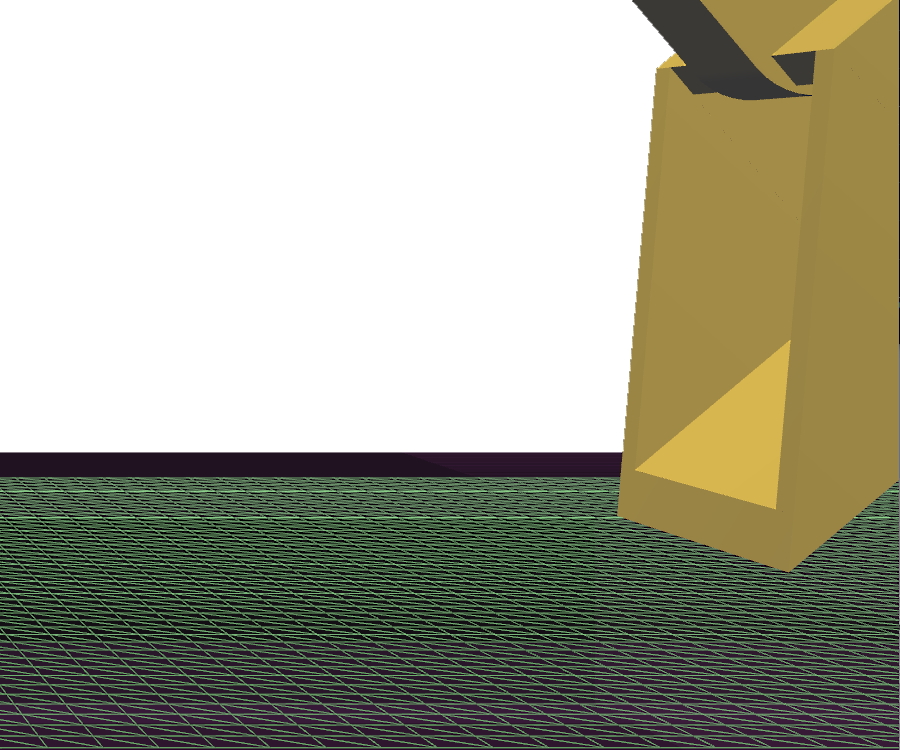}
    \caption{Simulated excavation for comparing the multiscale terrain model (left) and the microscale reference model (right, particles colour coded by particle height position) at time $8$ s and using material \texttt{dirt-1}.
Second and third row show sideview of the high-resolution model (particles colour coded by velocity, blue to red by 0 to 1.5 m/s) and the real-time model, at time 0, 2, 4, 6, 8, 10 and 12 s.  See supplementary \href{https://www.algoryx.se/papers/terrain/}{Video 4}.}
    \label{fig:NDEM_agxTerrain_excavator}
\end{figure}
The excavator model consists of four dynamic bodies and 4 joints: a bucket on the end of an articulated arm, with an outer and inner boom, attached to a revolving base on a static foundation.
The test follows the setup of \cite{obermayr:2014} where only the boom articulation joint is actuated, i.e., the inner arm is held fix.
The bucket is $1.2$ m wide, $1.2$ m long, $0.9$ m high. 
It weighs $100$ kg and has a thickness of $0.05$ mm.
The outer arm is $3.24$ m long and weighs $300$ kg.
A hinge motor drives the arm to rotate at target speed $0.1$ rad/s which translates to $0.4$ m/s at the bucket cutting edge.
For the penetration resistance, the blade's edge is discretized by $24$ teeth with $50$ mm maximum radius and length, and $10$ mm minimum radius.
The penetration force scaling is then calibrated to $2$ for all materials except frictionless soil for wich it is set to zero.
The force exerted on the bucket during the digging can be seen in Fig.~\ref{fig:NDEM_agxTerrain_excavator_frictional} and \ref{fig:NDEM_agxTerrain_excavator_cohesive}.
Observe that the bucket moves in the opposite direction to the bulldozing test, i.e., from right to left.
The bucket start penetrating the soil and fil the bucket at $ x = 2.0 $.
The horizontal force peak around $ x = 0 $ m.
Up to this point, the force in the vertical direction has been dominated by the penetration resistance, see Fig.~\ref{fig:NDEM_agxTerrain_excavator}.
The weight of material inside and in front of the bucket increase gradually, as do the resistance to shearing the soil along the failure plane.
From $ x = 0.5 $ m to $ x = 0 $ m these are in balance and after this the soil weight and the shear resistance dominate the vertical component of the digging force.
At $ x = -1.0 $ m the bucket breaks loose from the ground.
The weight of the material in the bucket can be observed as the residual vertical force $ F_\text{z} $ at $ x = -2.5 $.
There is good agreement between the models for the peak horizontal force and the residual vertical force.
Overall, the time-averaged forces agree within 25 \%.
There is a trend for the multiscale model to overestimate the vertical digging resistance during breakout around $x = -0.5$ m.
From Fig.~\ref{fig:NDEM_agxTerrain_excavator_force} it is not clear wether the penetration force or the separation force is to blame.
As in the bulldozing test, there are significant deviations between the models for the strongly cohesive soils, in particular \texttt{wet-sand-1}.
With the reference model, the force fluctuates heavily between $x = 1.0$ m and $x = - 1.0$ m, see Fig.~\ref{fig:NDEM_agxTerrain_excavator_cohesive}.
The reason for this is the strong cohesion.  
It prevents the soil from flowing freely and gradually fill the interior of the bucket.
Instead, a strong soil beam is formed.
This can be seen in the cross-section image in Fig.~\ref{fig:NDEM_agxTerrain_excavator_cohesive} for \texttt{wet-sand}.
The force fluctuations grow large when the beam hit the interior back wall of the bucket. 
The beam starts to compress, buckle and fail in an irregular manner.
The aggregate in the multiscale model does not represent such modes of deformations and failure and does not produce large force fluctuations. 
The height profiles after an excavation cycle are shown in Fig.~\ref{fig:surface_exc_frictional} and \ref{fig:surface_exc_cohesive}.
For \texttt{gravel-1} and \texttt{sand-1} the width and height of the trench, side berms and the pile are in relatively good agreement. The depth of the trench agrees within 10 mm and the other features within 10 \% on average.
For \texttt{dirt-1} the side berms and pile is roughly 50 \% larger for the reference model,
and for \texttt{wet-sand} and \texttt{cfs-weak} the pile is more than twice as tall. 
The behaviour of the different soil materials under excavation is shown in supplementary \href{https://www.algoryx.se/papers/terrain/}{Video 2}.
It is clear that the strongly cohesive soils are much more cohesive in the reference model compared to the multiscale model.


The active zone model, Eq.~(\ref{eq:failure_plane}), is validated using the reference model.
The induced particle motion when digging in flat and sloping soil of the type \texttt{dirt-1} with an excavator bucket can be seen in Fig.~\ref{fig:excavator_bucket_active_zone_with_angles} and in the supplementary \href{https://www.algoryx.se/papers/terrain/}{Video 5}.
\begin{figure}[h]
    \centering
    \includegraphics[width=0.85\textwidth]{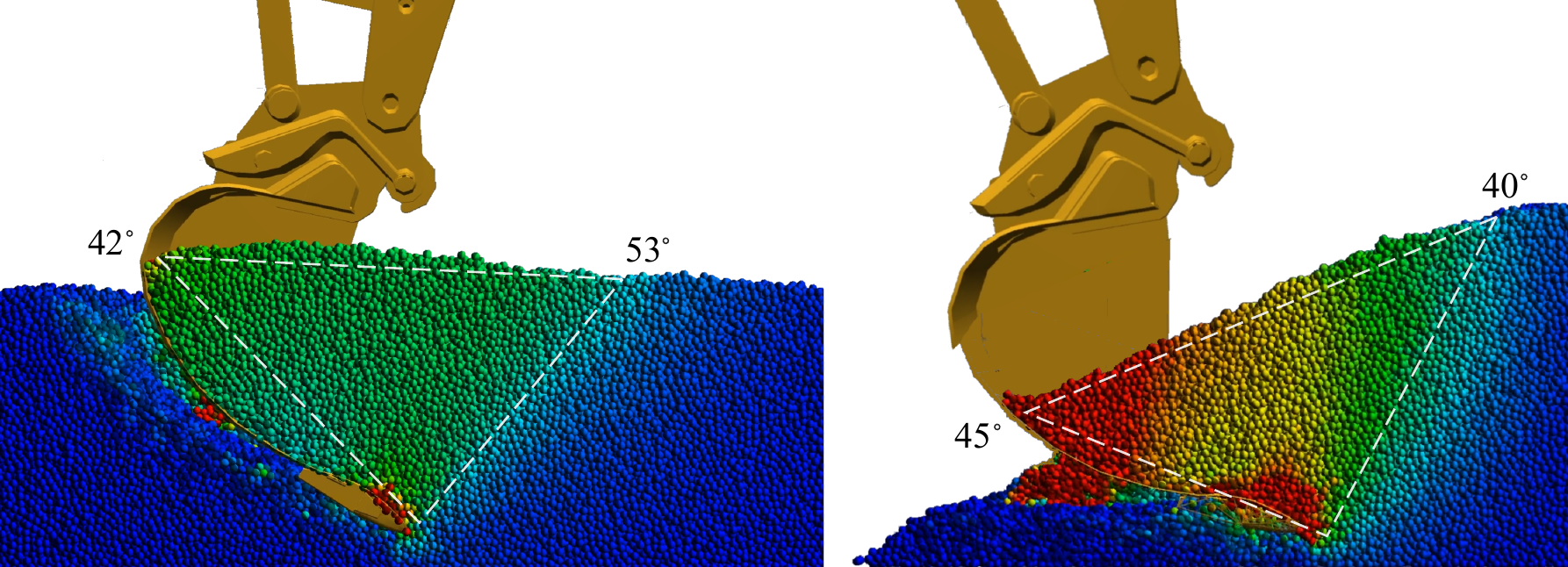}
    \caption{Analysis of the active zone when digging in a flat (left) and sloped (right) soil bed with a complex bucket.  The particles are colour coded by their velocity, with blue to red ranging from $0$ to $1$ m/s. See supplementary \href{https://www.algoryx.se/papers/terrain/}{Video 5}.}
    \label{fig:excavator_bucket_active_zone_with_angles}
\end{figure}
The shape of the wedge formed by the mobilized soil is indicated.
The angle between the dynamic separation plate and the soil surface is estimated to $\beta_\text{flat} = 42^\circ $ and $\beta_\text{sloped} = 45^\circ $ in the two configurations.
Eq.~(\ref{eq:failure_plane}) imply failure planes with inclination $\theta_\text{flat}^\text{model} = 47^\circ$ and $\theta_\text{sloped}^\text{model} = 45^\circ$, respectively.
From the reference model we estimate the failure angles at $\theta_\text{flat}^\text{ref} = 53^\circ$ and $\theta_\text{sloped}^\text{ref} = 40^\circ$, respectively, which correspond to a $12$ \% error.
%

%

We analysed also the computational speed of the multiscale model and the reference model, aware that the results depend on implementation, optimization efforts and the hardware specification\footnote{The tests were run on a desktop computer with Intel Core i7-7800X CPU at 3.5 GHz and 32 GB RAM.} 
The analysis was performed on the excavation test system shown in Fig.~\ref{fig:NDEM_agxTerrain_excavator} and the result is summarized in Table \ref{table:performance}. The number of rigid bodies $N^\text{rb}$ and particles $N^\text{p}$ are given for the multiscale model and reference model. The number of equations, for the velocity update and constraint multiplier, are divided by what is treated by the direct solver, $N^\text{rb}_\text{eq}$, and by the iterative solver, $N^\text{p}_\text{eq}$.
The multiscale model has a real-time factor of $1.5$, i.e., the computational time is $11$ ms per simulation time-step $16.7$ ms.
That means that there is room for simulating a more complex vehicle at real-time speed.
The reference model has a real-time factor of $0.001$, meaning that each $1$ ms time-step on average take $1$ s to compute.
In both cases, solving the MCP is what dominate the computational time.
The price of introducing the aggregate bodies (the main aggregate and a back deformer)
and penetration constraints is an increased number of equations ($N^\text{rb}_\text{eq}$) processed by the direct solver, namely 36 additional equations ($77$\% increase).
The relative speedup of 1500 is an effect of the 16.7 times larger time-step, 67 times fewer equations for the iterative solver, and 50 times less iterations.
To check the sensitivity of the multiscale model to the number of iterations and spatial resolution this was tested, but neither had strong effect on the reaction force.																																										 
%
\begin{table}[ht] 
    \caption{Performance analysis of the excavation test involving $N^\text{rb}$ rigid bodies and $N^\text{p}$ particles.    
    . The direct MCP solver process $N^\text{rb}_\text{eq}$ equations, and the iterative solver $N^\text{p}_\text{eq}$ equations.
    The multiscale model is $1500$ times more efficient due to the smaller number of equations for the iterative solver, fewer iterations, and larger time-step.}
     \footnotesize
      \centering
      \begin{tabular}{|l|cccccc|l|}
      \hline
                            & $N^\text{rb}$    & $N^\text{p}$     & $N^\text{rb}_\text{eq}$    & $N^\text{p}_\text{eq}$    & $N^\text{p}_\text{it}$   & time-step [ms] &  real-time \\
      \hline
      multiscale 	        & 8                & 950 	          & 85                         & 1.8e4	                   & 10                        & 16.7              & 1.5 \\
      reference 	        & 4 	           & 50e3             & 48                         & 1.2e6                     & 500                       & 1                 & 0.001\\
      \hline
      ratio 	            & 2                & 0.019            & 1.77	                   &  0.015                    &  0.02                     & 16.7               & 1500\\
      \hline
  \end{tabular}
  %
  %
  \label{table:performance}
\end{table}   

Finally, the multiscale model is demonstrated in use with full vehicle models in complex earthmoving scenarios.
Fig.~\ref{fig:wl_active_zone} shows a wheel loader digging in a steep wall of soil. See supplementary \href{https://www.algoryx.se/papers/terrain/}{Video 6}.
The active zone in the cutting direction, discretised in five parallel wedges, is visualized in the right image.
There are also active zones on each side of the bucket originating from the soil deformers.
The different size and inclination of the wedges reflect the nonuniform distribution of mass.
The active zone in the digging direction is resolved with particles, but not the side deformers because of low lateral velocity.
The contact points of the aggregate bodies are visualized with orange vectors.
In the left image the lift cylinders are actuated to raise the bucket.
However, the lift force cannot overcome the digging resistance and, consequently, the rear of the wheel loader is lifted from the ground.
This effect would not occur by the weight of the bucket and aggregate alone.
It is necessary to account also for the frictional-cohesive forces between the aggregates, the bucket, and the terrain, to capture the full resistance to breaking out from the wall.

\begin{figure}
    \centering
    \includegraphics[height=0.3\textwidth]{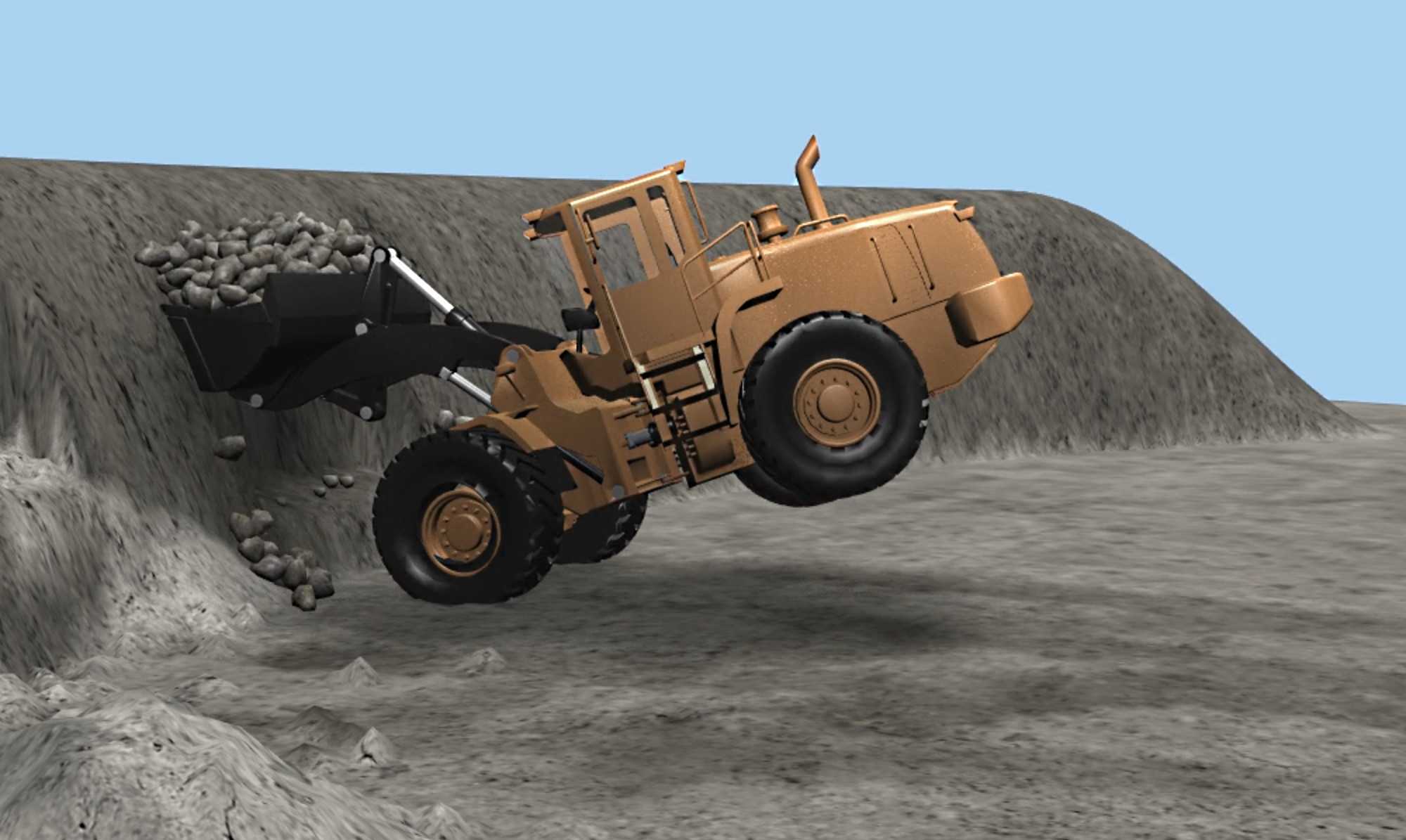}
    \includegraphics[height=0.3\textwidth]{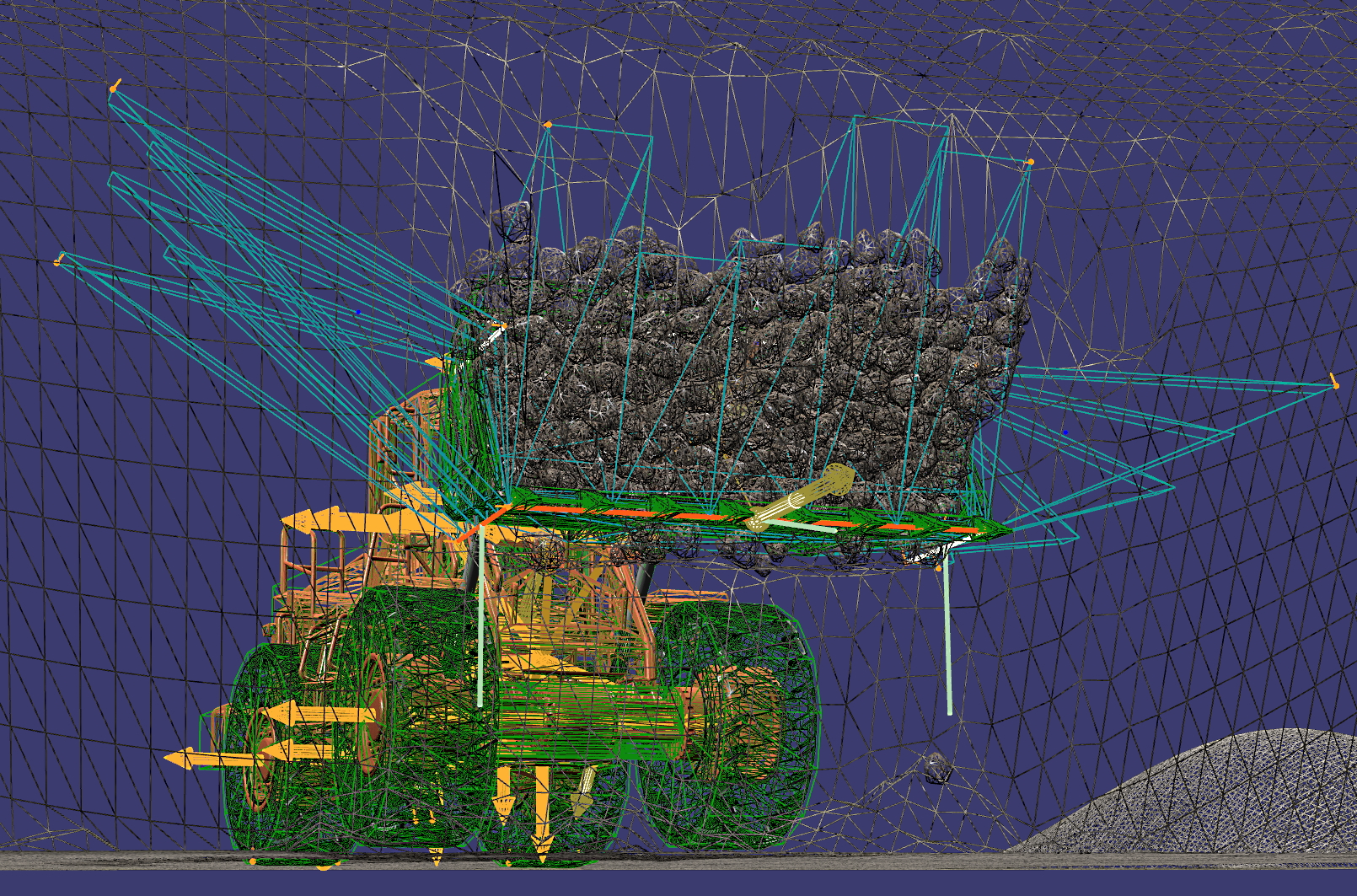}
    \caption{A wheel loader digging in a step pile of soil.  The resistance is too large for breaking out and instead the rear of the vehicle is raised from the ground. The right image shows the active zones, discretised in soil wedges, and the cohesive-frictional contacts between the aggregates and the terrain. See the supplementary \href{https://www.algoryx.se/papers/terrain/}{Video 6}.}
    \label{fig:wl_active_zone}
\end{figure}

\begin{figure}
    \centering
    \includegraphics[height=0.32\textwidth,trim={50mm 15mm 25mm 10mm},clip]{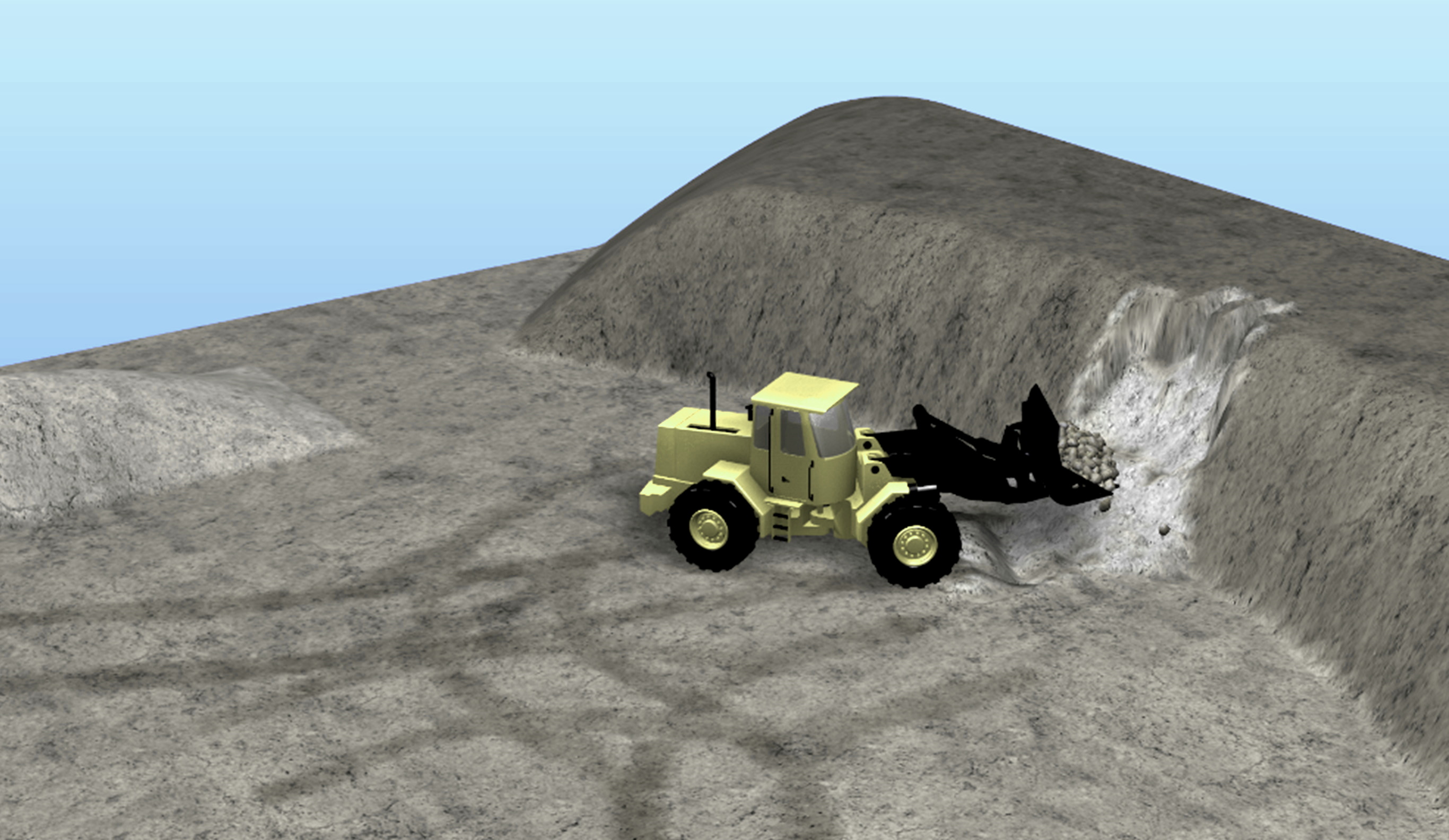}
    \includegraphics[height=0.32\textwidth]{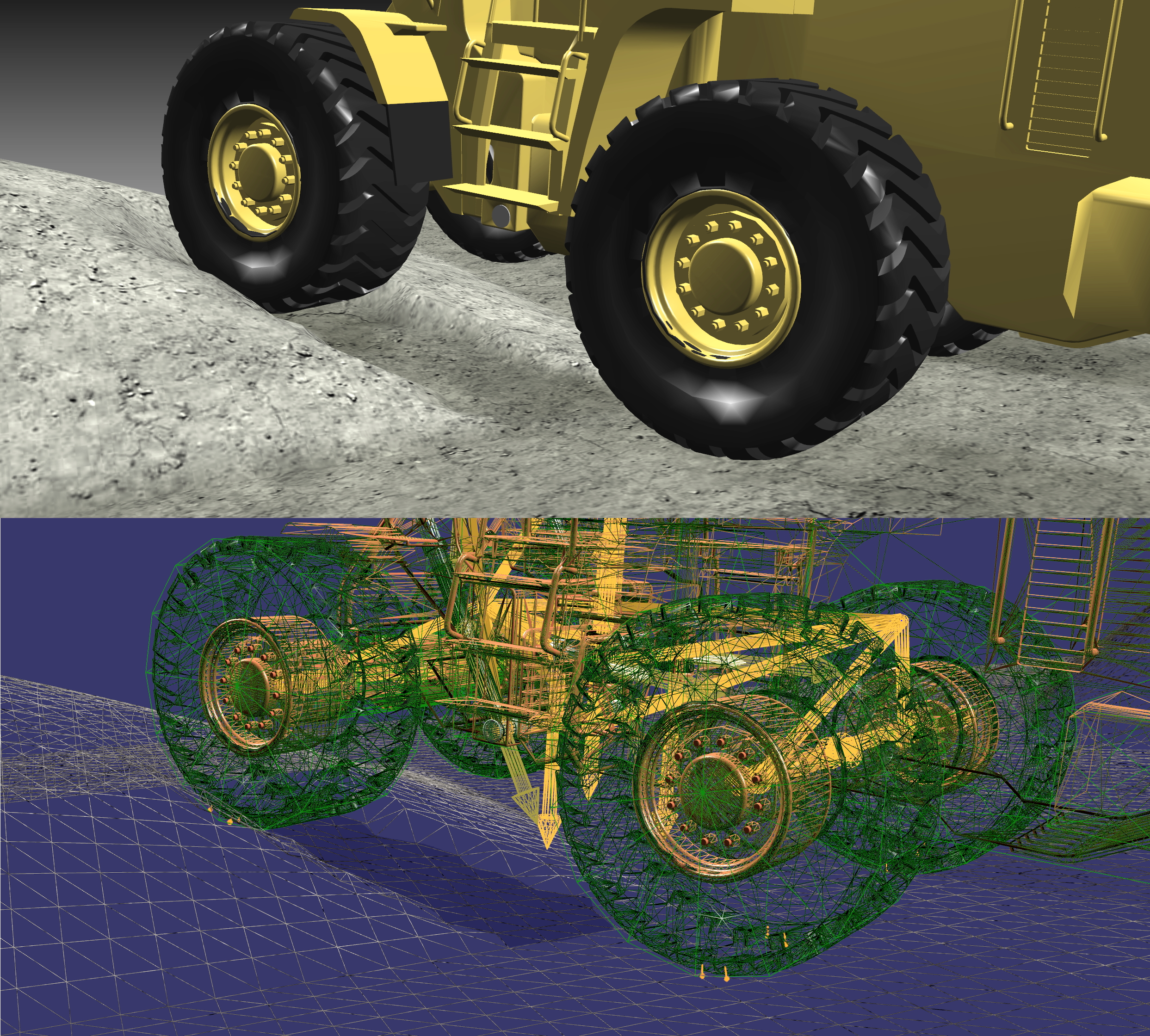}
    \caption{Demonstration of variable soil compaction and swelling. The subsoil stress from the tires cause compaction (dark grey).  Particles that are converted back to resting solid become more loosely packed (light grey) than nominal bank state. }
    \label{fig:variable_compaction}
\end{figure}

Variable soil compaction is illustrated in Fig.~\ref{fig:variable_compaction} and visualized by the intensity of the grey terrain.
Medium grey represents nominal compaction at the bank state, at which the bulk strength parameters are defined.
Light grey represent soil that has dilated due to shear deformations, e.g., have been dug or pushed with the bucket.
That soil has lower compaction and weaker strength according to the swell factor and dilatancy angle, respectively.
Dark grey represent soil that has been compacted, e.g., due to compressive stress from the tires. 
It has higher compaction and strength than the bank state value.
The right images show the wheel loader driving into the left pile of loose soil that is easily compacted.

\begin{figure}
    \centering
    \includegraphics[width=0.8\textwidth]{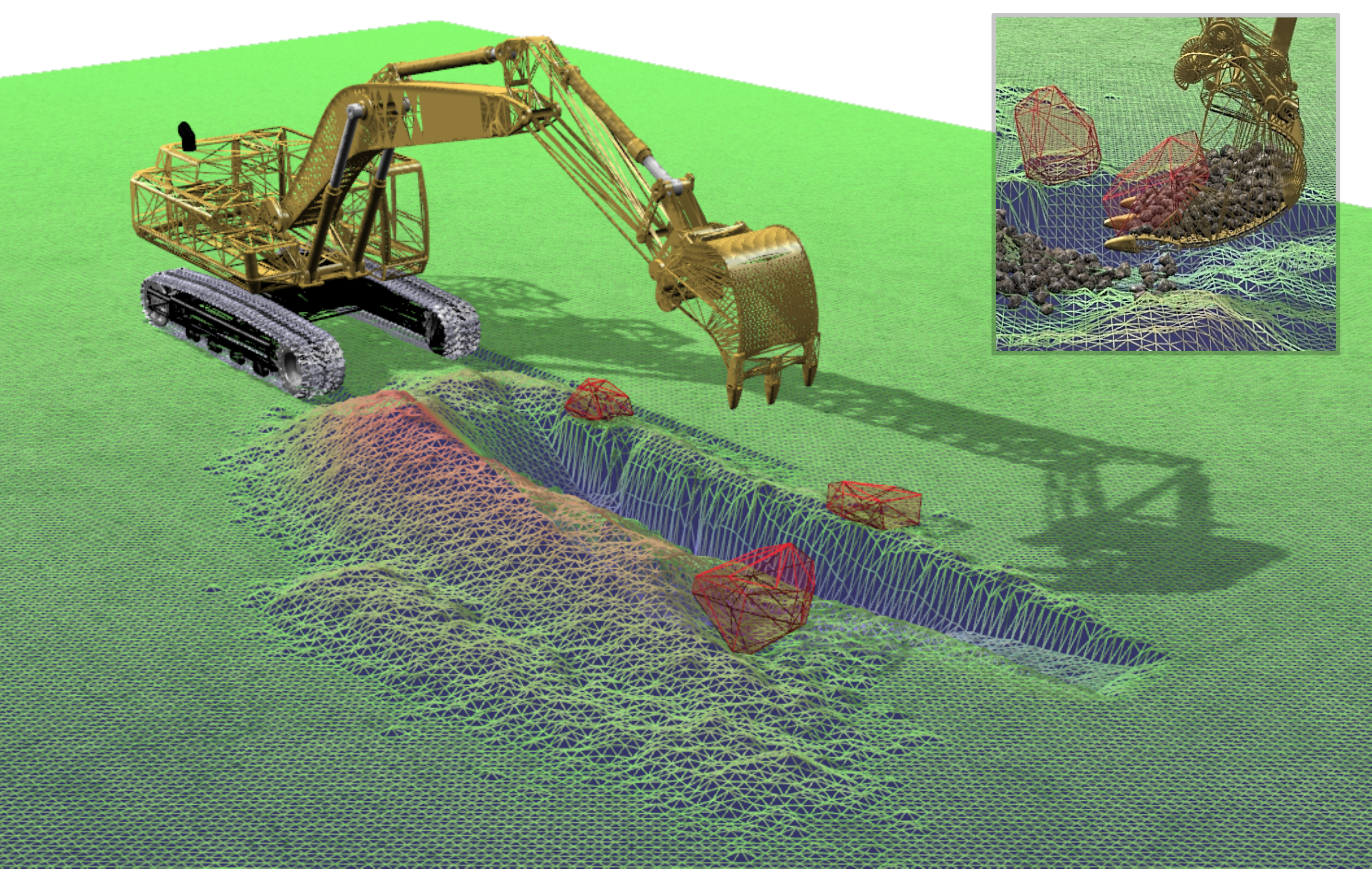}
    \caption{Demonstration interaction with other rigid bodies (red rocks) while digging a trench with a tracked excavator.  See the supplementary \href{https://www.algoryx.se/papers/terrain/}{Video 7}.}
    \label{fig:excavator-trenching-rocks}
\end{figure}

The involvement of additional rigid bodies interacting both with the earthmoving machine and the terrain is demonstrated in Fig.~\ref{fig:excavator-trenching-rocks} with a fullsized tracked excavator digging a deep trench.
The rocks contact directly with the excavator bucket, via the direct solver, but also with the particles, through the iterative solver.
The rocks force the soil to distribute around the rock inside the bucket or pile up around the rocks on the ground.
The terrain is initialized with a high state of compaction such that the trench can sustain steep side walls.
See the supplementary \href{https://www.algoryx.se/papers/terrain/}{Video 7}.

\begin{figure}
    \centering
    \includegraphics[width=0.75\textwidth,trim={0mm 0mm 0mm 0mm},clip]{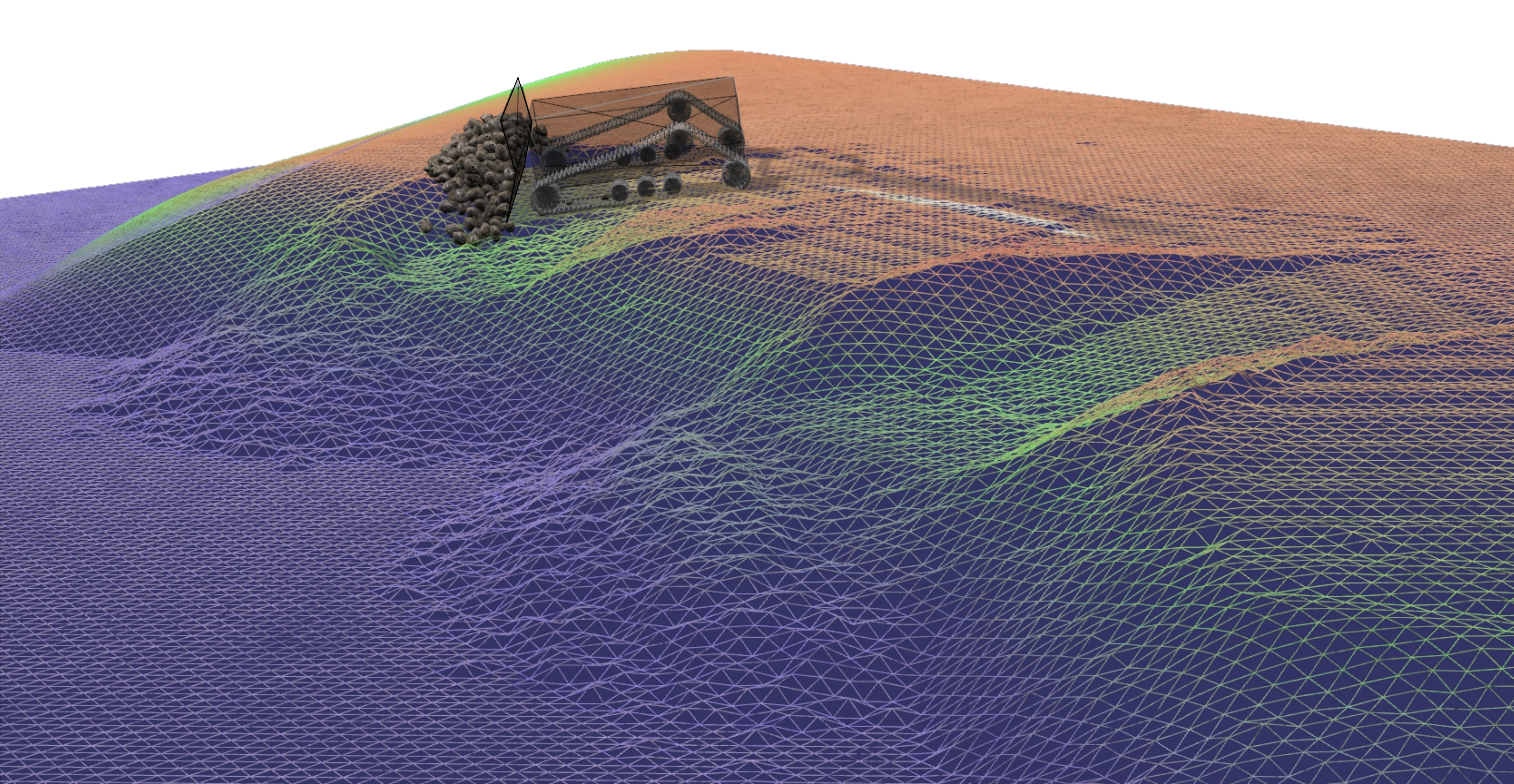}
    \caption{Demonstration of high-precision bulldozing in a large terrain using a tracked vehicle. See the supplementary \href{https://www.algoryx.se/papers/terrain/}{Video 8}.}
    \label{fig:bulldozer-terrain_2}
\end{figure}

The capability of representing large terrains, everywhere deformable, is demonstrated with the bulldozing example in Fig.~\ref{fig:bulldozer-terrain_2} and the supplementary \href{https://www.algoryx.se/papers/terrain/}{Video 8}.
Observe that the terrain is cut precisely at the dozer's cutting blade, with a precision much finer than the coarse particles, and match the vertical motion of the blade. 
The effect of bulldozer chassis and blade oscillations can be observed as wave pattern in the cut terrain surface.

\section*{Discussion}
The results show good performance of the multiscale model and mostly fair agreement with the the reference model, needing little calibration.
There are, however, a number of limitations with the method as well.
The most significant deviation occur for strongly cohesive soils, where the particle cohesion appear much stronger in the reference model than in the multiscale model.
Furthermore, the shape of the active zone is a crude approximation. 
It is clear from both the underlying theory and numerical studies that the shear failure surface is not a plane and the active zone is not well approximated by a wedge, or several parallel wedges. 
The deviation in the digging resistance at breakout indicate that the slope of the failure surface is not correct, or that the assumption of shear band localization is false.
The active zone model in the present paper requires some manual work with defining edges, direction vectors, teeth etc. 
Ideally, the dependency on such geometric features should emerge from a model that merely takes the 3D geometry of a digging tool as input.
The damping coefficient of the rigid aggregate has not been derived from the underlying theory or scrutinized using the microscale reference model.
An alternative to improving the active zone model by analytical means is to take a data-driven approach.
Following Wallin \cite{wallin:2020:ddm}, it is possible to train a model, from resolved reference simulations, to rapidly predict the digging resistance and the flow field in a granular bed from a digging tool of particular 3D shape without explicitly defining any cutting edges, direction vectors or teeth. 
Potentially this can be used for predicting the shape and mechanical strength of the active zone with higher precision and generality, and possibly also the soil displacements.
The drawback of the data-driven approach is the need for running many simulations in advance, covering a wide range of soils, terrain shapes, and tool trajectories to have a useful model. 

The computational bottleneck in the tested implementation is the mesoscale particle simulation. 
It is run synchronously on CPU with the macroscale simulation using fix and identical time-step.
This is not necessary but made so for simplicity because the simulation software, AGX Dynamics, is designed for strong coupling between the multibody dynamics and the (nonsmooth) DEM simulation.
It might be worth investigating alternative, possibly asynchronous, methods for simulating the active soil, e.g., smooth DEM on GPU.

The mesoscale model support plastic compaction of the solid terrain but not shear deformations.
That is a major limitation for simulating deep ruts and vehicles or other objects sinking or being buried in the terrain.

Finally, the presented method relies on a having a soil libary where particle paramaters are pre-calibrated to match the bulk mechaical paramaters.
The current library, involving 15 different soils, can easily be extended to a wider range of more soils, e.g., to include the over 100 virtual soil samples that was mapped in \cite{wiberg:2020:dem}.
From such data-sets it is possible to identify mapping functions $\bm{p}_\text{p} \to \bm{p}_\text{b}$, such that new soils can be introduced on-the-fly.
That is important for making machine learning models robust and transferable from one domain to another, e.g., from the simlation domain to the physical domain, using domain randomization \cite{Tobin2017}. 
However, the current model does not support inhomogeneous soil or mixing of two or more soils. That extension is left for future development.

A weakness in the presented study is that the validation and calibration is not based on field measurements but rely on reference simulations of relatively high resolution. 
Field experiments are left for future work, for example, using also the methodology in \cite{Zhao2020} for determining the soil’s bulk parameters using the earthmoving equipment.

It is interesting to also compare the investigated method with other multiscale methods for vehicle-terrain simulation. 
A method for hierarchical multiscale vehicle-terrain simulation is investigated in \cite{Yamashita2019}. 
It combines vehicle multibody dynamics with a finite element terrain model, where parallel DEM simulations of relatively small representative volume elements at the quadrature points in place of an assumed elastoplastic constitutive law. 
This is an attractive alternative to relying on an idealized model for predicting the failure surface around a digging tool, or the sub-soil deformations under tyres, but real-time performance appear remote at the present time.																																											 
\section*{Conclusion}
It has been found possible to simulate earthmoving operations in real-time with a model that captures the rigid multibody dynamics of the equipment, the reaction forces from the terrain, and much of its deformations and flow dynamics.
With a multiscale model the terrain's active zones are represented simultaneously as a rigid body, as particles and as a continuum. 
A direct solver is applied to the multibody system for high numerical precision and an iterative solver to the particle system for scalability. 
The models are made dynamically consistent through a soil library that is calibrated in advance using a reference model simulated at high-resolution.
The performance, realism, and capability to represent a wide range of materials and scenarios make the solution suitable for simulation-based development control and automation of earthmoving equipment.


\begin{backmatter}

\section*{Availability of data and materials}
The datasets used and/or analysed during the current study are available from the corresponding author on reasonable request. 

\section*{Competing interests}
TB and MS are affiliated to Algoryx Simulation that develop the software used in the study.

\section*{Funding}
Algoryx Simulation, eSSENCE (The Swedish e-Science Collaboration), and Umeå University.

\section*{Author's contributions}
The theory and algorithms was developed jointly by the authors. The implementation was made primarily by TB and SN. The simulation analysis was carried out primarily by MS. The authoring of the paper was lead by MS.

\section*{Acknowledgements}
Anders Backman, Viktor Knutsson, Fredrik Nordfelth and Martin Nilsson at Algoryx Simulation has made important contributions in the software implementation, testing and optimization.



\def\cprime{$'$}

\end{backmatter}
\end{document}